\documentclass[a4paper,oneside,reqno]{amsart}

\usepackage[english]{babel}
\usepackage[utf8]{inputenc} 
\usepackage[scaled]{helvet} 
\usepackage{courier} 
\usepackage{eulervm} 
\usepackage[bb=boondox,bbscaled=1.05,scr=dutchcal]{mathalfa}	%
\usepackage{dsfont}
\normalfont

\usepackage{tabularx,longtable,booktabs} 
\usepackage{caption,subcaption} 
\usepackage{fullpage} 
\usepackage[english]{varioref} 
\usepackage[dvipsnames]{xcolor} 
\usepackage{bookmark} 
\usepackage{tikz}
\usetikzlibrary{decorations.markings}
\usepackage{enumitem} 

\setlist[itemize,1]{leftmargin=2em}
\setlist[enumerate,1]{leftmargin=2em}

\usepackage{textcomp}															

\usepackage{hyperref} 
\definecolor{webbrown}{rgb}{0.65, 0.16, 0.16} 
\hypersetup{
colorlinks=true, linktocpage=true, pdfstartpage=1, pdfstartview=FitV,breaklinks=true, pdfpagemode=UseNone, pageanchor=true, pdfpagemode=UseOutlines,plainpages=false, bookmarksnumbered, bookmarksopen=true, bookmarksopenlevel=1,hypertexnames=true, pdfhighlight=/O,urlcolor=webbrown, linkcolor=RoyalBlue, citecolor=ForestGreen}

\usepackage[textsize=footnotesize]{todonotes}	
\presetkeys%
		{todonotes}%
		{color=Apricot}{}%
\usepackage[style=alphabetic,maxalphanames=4,giveninits,sorting=nyt,%
maxbibnames=99,backend=biber]{biblatex}
\renewbibmacro{in:}{}
\usepackage{csquotes}

\newcommand\encadremath[1]{\vbox{\hrule\hbox{\vrule\kern8pt 
\vbox{\kern8pt \hbox{$\displaystyle #1$}\kern8pt} 
\kern8pt\vrule}\hrule}}
\def\enca#1{\vbox{\hrule\hbox{
\vrule\kern8pt\vbox{\kern8pt \hbox{$\displaystyle #1$}
\kern8pt} \kern8pt\vrule}\hrule}}

\newcommand\figureframex[3]{
\begin{figure}[bth]
\hrule\hbox{\vrule\kern8pt 
\vbox{\kern8pt \vbox{
\begin{center}
{\mbox{\epsfxsize=#1.truecm\epsfbox{#2}}}
\end{center}
\caption{#3}
}\kern8pt} 
\kern8pt\vrule}\hrule
\end{figure}
}
\newcommand\figureframey[3]{
\begin{figure}[bth]
\hrule\hbox{\vrule\kern8pt 
\vbox{\kern8pt \vbox{
\begin{center}
{\mbox{\epsfysize=#1.truecm\epsfbox{#2}}}
\end{center}
\caption{#3}
}\kern8pt} 
\kern8pt\vrule}\hrule
\end{figure}
}

\makeatletter
\@addtoreset{equation}{section}
\makeatother
\newtheorem{theorem}{Theorem}[section]

\newtheorem{remark}{Remark}[section]
\newtheorem{proposition}{Proposition}[section]
\newtheorem{lemma}{Lemma}[section]
\newtheorem{corollary}{Corollary}[section]
\newtheorem{definition}{Definition}[section]
\newtheorem{example}{Example}[section]
\def\br{\begin{remark}\rm\small}
\def\er{\end{remark}}
\def\bt{\begin{theorem}}
\def\et{\end{theorem}}
\def\bd{\begin{definition}}
\def\ed{\end{definition}}
\def\bp{\begin{proposition}}
\def\ep{\end{proposition}}
\def\bl{\begin{lemma}}
\def\el{\end{lemma}}
\def\bc{\begin{corollary}}
\def\ec{\end{corollary}}
\def\beaq{\begin{eqnarray}}
\def\eeaq{\end{eqnarray}}
\def\bex{\begin{example}}
\def\eex{\end{example}}

\newcommand{\beq}{\begin{equation}}
\newcommand{\eeq}{\end{equation}}
\newcommand{\bea}{\begin{eqnarray}}
\newcommand{\eea}{\end{eqnarray}}

\newcommand{\examplename}{Example}
\newcommand{\exampleref}[1]{\hyperref[#1]{\examplename~\ref*{#1}}}

\renewcommand{\and}{{\qquad {\rm and} \qquad}}

\newcommand{\Tr}{\operatorname{Tr}}
\newcommand{\tr}{\operatorname{tr}}

\newcommand{\Res}{\mathop{\,\rm Res\,}}
\newcommand{\ord}{\operatorname{ord}}

\newcommand{\CC}{{\mathbb C}}
\newcommand{\ZZ}{{\mathbb Z}}
\newcommand{\RR}{{\mathbb R}}

\newcommand{\td}[1]{{\tilde{#1}}}

\newcommand{\mathcalP}{{\mathcal P}}

\newcommand{\ii}{{\mathrm{i}}}
\newcommand{\e}{{\,\rm e}\,}

\newcommand{\Pint}{{\int\kern -1.em -\kern-.25em}}

\renewcommand{\Re}{{\mathrm{Re}}}
\renewcommand{\Im}{{\mathrm{Im}}}

\newcommand{\curve}{{\Sigma}}
\newcommand{\Field}{{\mathbb C}}

\newcommand{\genus}{{\mathfrak g}}

\newcommand{\acycle}{{\mathcal A}}
\newcommand{\bcycle}{{\mathcal B}}

\newcommand{\modsp}{{\mathcal M}}

\newcommand{\Newt}{{\mathcal N}}
\newcommand{\dNewt}{{\partial\mathcal N}}
\newcommand{\Newtint}{\displaystyle{\mathop{{\mathcal N}}^{\circ}}}

\begin{document}

\renewcommand{\hbar}{\hslash}

\title[Existence of Boutroux curves, $g$-functions and spectral networks from Newton's polygon]{Existence of Boutroux curves, $g$-functions and spectral networks from Newton's polygon}

\author[B. Eynard]{B.~Eynard}%
\address[B. Eynard]{
	Universit\'e Paris-Saclay, CNRS, CEA, Institut de Physique Th\'eorique, Gif-sur-Yvette, France \emph{\&} %
	CRM, Centre de Recherches Math\'ematiques de Montr\'eal, Universit\'e de Montr\'eal, QC, Canada.
	}%
\email{bertrand.eynard@ipht.fr}
\author[S. Oukassi]{S.~Oukassi}%
\address[S. Oukassi]{
	Universit\'e Paris-Saclay, CNRS, CEA, Institut de Physique Th\'eorique, Gif-sur-Yvette, France.  %
	}%
\email{soufiane.oukassi@ipht.fr}

\begin{abstract}
We prove the existence of an algebraic plane curve of equation $P(x,y)=0$, with prescribed asymptotic behaviors at punctures, and with the Boutroux property, namely, periods have vanishing real part, i.e, $\Re(\int_\gamma y dx)=0$ for every closed loop $\gamma$. 
This has applications in the Riemann-Hilbert problem, in random matrix theory, in spectral networks, in WKB analysis and Stokes phenomenon, in algebraic and enumerative geometry, and many applications in mathematical physics.
From Newton's polygon we can define an affine space such that there exists always a Boutroux curve. This result is applied to random matrix and asymptotic theory, in which a key ingredient is called the $g$-function, the function $g(x)=\int_o^x Y dX$ is a $g$-function precisely if and only if the algebraic plane curve is a Boutroux curve.  
\end{abstract}

\maketitle



\section{Introduction}
\label{sec:Intro}
In all what follows, $P(x,y)\in\CC[x,y]$ is a bivariate complex polynomial.
$P'_x(x,y)$ and $P'_y(x,y)$ denotes its partial derivatives
\beq
P'_x(x,y) = \frac{\partial}{\partial x}P(x,y), \quad
P'_y(x,y) = \frac{\partial}{\partial y}P(x,y).
\eeq

\subsection{Purpose and results}

Let $P(x,y)\in\CC[x,y]$, we define the Newton's polygon as a convex polytope $\Newt\subset \ZZ_+\times \ZZ_+$
\begin{equation}
    \Newt=\{(i,j) \ | \ \mathcal P_{i,j}\neq 0\}. 
\end{equation}
It is a vector space $\CC[\Newt] \sim \CC^{\#\Newt} $.

We define the interior of the Newton's polygon (but shifted by $(-1,-1)$):
\beq
\Newtint=\{(i-1,j-1) \ | \ (i,j) \ \text{strictly interior of convex envelope} \}.
\eeq
Let $\mathcalP\in\CC[\Newt] $ a once for all fixed bivariate polynomial. We want to study the set of bivariate polynomials that differ from $\mathcalP$ just from the interior, i.e. the affine space:
\beq
\modsp=\mathcalP+\CC[\Newtint].
\eeq
Fixing the exterior part (i.e. $\mathcalP$) is equivalent to fixing  the asymptotic behaviors of solutions $y=Y(x)$  of $P(x,y)=0$ at points where $x$ and/or $y$ tend to $\infty$ (called punctures).

Our goal is to prove that there exists $P_{\text{Boutroux}}\in\modsp$, that has the Boutroux property:
\beq
\forall \ \gamma = \text{Jordan loop} \quad
\quad  \Re \oint_\gamma Y dx=0.
\eeq
Boutroux curves have many applications:

- For example in asymptotic theory, the Riemann-Hilbert method of \cite{deift1992steepest,deift1999} strongly relies on the existence of a so-called $g$\textit{-function}, whose differential $d g=y dx$ has prescribed asymptotic behaviors and has the Boutroux property.
In some sense this article provides a {\bf{theorem of existence of $g$-functions}}.

- Also in geometry, cutting surfaces along some ``horizontal trajectories" is a way to make combinatorial models of moduli spaces of surfaces. This was used by Strebel, Harrer-Zagier, Kontsevich, Penner, Thurston, and many others \cite{Strebel,Thurston,Kon92,Zagier1986,penner2003cell,penner2003decorated}.

- A seminal work of Gaiotto, Moore, and Neitzke  relates WKB asymptotic expansion to ``spectral networks" \cite{gaiotto2011wallcrossing}, and is again closely related to Boutroux curves.

 All these authors have considered foliations of surfaces by cutting along ``horizontal trajectories". Horizontal trajectories of the differential form $\frac{1}{2\pi\ii}ydx$ give a good foliation, typically when it has the Boutroux property. The existence of this differential, and thus the existence of this foliation, is what gives the bijection between the combinatorial moduli space and the geometric moduli space. During an IHES seminar, M.~Kontsevich was quoted saying that if proved, ``this theorem of existence would be the most useful tool possible".

\medskip

Our method is to obtain the Boutroux curve by minimizing some real function called ``Energy" $F:\modsp\to\RR $,
which can be interpreted as the ``area" of the curve.
In other words, the Boutroux curve will be the ``minimal surface".

The plan of the article is:

 Section \ref{sec:Intro} is a brief introduction to the property of Boutroux curves, and also to different applications of this property ranging from the existence of $g$-functions to foliations of surfaces by the so-called spectral networks.

 Section \ref{sec:AlgCurves}: we recall basic notions about Newton's polygon, algebraic Riemann surfaces and plane curves. We shall in particular introduce  ``times" and ``periods".

 Section \ref{sec:energy}: we define the energy as a regularized area of the surface, by removing some small discs around punctures and adding  appropriate correction terms. Then proving that the energy is bounded from below, continuous and with tight compact level sets. This will imply the existence of a minimum (the intersection of all decreasing compact level sets is a non-empty compact).
In addition we rewrite the energy as a function of times and periods (this requires choosing a basis of cycles on the curve, called a ``marking"). 

Section \ref{sec:BC}: we can finally prove that the minimum of the energy is a Boutroux curve.

Section \ref{sec:spnetwork1}: we associate a spectral network to a Boutroux curve. This is in fact done in two ways. The first kind of the spectral network is similar to the notion of ``Strebel graph", and provides a canonical atlas of the curve, made of rectangular pieces.

Section \ref{sec:spnetwork2}: the second kind of spectral network associated to a Boutroux curve, is the one useful for random matrices, and spectral networks as in \cite{gaiotto2011wallcrossing}.

Section \ref{sec:apps}: we study examples of applications, like Strebel graphs, and random matrices.

Section \ref{sec:conc}: we gather a number of concluding remarks.

\section{Newton's polygon and algebraic curves}
\label{sec:AlgCurves}

\subsection{Newton's polygon}

From now on we choose $\mathcalP= \sum_{i,j} \mathcalP_{i,j} x^i y^j \in \CC[x,y] $ a bivariate polynomial, fixed once for all.
Let $\Newt=\{(i,j) \ | \ \mathcal P_{i,j}\neq 0\} $. 

We require that $\Newt$ has at least three points non aligned. 

We define $\mathcalP_d(x)$ the coefficient of $y^d$ the highest degree term in $y$ of $\mathcalP$.

Our goal is to study the space of bivariate polynomials that differ from $\mathcalP$ only by ``interior" coefficients.

\bd[Newton's polygon]
The Newton's polytope
\beq
\Newt:=\{(i,j)\in \ZZ^2 \ | \ \mathcal P_{i,j}\neq 0\} .
\eeq
is a set of points in $\mathbb Z_+\times \mathbb Z_+$. 
We define its completion with all integer points enclosed within its convex envelope:
\beq
\bar\Newt := \{(i,j) \in \ZZ\times \ZZ \,|\,(i,j)\in\,\text{inside or on the boundary  of the convex envelope of }\,\Newt\}.
\eeq

We define its \textbf{interior} $\Newtint\subset\mathbb Z\times \mathbb Z$, shifted by $(-1,-1)$:
\beq
\Newtint := \{(i,j)\in \bar\Newt \ |\,(i+1,j+1)\in\,\text{strictly interior of the convex envelope of }\,\Newt\}.
\eeq
and its \textbf{boundary} (the integer points of the convex envelope)
\beq
\dNewt:=\bar{\Newt}\setminus(\Newtint+(1,1)),
\eeq
and we define
\beq
\Newt''':=\{(i,j)\in \bar\Newt \ | \ (i+1,j+1) \in \dNewt \} = \text{``3rd kind points"}
\eeq
\beq
\Newt'':=\{(i,j)\in \bar\Newt \ | \ (i+1,j+1) \  \notin \bar{\Newt}\}= \text{``2nd kind points"}.
\eeq
The points of $\Newtint$ are also called ``1st kind".
\begin{itemize}
\item 1st kind $\Newtint$ := interior : $(i+1,j+1)\in$ strict interior 
\item 3rd kind $\Newt'''$ := boundary : $(i+1,j+1)\in$ boundary
\item 2nd kind $\Newt''$ := exterior : $(i+1,j+1)\in$ exterior
\end{itemize}
\ed
To recall why they are called 1st, 2nd or 3rd kind, cf lectures notes \cite{eynardlecturesRS}.

We want now to study the moduli space of polynomials sharing the same exterior and boundary as $\mathcalP$, i.e. differ only by interior points.

\bd[Moduli space of a Newton polygon]
\label{def:modsp}

If $\deg \mathcalP_d(x)=0$, we let
\beq
\CC[\Newtint] := \left\{Q(x,y)\in \Field[x,y] \,|\, Q=\sum_{(i,j)\in \Newtint} Q_{i,j} x^i y^j \right\} .
\eeq
It is a complex vector space of dimension $\#\Newtint$, or a real vector space of dimension $2\#\Newtint$.

In the general case, if $\deg \mathcalP_d(x)>0$, let
\bea
\CC[\Newtint] &:=& 
\left\{Q(x,y)\in \Field[x,y] \,|\, Q=\sum_{(i,j)\in \Newtint} Q_{i,j} x^i y^j \right\} \cr
&& \mathop{\cap}_{x_0=\text{zero of }\mathcalP_d} \left\{ Q(x,y)=\sum_{(i,j)\in \Newtint (\mathcalP(x+x_0,y))} \td Q_{i,j} (x-x_0)^i y^j \right\}. \cr
\eea

In both cases we define the moduli-space
\beq
\modsp(\mathcalP):=\mathcalP+\CC[\Newtint],
\eeq
which is a complex affine space.
It is equipped with the canonical topology of $\CC^{\dim \modsp} $.
\ed

Since we work with a once for all fixed $\mathcalP$, for easier readability we shall drop $\mathcalP$ from the notations and write
\beq
\modsp=\modsp(\mathcalP).
\eeq
\br
It may seem an ``overkill" to call $\modsp$ a ``moduli-space", because it is just an affine space. However, we shall later decompose it into strata by genus $\modsp=\cup_{\genus} \modsp^{(\genus)}$, which correspond to usual notions of moduli spaces.
\er
\br[\textbf{Hypothesis 1}: $\CC[\Newtint]\neq 0$]
\label{hyp:Nnotvoid}
From now on, we shall only consider the case with $\CC[\Newtint]\neq  0$.
Proving the Boutroux curve when $\CC[\Newtint]= 0$ is trivial.
\er

We shall often illustrate our proposal with the following examples:\\
\bex[Weierstrass curve]

$\mathcalP(x,y)=y^2-x^3+g_2 x+g_3$ has the following Newton's polygon
$$
\includegraphics[scale=0.85]{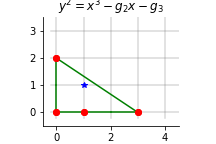}
$$
where red points represents non-zero coefficients $\mathcalP_{i,j}$, the dots $.$ represent zero coefficients, and $*$ in position $(1,1)$ is the only interior point to the polygon. Therefore $\Newtint=\{(0,0)\}$,
$\partial \Newt=\{(0,0),(0,1),(0,2),(1,0),(2,0),(3,0)\}$, $\Newt'''=\emptyset$, $\Newt''=\{(0,2),(0,1),(1,0),(2,0),(3,0),(1,1)\}$.
 $\modsp\sim \CC$ is the 1-dimensional affine space  $\modsp=\{P_{0,0}\}$.
In other words  $\modsp$ corresponds to choices of $P_{0,0}=g_3$.
\eex
\bex[Strebel-3]

$\mathcalP(x,y)=y^2(x-z_1)^2(x-z_2)^2(x-z_3)^2-Ax^2-Bx-C$ has the following Newton's polygon.
$$
\includegraphics[scale=0.5]{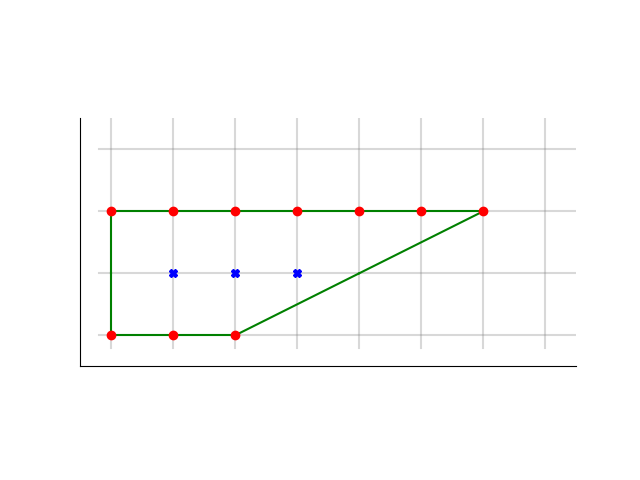}
$$
There are  3 interior points  $\Newtint=\{(0,0),(1,0),(2,0)\}$.
However since $\mathcal P_d=\prod_{i=1}^3 (x-z_i)^2$ is of degree 6, with $3$ double zeros, Definition \ref{def:modsp} gives
\beq
\CC[\Newtint]=\{0\},
\eeq
which has dimension 0, and $\modsp=\{\mathcalP\}$.
\eex
\bex[ Strebel-4]

$\mathcalP(x,y)=y^2(x-z_1)^2(x-z_2)^2(x-z_3)^2(x-z_4)^2-Ax^4-Bx^3-Cx^2-Dx-E$ has the following Newton's polygon.
$$
\includegraphics[scale=0.5]{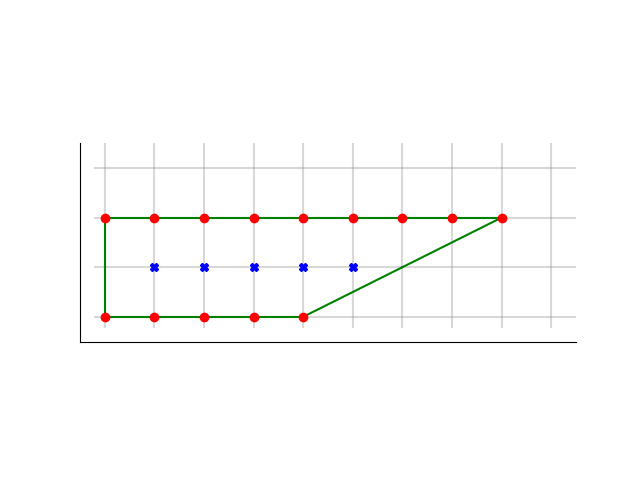}
$$
There are  5 interior points  $\Newtint=\{(0,0),(1,0),(2,0),(3,0),(4,0)\}$.
However since $\mathcal P_d=\prod_{i=1}^4 (x-z_i)^2$ is of degree 8, Definition \ref{def:modsp} gives
\beq
\CC[\Newtint]=(x-z_1)(x-z_2)(x-z_3)(x-z_4) \CC,
\eeq
which has dimension 1.
\beq
\modsp = \mathcalP + (x-z_1)(x-z_2)(x-z_3)(x-z_4) \CC
\quad , \quad \dim \modsp=1.
\eeq
\eex

\subsection{Riemann surface}
\label{sec:RiemannSurfaces}

For $P\in \modsp$, the zero-locus of $P$ defines a subset of $\CC\times \CC$, which is locally a Riemann surface
\beq
\td\curve = \td\curve_P := \{(x,y)\in\mathbb C\times \mathbb C \,\, | \,\, P(x,y)=0\},
\eeq
(most of the time, we shall drop  the $P$ index when no confusion is possible).
This surface might be not connected (if $P$ is factorizable), it is not compact (there are punctures, where $x$ and/or $y$ tend to $\infty$), and in fact it is not even a surface, as it may have self intersections points with neighborhoods not homeomorphic to a Euclidean disc (rather union of discs), called nodal points, viewed as ``pinchings" in the figure below.

\begin{center}
\begin{tikzpicture}
\draw[->,gray, thick] (0,0) -- (10,0) node[right]{$x$ $\mathbb{C}$ };
\draw[->,gray, thick] (0,0) -- (0,6) node[right]{$y$ $\mathbb{C}$};
\draw (1,1) .. controls (1,1) and (5,4) .. (1,5);;
\draw (1.2,1) .. controls (0,0) and (4,4) .. (8.7,1);;
\draw (9,1) .. controls (5.1,4.1) and (9.1,5.2) .. (9,5);;
\draw (1.7,5) .. controls (4,4) and (8,4.6) .. (8.5,5);;
\draw (2.11,2.11) .. controls (3,2) and (3,1.9) .. (4.5,2.5);;
\draw (2.4,2.05) .. controls (2.9,3) and (2.9,3) .. (4.2,2.40);;
\draw (6.4,2.1) .. controls (6.3,2.9) and (6.9,2.8) .. (7.6,2.45);
\draw (6.3,2.1) .. controls (7.3,1.9) and (7.9,1.8) .. (7.6,2.45);
\draw (3.2,3) .. controls (6.5,2.6) and (6.5,3.2) .. (7.31,3.3);;
\draw (3.6,2.95) .. controls (5.2,4.2) and (4.5,3.8) .. (7.31,3.3);;
\node at (5,6) {\huge$\tilde\Sigma$};
\end{tikzpicture}
\end{center}
Let $\Sigma$ the normalization of $\td\Sigma$ (possibly disconnected), a compact Riemann surface, equipped with two meromorphic functions, $X:\curve\to \CC$ and $\,Y:\curve\to \CC$, such that
\beq
\td\curve  = \{(x,y)\in\mathbb C\times \mathbb C \,\, | \,\, P(x,y)=0\}
= \{(X(p),Y(p))\,\,|\,\,p\in \curve\setminus\{\text{punctures}\}\}.
\eeq

$\bullet$ The map 
\bea\label{def:immersion}
\ii:\Sigma & \hookrightarrow & \CC\times \CC \cr
p & \mapsto & (X(p),Y(p))
\eea
is a meromorphic immersion, whose image is $\ii(\Sigma)=\td\Sigma$.

$\bullet$ The punctures are the locus where either $x$ or $y$ tends to $\infty$, i.e. the poles of $X$ and/or $Y$.
It is well known (See appendix \ref{Apppunctures} or literature \cite{eynardlecturesRS,}) that there is a 1-1 correspondence between punctures and boundaries of the convex envelope of the Newton's polygon.
A pole $\alpha$ of $X$ and $Y$ of degree $a_\alpha=\deg_\alpha X, b_\alpha=\deg_\alpha Y $, is associated to a boundary of $\partial\Newt$ of slope $-a_\alpha/b_\alpha$.

$\bullet$ At all points $(X,Y)\in\td\curve$ where the vector $\nabla P=(P'_x(X,Y),P'_y(X,Y))\neq (0,0)$, the surface $\td\curve$ is smooth, it has a well defined tangent plane $T_{(X,Y)}\td\Sigma = (P'_y(X,Y),-P'_x(X,Y))\CC$.

$\bullet$ The meromorphic map
\bea
X:\Sigma & \to & \CC P^1 \cr
p & \mapsto & X(p)
\eea
is a holomorphic ramified covering of $\CC P^1$ by $\curve$.
Its ramification points occur when two (or more) branches meet, and thus at $p=$ zeros of $P'_y(X(p),Y(p))$, and/or possibly at punctures.
The degree of the covering is
\beq
d=\deg X = \deg_y \mathcal P(x,y) = \text{height of the Newton's polygon}.
\eeq

$\bullet$ Zeros of $P'_y(X,Y)$ can be either regular ramification points, or they can also be nodal points i.e. self-intersection points, and they can be higher ramified.

$\bullet$ For generic $P\in \modsp$, the zeros of $P'_y(X,Y)$ and $P'_x(X,Y)$ are distinct on $\curve$, $\td\curve$ has everywhere a tangent and is smooth. However for non-generic points these may coincide, and the surface is not smooth. We have a degenerate curve with nodal points of possibly higher degeneracy order.

\medskip
\bex[Weierstrass curve]

$P(x,y)=y^2-x^3+g_2 x+g_3$.

$\bullet$  For generic $g_2,g_3$, the curve $\curve$ is a torus. Every torus is conformally isomorphic to a parallelogram $\CC/(\ZZ+\tau\ZZ)$ whose modulus $\tau$ satisfies $\Im \tau>0$. The map $\ii:\curve \to \CC\times \CC$ is then worth
\bea
\ii \ : \qquad & X(z) =& -\nu^2 \wp(z,\tau), \cr
& Y(z) =& \frac{\ii}2\nu^3 \wp'(z,\tau),
\eea
where $\wp$ is the Weierstrass function (the unique ellitptic function biperiodic $\wp(z+1)=\wp(z+\tau)=\wp(z) $ and with a double pole  $\wp(z)\sim z^{-2} + O(z^2) $ at $z=0$). The parameters $(\nu,\tau)$ are functions of $(g_2,g_3)$, whose inverse map $(\nu,\tau)\mapsto(g_2,g_3)$ is:
\beq
g_2 = 15\nu^4 G_4(\tau), \quad
g_3 = -35\nu^6 G_6(\tau),
\eeq
with $G_4$ and $G_6$ the modular Eisenstein $G$-series.

There are three ramification points, at $z=\frac12$, $z=\frac\tau{2}$, $z=\frac12(1+\tau)$, corresponding to branch points $X(\frac12), X(\frac\tau{2}), X(\frac12(1+\tau))$.
There is one puncture (pole of $X$ and $Y$) at $z=0$, with $a_0=\deg_0 X=2$ and $b_0=\deg_0 Y=3$, at which $Y\sim  X^{\frac32}$, and notice that the boundary of the Newton's polygon has indeed a slope $-a_0/b_0=-\frac23$. 

$\bullet$  If $4g_2^3-27g_3^2=0$, then the torus is degenerate, $\curve=\CC P^1$ is then a sphere, and $\td\curve$ has a nodal point.
We parametrize the sphere $\curve=\CC P^1=\CC\cup\{\infty\}$ with a complex variable $z$, and up to an automorphism of the sphere, we can write the immersion map $\ii:\curve\to\CC\times \CC$ as
\beq
\begin{split}
X(z) &= z^2-2u \cr
Y(z) &= z^3-3uz, \cr
\end{split}
\eeq
with $u = -\frac32 g_3/g_2$, i.e.
\beq
g_2=3 u^2, \quad g_3 = -2u^3.
\eeq
The nodal point is $\beta = \ii(\sqrt{3u}) = \ii(-\sqrt{3u})$, with $x_\beta = X(\beta) = u$ and $y_\beta=Y(\beta)=0$.

There is one branch point, at $z=0$.
There is one puncture (pole of $X$ and $Y$) at $z=\infty$, with $a_\infty=\deg_\infty X=2$ and $b_\infty=\deg_\infty Y=3$, at which $Y\sim  X^{\frac32}$, related to the unique boundary of the Newton's polygon, which has  slope $-a_\infty/b_\infty=-\frac23 $. 
\eex
\medskip

\bd[Nodal points]
A nodal or branch point $\beta=(x_\beta,y_\beta)\in \td\curve$, is a point at which $P'_y(x_\beta,y_\beta)=0$.
Let
\beq
(\beta^{(1)},\dots,\beta^{(\ell_\beta)}) = \ii^{-1}(\beta),
\eeq
its pre-images (the labeling doesn't matter) on $\curve$. 
These are smooth points on $\curve$.
If $\ell_\beta=1 $, we say that it is a pure ramification point $\beta^{(1)}$ corresponding to a branch point $x_\beta=X(\beta^{(1)})$,
and if $\ell_\beta\geq 2 $, we say that it is a nodal point.
\ed

\bl
\label{lem:existPregularinModsp}
In $\modsp$, there exists some $P$ that have no nodal points, and have only simple ramification points.
More precisely, the set of $P$ that have no nodal points, and have only simple ramification points, is open dense in $\modsp$.
\el

\proof{
The subset of $\modsp $ that have degenerate ramification points or nodal points, is a  union of algebraic submanifolds, given by the vanishing of the discriminant, i.e. the condition that $P,P'_y,P''_{yy}$, or $P,P'_y,P'_{x}$ have a common zero.
It is thus the complement of an algebraic set, it is open dense.
}

\subsection{Canonical local coordinates}
\label{sec:canonicalcoord}

\bd\label{def:canonicalcoord}
Let $p\in\curve$.
\begin{itemize}
\item If $X(p)=\infty$, we define $a_p:=\deg_p X=-\ord_p X$, and $X_p=0$. We have $a_p>0$.

\item If $X(p)\neq\infty$, we define $a_p:=-\ord_p (X-X(p))$, and $X_p=X(p)$. We have $a_p<0$.
\end{itemize}

We define the canonical local coordinate at $p$:
\beq
\zeta_p(z) := (X(z)-X_p)^{\frac{-1}{a_p}}.
\eeq
$\zeta_p(z)$ vanishes linearly (order 1) at $z=p$.

The canonical local coordinate is defined modulo a root of unity.
Let
\beq
\rho_k = e^{\frac{2\pi\ii}{k}}.
\eeq
Other local coordinates are
\beq
\zeta_p(z) \rho_{a_p}^j \quad j=1,\dots,|a_p|.
\eeq
Choosing a canonical local coordinate is equivalent to choosing one of the rays (there are $|a_p|$ of them) starting from $p$ in the direction $X(z)-X_p\in \RR_-$.
\ed

\subsection{Genus and cycles}

The compact Riemann surface $\Sigma$ is possibly disconnected $\curve=\curve_1\cup \dots \curve_m$, and each connected component has some genus $\genus_i$.
Let us define the total genus
\beq
\genus:=\sum_{i=1}^m \genus_i.
\eeq
It is well known (and we shall recover it below) that the genus is at most the number of interior points to the Newton's polygon:
\beq
0\leq \genus \leq \dim\CC[\Newtint] \leq \#\Newtint.
\eeq
The Homology space $H_1(\Sigma,\ZZ)$ has dimension
\beq
\dim H_1(\Sigma,\ZZ) =  2\genus,
\eeq
which means that there exists $2\genus$ independent non-contractible cycles, and it is possible (but not uniquely) to choose a symplectic basis:
\beq
\acycle_1,\dots,\acycle_{\genus}, \ \bcycle_1,\dots,\bcycle_{\genus},
\eeq
such that
\beq
\acycle_i\cap \acycle_j = 0
\ \ , \quad
\bcycle_i\cap \bcycle_j = 0
\ \ , \quad
\acycle_i\cap \bcycle_j = \delta_{i,j}.
\eeq

Such a choice of symplectic basis of cycles is called a Torelli marking of $\curve$.

Cycles are defined as linear combinations of homotopy classes of Jordan loops. However, here so far we are considering cycles on $\curve$, and we are going to integrate 1-forms (for example $Y dX$) that have poles, and one should consider the homotopy classes of Jordan loops on $\curve\setminus \text{poles}$. We could also consider removing nodal and ramification points.
A way to avoid this, is just to choose Jordan loops rather than cycles.

So here we need the following notion of marking:
\bd[Marking]
We call a marking of $\curve$, a choice of $2\genus$ \textbf{Jordan loops} on \\$\curve\setminus \{\text{punctures, ramification points, nodal points}\}$, satisfying
\beq
\acycle_i\cap \acycle_j = 0, \quad
\bcycle_i\cap \bcycle_j = 0, \quad
\acycle_i\cap \bcycle_j = \delta_{i,j}.
\eeq
Their projection to $H_1(\Sigma,\ZZ)$ is a Torelli marking of $\curve$.

\ed

\br We insist that $\acycle_i$ and $\bcycle_i$ are not cycles, they are Jordan loops.
\er

\bl[Continuous Jordan cycles]\label{Lemmacontinuouscylces}
Let $P\in \modsp$ such that $\curve_P$ has total genus $\genus$,  and let $\{\acycle_i,\bcycle_i\}_{i=1,\dots,\genus}$ a marking of symplectic Jordan loops on $\curve_P\setminus \{\text{punctures, branch points, nodal points}\}$, whose projection to $H_1(\curve,\ZZ)$ forms a symplectic basis.

There exists some neighborhood $U$ of $P$ in $\modsp$, such that for each $Q\in U$, there exists a unique family of symplectic Jordan loops  on $\curve_Q\setminus \{\text{punctures, branch points, nodal points}\}$, whose projection by $\ii$ are continuous on $U$, and whose projection by $X$ is constant over $U$. 
We shall call it a ``continuous choice of Jordan cycles" in $U$.
\el
\proof{
Away from punctures or ramification or nodal points, $X$ is locally a homeomorphism, and the restriction of $\ii$ is locally continuous on $U$.
Use $X$ to push the Jordan loops to the base and pull them back on any $Q\in U$.
}

\br Notice that in a neighborhood of $P$, there can be some $Q$ with higher genus, and thus the family of symplectic Jordan loops is not a basis, it is only an independent family. One can obtain a basis by completing it with other cycles. For example one could  add new cycles corresponding to unpinching the nodal points.
However, this will not be needed in this article.
\er


\subsubsection{Holomorphic forms}

Let $\Omega^1(\curve)$ the space of holomorphic differential 1-forms on $\curve$.

The following is a classical theorem going back to Riemann
\bt[Riemann] We have
\beq\dim \Omega^1(\curve) =\genus.
\eeq
Having made a choice of Torelli marking of $\Sigma$, there exists a unique basis $\omega_1,\dots,\omega_{\genus}$ of $\Omega^1(\Sigma)$, such that
\beq
\oint_{\acycle_i}\omega_j=\delta_{i,j}.
\eeq
This is used to define the Riemann matrix of periods
\beq
\tau_{i,j} = \oint_{\bcycle_i}\omega_j.
\eeq
$\tau$ is a  ${\genus}\times{\genus}$  Siegel matrix, i.e. a complex symmetric matrix, whose imaginary part is positive definite:
\beq
\tau^t=\tau, \quad
\Im\ \tau>0.
\eeq
\et

We can also obtain $\Omega^1(\curve)$ algebraically from the Newton's polygon,
the following is a classical theorem
\bt
For any $\ Q\in \CC[\Newtint]$,
the following differential form
\beq
\frac{Q(X,Y) \ dX}{P'_y(X,Y)}  
\eeq
is holomorphic at all the punctures. 
Its only poles could be at the zeros of $P'_y(X,Y)$ if these are not compensated by zeros of $dX$, 
i.e. these can be only nodal points.
We define
\beq
H'^1(\curve) = \frac{dX}{P'_y(X,Y)} \CC[\Newtint].
\eeq

In the generic case, all zeros of $P'_y(X,Y)$ are simple and are zeros of $dX$, so that this 1-form has no pole at all, it is holomorphic.

$\bullet$ In the generic case we have
\beq
\Omega^1(\Sigma) = H'^1(\Sigma), \qquad
\genus=\dim \Omega^1(\Sigma)= \dim\CC[\Newtint].
\eeq

$\bullet$ In the non-generic case we only have
\beq
\Omega^1(\Sigma) \subset H'^1(\Sigma), \qquad
\genus=\dim \Omega^1(\Sigma)\leq \dim\CC[\Newtint].
\eeq
In all cases there exists a rectangular matrix $\mathcal K_{k;(i,j)}$ of size $\genus\times \dim\CC[\Newtint]$, such that the normalized holomorphic differentials can be written
\beq
\forall \ k=1,\dots,\genus , \qquad
\omega_k = \sum_{(i,j)\in \Newtint} \mathcal K_{k;(i,j)} \frac{X^i Y^j \ dX}{P'_y(X,Y)}.
\eeq
Let the $\dim\CC[\Newtint]\times \genus$ rectangular matrix
\beq
 \hat{\mathcal {K}}_{(i,j);k} = \oint_{\acycle_k}   \frac{X^i Y^j \ dX}{P'_y(X,Y)}.
\eeq
By definition we have $\mathcal {K} \hat{\mathcal {K}} = \text{Id}_{\genus}$, i.e.
\beq
\sum_{(i,j)\in \Newtint} \mathcal {K}_{k;(i,j)}   \ \hat{\mathcal {K}}_{(i,j);k}   =\delta_{k,l}.
\eeq
This shows that when $\genus=\dim\CC[\Newtint]$, $\mathcal K$ is invertible.

In all cases we have
\beq
\Omega^1(\curve) = \frac{dX}{P'_y(X,Y)} \mathcal K(\CC[\Newtint]).
\eeq

\et
\proof{Classical theorem, see \cite{farkas2012riemann,fay1973theta,TataLectures,eynardlecturesRS}.}

\medskip
\bex[Weierstrass curve]
$P(x,y)=y^2-x^3+g_2 x+g_3$, for generic $g_2,g_3$. $\curve=\CC/(\ZZ+\tau\ZZ)$ is a torus, with the immersion map $\ii:\curve \to \CC\times \CC$ given by
\beq
\begin{split}
X(z) &= -\nu^2 \wp(z,\tau) \cr
Y(z) &= \frac{\ii}2\nu^3 \wp'(z,\tau). \cr
\end{split}
\eeq
We have
\beq
\frac{dX}{P'_y(X,Y)} = \frac{dX}{2Y} = \frac{\ii \nu^2 \wp'(z,\tau)dz}{\nu^3 \wp'(z,\tau)} = \frac{\ii}{\nu} dz.
\eeq
$dz$ is indeed a holomorphic form, it has no pole in the parallelogram $(0,1,1+\tau,\tau)$, and it is biperiodic $dz = d(z+1)=d(z+\tau)$.
We choose the Jordan loops $\acycle=[p,p+1]$ and $\bcycle=[p,p+\tau]$ with $p$ a generic point.
The matrix $\hat{\mathcal K}$ is a $1\times 1$ matrix, worth
\beq
\hat{\mathcal K}=\oint_{\acycle} \frac{dX}{P'_y(X,Y)} = \frac{\ii}{\nu} \int_p^{p+1} dz = \frac{\ii}{\nu}.
\eeq
Its inverse is
\beq
\mathcal K = -\ii\nu.
\eeq
The normalized holomorphic differential is
\beq
\omega=dz = -\ii\nu \frac{dX}{P'_y(X,Y)} .
\eeq
Its $\bcycle$-cycle integral is
\beq
\oint_{\bcycle} \omega = \int_p^{p+\tau} dz = \tau.
\eeq
Riemann's theorem ensures that $\Im\tau>0$.
\eex

\bd[Cells of fixed genus] We define
\beq
\modsp^{(\genus)}:= \{ P\in \modsp \ | \ \curve=\curve_1\cup\dots\cup \curve_m , \ \ \genus=\sum_{i=1}^m \genus_i  \}.
\eeq
\ed

\bp\label{prop:cellMgfromdimnopole}
Alternatively

\beq
\modsp^{(\genus)}= \left\{ P\in \modsp \ | \ \genus  = \dim  \left\{Q\in \CC[\Newtint] \ \Big| \frac{Q(X,Y)dX}{P'_y(X,Y)} \ \text{has no pole} \right\}  \right\}.
\eeq
\begin{itemize}
\item $\modsp^{(\genus)}$ is an algebraic subset of $\modsp$.

\item Each $\modsp^{(\genus)}$ has a finite number of connected components.
\end{itemize}

\ep
\proof{
Indeed if the form $QdX/P'_y$ has no pole, then it belongs to $\Omega^1(\curve)$, and vice-versa, i.e.
\beq
\Omega^1(\curve)=\left\{Q\in \CC[\Newtint] \ \Big| \frac{Q(X,Y)dX}{P'_y(X,Y)} \ \text{has no pole at nodal points} \right\}. 
\eeq
It has thus dimension $\genus$.
The fact that $\modsp^{(\genus)}$ is an algebraic subset of $\modsp$, comes from the fact that requiring that $dX/P'_y(X,Y)$ having zeros of a certain order, can be formulated with resultants of $P,P'_y,P'_x$ and higher order derivatives having to vanish, i.e. some polynomials relations of the $P_{i,j}$'s. 
Each algebraic equations has a finite number of solutions.
}

\subsubsection{Non-generic case}

\bd
Let $\beta$ a nodal point, with preimages $\ii^{-1}(\beta)=\{\beta^{(1)},\dots,\beta^{(\ell_\beta )}\} $, 
Let
\bea
H'^1_{\beta}(\curve)  
&:=& \Big\{ Q(X,Y)\frac{dX}{P'_y(X,Y)}\ | Q\in \CC[x,y] \text{ having pole(s) at most}\cr
&& \text{at preimage(s) of }\beta  \text{ and no other pole} \Big\} \ / \ \Omega^1(\curve).  
\eea
We define the algebraic genus of the nodal point $\beta$ as
\beq
\genus_\beta := \dim H'^1_{\beta}(\curve).
\eeq

\ed

\bt
We have
\beq
H'^1(\curve) = \Omega^1(\curve) \underset{{\beta=\text{nodal points}}}{\oplus} H'^1_{\beta}(\curve),
\eeq
whose dimensions are
\beq
\dim\CC[\Newtint] = \genus + \sum_{\beta=\text{nodal points}} \genus_\beta.
\eeq
\et

\proof{
$H'^1(\curve)$ is the space of forms having no poles at punctures. The only places where an element of $H'^1(\curve)$ could have poles is where $P'_y(X,Y)$ vanishes at an order higher than that of $dX$, i.e. nodal points. 
Either this form has no poles at all, and is in $\Omega^1(\curve)$, or must be in some $H'^1_\beta(\curve)$.
By definition all the $H'^1_\beta(\curve)$ are disjoint for different $\beta$, so we have a direct sum.
 }

\medskip
\bex[Degenerate Weierstrass curve]
$P(x,y)=y^2-x^3+g_2 x+g_3$, with $4g_2^3-27g_3^2=0$.
The immersion map $\ii:\curve\to\CC\times \CC$ is
\beq
\begin{split}
X(z) &= z^2-2u \cr
Y(z) &= z^3-3uz \cr
\end{split}
\eeq
with 
$g_2=3 u^2, \quad g_3 = -2u^3$.
The nodal point is $\beta=(u,0)$ and $\ii^{-1}(\beta) =\{\sqrt{3u},-\sqrt{3u}\}$, so that $\ell_\beta=2$.
Since $\curve=\CC P^1$, we have  $\dim\Omega^1(\curve)=\genus=0$, and we have $H'^1(\curve) = \frac{dX}{P'_y(X,Y)}, \CC$ a one dimensional space.
We have
\beq
\frac{dX}{P'_y(X,Y)} = \frac{2z dz}{2 Y} = \frac{dz}{z^2-3u}.
\eeq
It has simple poles at $z=\pm \sqrt{3u}$, each with degree one.
The algebraic genus of $\beta$ is  thus
\beq
\genus_\beta=1.
\eeq
The Newton's polygon has one interior point, we have
\beq
1 = \#\Newtint = \dim\CC[\Newtint] = \genus+\genus_\beta = 0+1.
\eeq

\eex
\subsection{Period coordinates}

\bd[Period coordinates]
Let $P\in \modsp^{(\genus)}$.
Having chosen a symplectic marking of Jordan loops in a neighborhood of $P$ (lemma \ref{Lemmacontinuouscylces}) we define:
\beq
\acycle-\text{periods :}\qquad \forall \ i=1,\dots,\genus , \quad \eta_i := \frac{1}{2\pi\ii} \oint_{\acycle_i} Y dX 
\eeq
\beq
\bcycle-\text{periods :}\qquad \forall \ i=1,\dots,\genus , \quad \td\eta_i=\eta_{i+\genus} := \frac{1}{2\pi\ii} \oint_{\bcycle_i} Y dX
\eeq

\beq
\text{Real periods :}\qquad \forall \ i=1,\dots,2\genus , \quad \epsilon_i:=\Re\ \eta_{i}  \ \ , \ \ \zeta_i:=\Im\ \eta_i.
\eeq
\beq
\quad\qquad \forall \ i=1,\dots,\genus , \quad \td\epsilon_i:=\Re\ \td\eta_{i} = \epsilon_{\genus+i}  \ \ , \ \ \td\zeta_i:=\Im\ \td\eta_i = \zeta_{\genus+i}.
\eeq

\ed

\bt[periods = local coordinates]
We have the following:

\begin{itemize}
\item The periods $\eta_1,\dots,\eta_\genus$ are local complex coordinates on $\modsp^{(\genus)}$.

\item The periods $\epsilon_1,\dots,\epsilon_{2\genus}$ are local real coordinates on $\modsp^{(\genus)}$.
\end{itemize}

\et
\proof{
This is a well known theorem, however, since it plays an important role in this article, lets give a proof.
First from lemma \ref{Lemmacontinuouscylces} a marking with Jordan loops can be chosen continuous in some neighborhood of $P\in \modsp^{(\genus)}$.

The tangent space is generated by tangent vectors
\beq
\partial_k = -\sum_{(i,j)\in\Newtint} \mathcal K_{k;(i,j)} \frac{\partial}{\partial P_{i,j}}.
\eeq
We have
\bea
\partial_k \eta_l 
&=& -\frac{1}{2\pi\ii} \oint_{\acycle_l} \sum_{(i,j)\in\Newtint} \mathcal K_{k;(i,j)} \frac{\partial Y}{\partial P_{i,j}} \ dX \cr
&=& \frac{1}{2\pi\ii} \oint_{\acycle_l} \sum_{(i,j)\in\Newtint} \mathcal K_{k;(i,j)} \frac{X^i Y^j \ dX}{P'_y(X,Y)}  \cr
&=& \sum_{(i,j)\in\Newtint} \mathcal K_{k;(i,j)} \hat K_{(i,j);l)} \cr
&=& \delta_{k,l}.
\eea
This implies that $\eta_1,\dots,\eta_\genus$ are coordinates because the Jacobian matrix is invertible.

Then compute the differential
\beq
\begin{aligned}
-2\pi\ii \ d\td\eta_k 
&=&&  \sum_{(i,j)\in \Newtint} dP_{i,j}  \oint_{\bcycle_k} \frac{X^i Y^j}{P'_y(X,Y)}\ dx \cr
&=&&  \sum_{(i,j)\in \Newtint} dP_{i,j}  \oint_{\bcycle_k}  \sum_{l=1}^\genus \mathcal K^{-1}_{(i,j);l} \omega_l \cr
&=&&  \sum_{(i,j)\in \Newtint} dP_{i,j}   \sum_{l=1}^\genus \mathcal K^{-1}_{(i,j);l} \tau_{l,k} \cr
&=&& -2\pi\ii \sum_{l} \tau_{k,l} d\eta_l.
\end{aligned}
\eeq
Let us decompose $\tau$ in its real and imaginary part
\beq
\tau = R+\ii I,
\eeq
and recall that $I>0$ so that in particular $I$ is invertible.
We have
\bea
d\eta &=& d\epsilon + \ii d\zeta \cr
d\td\eta &=& \tau d\eta = (R+\ii I)(d\epsilon+\ii d\zeta) \cr
&=& (R d\epsilon - I d\zeta) +\ii (I d\epsilon+R d\zeta), \cr
\eea
i.e.
\beq
d\td\epsilon = R\  d\epsilon-I \ d\zeta,
\eeq
and thus
\beq
I d\zeta  = -d\td\epsilon + R d\epsilon,
\eeq
\beq
d\eta = (1+\ii I^{-1}R)d\epsilon - \ii \ I^{-1}  d\td \epsilon.
\eeq
\beq
d\epsilon=\Re\ d\eta, \qquad
d\td\epsilon = \Re \ \tau   d\eta.
\eeq
In other words the Jacobian of the change of variable $\eta\to (\epsilon,\td\epsilon)$ is invertible.
This implies that $(\epsilon,\td\epsilon)$ is also a set of local coordinates.
}

\subsubsection{Nodal point coordinates}

We should think of nodal points, as cycles that have been pinched (collapsed).
It is useful to associate also period coordinates to them.

\bd
Let $\beta$ a nodal point, with preimages $\ii^{-1}(\beta)=\{\beta^{(1)},\dots,\beta^{(\ell_\beta )}\} $, 
let $\zeta_i = \zeta_{\beta^{(i)}}$ the canonical local coordinate at $\beta^{(i)}$.
Let
\beq
\eta_{\beta^{(i)},k} = \Res_{\beta^{(i)}} \zeta_i^k YdX = 0, \qquad \td\eta_{\beta^{(i)},k} = \frac{1}{k} \Res_{\beta^{(i)}} \zeta_i^{-k} YdX.
\eeq
($\eta_{\beta^{(i)},k}=0$ because $YdX$ has no pole at nodal points).

This allows to consider the period-vector $(\eta_1,\dots,\eta_{\genus})$ of dimension $\genus=\dim \modsp^{(\genus)} $  as a period-vector $(\eta_1,\dots,\eta_{\genus},0,\dots,0)$ of dimension $\genus + \sum_\beta \genus_\beta = \dim \CC[\Newtint] = \dim \modsp$.
With this definition the period-vector is continuous in a neighborhood in $\modsp$ of any $P\in \modsp^{(\genus)}$ (with the topology of $\modsp$).
\ed

\subsection{Punctures coordinates}
\label{sec:punctures}

We use the canonical local coordinates of Definition \ref{def:canonicalcoord}.

\bd[Times at punctures]
Let $\alpha$ a puncture.

The 1-form $Y dX$ has a local Laurent series expansion near $\alpha$:
\beq
Y dX \sim \sum_{k=0}^{r_\alpha} t_{\alpha,k} \ \zeta_\alpha^{-k-1} d\zeta_\alpha + \text{analytic at }\alpha.
\eeq
We have
\beq
t_{\alpha,k} = \Res_\alpha \zeta_\alpha^k Y dX.
\eeq

The coefficients $t_{\alpha,k}$ are called the ``times" of $P$.

We have
\beq
Y \sim \frac{-t_{\alpha,r_\alpha}}{a_\alpha} \zeta_\alpha^{-b_\alpha}, \quad 
b_\alpha = r_\alpha-a_\alpha.
\eeq
In other words
\beq
Y \sim \frac{-t_{\alpha,r_\alpha}}{a_\alpha} \ (X-X_\alpha)^{b_\alpha/a_\alpha}.
\eeq
There is only a finite number of non-vanishing times:
\beq
\{ t_{\alpha,k} \ | \alpha=\text{punctures} \ , \ k=0,\dots,r_\alpha \}.
\eeq

\ed

\bp[Times and exterior coefficients]
The times $t_{\alpha,k}$ are algebraic functions of the coefficients  $P_{i,j}$ of $P$, and it is well known (see proposition \ref{prop:reconstructionP} of appendix \ref{Apppunctures}) that they are algebraic functions of only the exterior coefficients of $P$.
In other words they are  algebraic functions of the coefficients  $\mathcal P_{i,j}=P_{i,j}$ of $\mathcal P$ and are the same for all $P\in \modsp $.

Vice--versa, the exterior coefficients of $\mathcal P$ are polynomials of the times.
\ep

\proof{
Done in appendix \ref{apx:reconstructionP}.}

It is well known that the exponent $-b_\alpha/a_\alpha$ is a slope of the convex envelope of the Newton's polygon, and there exists a line of equation
\beq
D_\alpha = \{(i,j) \ | \ a_\alpha i+ b_\alpha j =m_\alpha\}
\eeq
tangent to the Newton's polygon.
See appendix \ref{Apppunctures}.

\br[\textbf{Hypothesis: real residues}]\label{hyp:realresidues}
From now on, we shall assume that $\mathcal P$ has been chosen so that:
\beq
\forall \alpha \ , \ \quad t_{\alpha,0} = \Res_\alpha Y dX \in \RR.
\eeq
In fact this is necessary for having a chance to satisfy Boutroux property.
Indeed if $\gamma$ is a small circle around $\alpha$ we have
\beq
\Re \oint_\gamma Y dX = \Re \left( 2\pi\ii \Res_\alpha Y dX\right) = -2\pi \ \Im \ t_{\alpha,0} = 0 \ \ \text{ if Boutroux}.
\eeq
\er

\bex[Weierstrass curve]
$P(x,y)=y^2-x^3+g_2 x+g_3$.
There is one puncture $\alpha=\infty$, at which both $x$ and $y$ become infinite with the asymptotic behavior:
\beq
y\sim x^{\frac32} \ \left(1-g_2 x^{-2} - g_3 x^{-3}\right)^{\frac12}.
\eeq
Using the canonical local coordinate $\zeta  = \zeta_\infty= x^{-\frac12}$, i.e. $a_\alpha=2$, we have
\beq
x= \zeta^{-2}, \qquad
dx = -2 \zeta^{-3} d\zeta,
\eeq
and
\beq
y  = \zeta^{-3} \left( 1- \frac{g_2}{2} \zeta^{4} - \frac{g_3}{2} \zeta^6 + O(\zeta^8) \right).
\eeq
The exponents $a_\alpha=2$ and $b_\alpha=3$ (degrees of poles of $x$ and $y$ in function of $\zeta$), are related to the boundary of the Newton's polygon with normal vector $(2,3)$.
We have
\beq
Y dX = -2 \zeta^{-6}\left( 1- \frac{g_2}{2} \zeta^{4} + O(\zeta^6) \right) d\zeta
= -2\zeta^{-5-1}d\zeta + g_2 \zeta^{-1-1}d\zeta + O(1) d\zeta.
\eeq
which gives $r_\infty=5$ and  the  times
\beq
t_{\infty,5}=-2, \quad 
t_{\infty,1}=g_2,
\eeq
and all the other  times are vanishing.
The times are independent of $g_3$, and thus are the same for all $P\in \modsp$.

For the non-degenerate case we have
\beq
\eta=3\ii \nu^5 G'_4(\tau), \quad
\td\eta = \tau\eta +12\ii \nu^5 G_4(\tau).
\eeq
\eex

\bigskip

\bd[Conjugate times]
For $k>0$, let
\beq
\td t_{\alpha,k} = \frac{1}{k} \Res_{\alpha} \zeta_\alpha^{-k} Y dX.
\eeq
We have the Laurent series expansion:
\beq\label{eq:Laurentseriesalpha}
Y dX \mathop{\sim}_\alpha \sum_{k=0}^{r_\alpha} t_{\alpha,k} \zeta_\alpha^{-k-1}d\zeta_\alpha + \sum_{k=1}^{\infty} k \td t_{\alpha,k} \zeta_\alpha^{k-1}d\zeta_\alpha.
\eeq

\ed

The coordinates $\td t_{\alpha,0}$ conjugated to $t_{\alpha,0}$ are slightly more tricky to define.
We first need:

\bd[Fundamental domain]\label{def:fondDomain}
Let $p_i$ a generic point in the $i^{th}$ connected component $\curve_i$ of  $\curve$.
For each connected component $\curve_i$, let us consider a set of disjoints smooth Jordan arcs $e_\alpha$ from $p_i$ to all punctures $\alpha$ that are in $\curve_i$.

$\curve\setminus \cup_\alpha e_\alpha$ is typically a finite union of disjoint surfaces of total genus $\genus$. 
On $\curve\setminus \cup_\alpha e_\alpha$ it is possible to choose a set of $2\genus$ smooth closed Jordan loops  $\acycle_1,\dots,\acycle_\genus,\bcycle_1,\dots,\bcycle_\genus$ (and we denote $\acycle_{\genus+i}=\bcycle_i$) starting and ending at $p_i$, such that $\curve\setminus ( \cup_\alpha e_\alpha\cup_{i=1}^{2\genus}\acycle_i ) $ is simply connected, and we can choose them so that they form a symplectic marking of cycles.

Let $\Upsilon$ the graph of all these edges.
It is a graph whose vertices are at $p_i$ and at the punctures.
Each edge has at least one boundary being $p_i$.
Let 
\beq
D:=\curve\setminus\Upsilon.
\eeq
$D$ is a finite union of simply connected open domains of $\curve$:
$D$ may be disconnected (if $\curve$ was) and is a finite union of topological discs (as many as the connected components).
The boundary of $D$ is made of edges of $\Upsilon$, and each edge $e$ of $\Upsilon$ appears twice as a boundary of $D$, with two opposite orientations.
We call $e_+$ (resp. $e_-$) the edge of $\partial D$ corresponding to the edge $e$ of $\Upsilon$ whose orientation with respect to $D$ (having $D$ on its left) is the same (resp. opposite) as $e$ in $\Upsilon$. 
We have
\beq
\partial D = \sum_{e\in \Upsilon} e_+ - e_-.
\eeq

Let $o_i$ a generic point inside the $i^{th}$ connected component of  $D$.
We define for $z$  is in the $i^{th}$ connected component of  $D$
\beq
g(z) := \int_{o_i}^z Y dX,
\eeq
where the integration path is the (unique up to homotopy) path from $o_i$ to $z$ in the fundamental domain $D$.

\ed

\bd[Conjugate times case $k=0$]
Let $D$ a fundamental domain, and let $\alpha$ a puncture.
We may assume that in a neighborhood of $\alpha$, the edge $e_\alpha$ is such that $\zeta_\alpha\in \RR_-$, so that if $z$ is a point close to $\alpha$ of coordinate $\zeta_\alpha = r e^{\ii\theta}$, we shall define the logarithm with cut on $\RR_-$:
\beq
\ln\zeta_\alpha(z)  := \ln r + \ii \theta \qquad \theta\in ]-\pi,\pi].
\eeq

In a neighborhood of $\alpha$ in $D$ we define the local ``potential"
\beq
V_\alpha := -\sum_{k=1}^{r_\alpha} \frac{1}{k}t_{\alpha,k} \zeta_\alpha^{-k}.
\eeq
It is such that $Y dX-dV_\alpha$ can have at most a simple pole at $\alpha$:
\beq
Y dX-dV_\alpha = t_{\alpha,0} \zeta_\alpha^{-1}d\zeta_\alpha + \text{holomorphic at }\alpha.
\eeq
We define:
\bea
g_\alpha(z) &:=& \int_{\alpha}^z (Y dX-dV_\alpha - t_{\alpha,0}\frac{d\zeta_\alpha}{\zeta_\alpha}) + V_\alpha(z) +  t_{\alpha,0} \ln \zeta_\alpha(z)  \cr
&=& -\sum_{k=1}^{r_\alpha} \frac{t_{\alpha,k}}{k} \zeta_\alpha^{-k} + t_{\alpha,0}\ln{\zeta_\alpha} + \sum_{k=1}^\infty \td t_{\alpha,k} \zeta_\alpha^k.
\eea
We define:
\beq
\td t_{\alpha,o} := g(z)-g_\alpha(z)
\eeq
which is independent of $z\in D$.

Moreover, since the sum of residues of a meromorphic 1-form has to vanish, then $\sum_\alpha t_{\alpha,0} =0$, and this implies that
\beq
\sum_\alpha t_{\alpha,0}\td t_{\alpha,o}
\eeq
is independent of the point $o_i$ used to define the function $g$.
\ed

\section{Energy as regularized area}
\label{sec:energy}
The Boutroux curve will be obtained by a variational principle: minimizing an ``energy".
The energy will be the ``area" and Boutroux curves will then be ``minimal surfaces".

\subsection{Regularized area}

Let us give a first definition of our energy here, and it will be shown later that it is equivalent to another definition.

Recall that on $\CC$ the Euclidean metric is related to the symplectic metric
\beq
|dx|^2 = \overline{ dx} \wedge dx = 2\ii \ d^2 x = 2\ii \ d\Re x \wedge d\Im x.
\eeq
In $\CC\times \CC$, we have the canonical symplectic form
$ \overline{dx} \wedge \overline{dy} \wedge dy \wedge dx $.
Its reduction to $\td\curve$, is the canonical metric on $\td\curve$
\beq
|y dx|^2 = 2\ii |y|^2 d^2 x,
\eeq
and its pullback by $\ii$ is the canonical metric $|Y dX|^2$ on $\curve$.
Because of punctures, the total area $\int_\curve |Y dX|^2$ is infinite. We need to ``regularize" it.

\bd[Energy = regularized area]
\label{def:regArea}
For each puncture $\alpha$, let us choose a small radius $R_\alpha>0$, and consider the disc $D_\alpha: \ |\zeta_\alpha|<R_\alpha$ in $\curve$ and its boundary $\mathcal C_\alpha$ the circle $|\zeta_\alpha|=R_\alpha$, i.e. $\mathcal C_\alpha=\{ R_\alpha e^{\ii\theta} | \theta\in ]-\pi,\pi]\}$.
We choose $R_\alpha$ small enough so that all $D_\alpha$ are topological discs and are all disjoint.
In particular, each of them encloses only one puncture, and doesn't enclose any nodal or branch point.

Then we define the ``\textbf{regularized area}":
\bea
4 F
& := & \frac{1}{2\pi i} \int_{\curve\setminus \cup_\alpha D_\alpha} |Y dX|^2 \cr
&& - \sum_\alpha \sum_{k=1}^{r_\alpha} \frac{1}{k} |t_{\alpha,k}|^2 R_{\alpha}^{-2k}  + 2\sum_\alpha |t_{\alpha,0}|^2 \ln R_\alpha \cr
&& + \sum_\alpha \sum_{k=1}^{\infty} k |\td t_{\alpha,k}|^2 R_{\alpha}^{2k} 
 -2 \Re \sum_\alpha \sum_{k=1}^{r_\alpha} t_{\alpha,k} \td t_{\alpha,k}.
\eea

\ed

\bl
$F$ is independent of the choice of radius $R_\alpha$.
\el

\proof{
Let $D$ a fundamental domain.
Let $\td R_\alpha<R_\alpha$.
The proof uses Stokes theorem, we compute the integral on the annulus $\td R_\alpha<|\zeta_\alpha|<R_\alpha$ in the fundamental domain:
\bea
\frac{1}{2\pi \ii} \int_{D\cap D_\alpha\setminus \td D_\alpha} \overline{Y dX} \wedge YdX
&=& \frac{1}{2\pi \ii} \int_{\partial (D\cap D_\alpha\setminus \td D_\alpha)} \overline{g_\alpha} \ YdX \cr
&=& \frac{1}{2\pi \ii} \int_{|\zeta_\alpha|=R_\alpha} \overline{g_\alpha} \ YdX  - \frac{1}{2\pi \ii} \int_{|\zeta_\alpha|=\td R_\alpha} \overline{g_\alpha} \ YdX \cr
&& + \frac{1}{2\pi \ii} \int_{[\td p_\alpha, p_\alpha]_{left}} \overline{g_\alpha} \ YdX 
 - \frac{1}{2\pi \ii} \int_{[\td p_\alpha, p_\alpha]_{right}} \overline{g_\alpha} \ YdX  \cr
\eea
where $p_\alpha$ (resp. $\td p_\alpha$) is the point of coordinate $\zeta_\alpha = R_\alpha e^{\ii \pi}$ (resp. $\zeta_\alpha = \td R_\alpha e^{\ii \pi}$).

For the last two terms, remark that
$g_\alpha(z_{right})-g_\alpha(z_{left})  = 2\pi \ii t_{\alpha,0}$ therefore
\bea
\frac{1}{2\pi \ii} \int_{[\td p_\alpha, p_\alpha]_{left}} \overline{g_\alpha} \ YdX 
 - \frac{1}{2\pi \ii} \int_{[\td p_\alpha, p_\alpha]_{right}} \overline{g_\alpha} \ YdX 
&=&  t_{\alpha,0} \int_{\td p_\alpha}^{p_\alpha}  YdX \cr
&=&  t_{\alpha,0} (g_\alpha(p_\alpha)-g_\alpha(\td p_\alpha)) . 
\eea
The first two terms, we use lemma \ref{lem:intCalpha} of appendix \ref{apx:lemmaintCalpha}, and we get
\bea
&& \frac{1}{2\pi \ii} \int_{D\cap D_\alpha\setminus \td D_\alpha} \overline{Y dX} \wedge YdX \cr
&=& t_{\alpha,0} (g_\alpha(p_\alpha)-g_\alpha(\td p_\alpha))  \cr
&& - \sum_{k=1}^{r_\alpha} \frac{|t_{\alpha,k}|^2}{k} R_\alpha^{-2k}
+ \sum_{k=1}^{\infty} k |\td t_{\alpha,k}|^2 R_\alpha^{2k} + 2 t_{\alpha,0}^2 \ln R_\alpha  -t_{\alpha,0} g_\alpha(p_\alpha) \cr
&& + \sum_{k=1}^{r_\alpha} \frac{|t_{\alpha,k}|^2}{k} \td R_\alpha^{-2k}
- \sum_{k=1}^{\infty} k |\td t_{\alpha,k}|^2 \td R_\alpha^{2k} - 2 t_{\alpha,0}^2 \ln \td R_\alpha  +t_{\alpha,0} g_\alpha(\td p_\alpha) \cr
&=& - \sum_{k=1}^{r_\alpha} \frac{|t_{\alpha,k}|^2}{k} R_\alpha^{-2k}
+ \sum_{k=1}^{\infty} k |\td t_{\alpha,k}|^2 R_\alpha^{2k} + 2 t_{\alpha,0}^2 \ln R_\alpha  \cr
&& + \sum_{k=1}^{r_\alpha} \frac{|t_{\alpha,k}|^2}{k} \td R_\alpha^{-2k}
- \sum_{k=1}^{\infty} k |\td t_{\alpha,k}|^2 \td R_\alpha^{2k} - 2 t_{\alpha,0}^2 \ln \td R_\alpha .
\eea

This proves the Lemma.
}

\bt[Continuity]
The energy is continuous on $\modsp$.
\beq
F\in C^0(\modsp,\RR).
\eeq

\et

\proof{
The immersion $\td\curve \setminus \cup_\alpha \ii(D_\alpha) $ in $\CC\times \CC$, being the locus of solutions of $P(x,y)=0$, is continuous on $\modsp$.
The integral over $\curve\setminus \cup_\alpha D_\alpha$ is the area of $\td\curve\setminus \cup_\alpha \ii(D_\alpha)$ with the metric $|y|^2d^2 x$  of $\CC\times \CC$, therefore it is  continuous.
The times $t_{\alpha,k}$ are constant on $\modsp$,
the conjugate times $\tilde t_{\alpha,k}$ are continuous and
the radius $R_\alpha$ are taken locally constant. Thus, $F$ is continuous.
}

\subsection{Minimum}

\bt[Bounded from below]
$F$ is bounded from below on $\modsp$.
\et

\proof{
We shall compare $F(P)$ to $F(\mathcal P)$. Recall that $P$ and $\mathcalP$ have the same times $t_{\alpha,k}$, but their conjugate times $\td t_{\alpha,k}$ can be different.
Choose the radius $R_{\alpha}$ small enough so that they can be used for both $P$ and $\mathcalP$.

We let $\mathcal X,\mathcal Y$ denote the functions $X,Y$ when $P=\mathcal P$. We have
\bea
4 F(P) - 4 F(\mathcal P)
&=&  \frac{1}{2\pi i} \int_{\curve\setminus \cup_\alpha D_\alpha} |Y dX|^2 - \frac{1}{2\pi i} \int_{\curve_{\mathcal P} \setminus \cup_\alpha D_\alpha} |\mathcal Y d \mathcal X|^2 \cr
&& + \sum_\alpha \sum_{k=1}^\infty \left|\frac{1}{\sqrt k}\bar t_{\alpha,k} R_\alpha^{-k} - \sqrt k \td t_{\alpha,k} R_\alpha^k \right|^2 \cr
&& - \sum_\alpha \sum_{k=1}^\infty \left|\frac{1}{\sqrt k}\bar t_{\alpha,k} R_\alpha^{-k} - \sqrt k \td t_{\alpha,k}(\mathcal P) R_\alpha^k \right|^2. \cr\eea
Therefore:
\beq
4 F(P)
\geq  4 F(\mathcal P)
 - \frac{1}{2\pi i} \int_{\curve_{\mathcal P}\setminus \cup_\alpha D_\alpha} |\mathcal Y d\mathcal X|^2  - \sum_\alpha \sum_{k=1}^\infty \left|\frac{1}{\sqrt k}\bar t_{\alpha,k} R_\alpha^{-k} - \sqrt k \td t_{\alpha,k}(\mathcal P) R_\alpha^k \right|^2 .
\eeq
Since the rhs is independent of $P$ this shows that $F$ is bounded from below on $\modsp$.
}

\bt
The level sets of $F$ are compact in the canonical topology of $\CC[\Newtint]$.
(we recall that the level set of level $L$ is the set $\{ Q \in \CC[\Newtint] \ | \ F(\mathcal P+Q)\leq L\}$.)
\et

\proof{
Since $F$ is continuous, its level sets are closed.

It remains to prove that they are bounded.
Let $L>\inf F$, so that the level set is not empty.

If $F(P)\leq L$, this implies:
\bea
&&  \frac{1}{2\pi i} \int_{\curve\setminus \cup_\alpha D_\alpha} |Y dX|^2   + \sum_\alpha \sum_{k=1}^\infty \left|\frac{1}{\sqrt k}\bar t_{\alpha,k} R_\alpha^{-k} - \sqrt k \td t_{\alpha,k} R_\alpha^k \right|^2 \cr
&\leq &
4L + \frac{1}{2\pi i} \int_{\curve\setminus \cup_\alpha D_\alpha} |\mathcal Y d \mathcal X|^2  - 4 F(\mathcal P) 
+ \sum_\alpha \sum_{k=1}^\infty \left|\frac{1}{\sqrt k}\bar t_{\alpha,k} R_\alpha^{-k} - \sqrt k \td t_{\alpha,k}(\mathcal P) R_\alpha^k \right|^2  = \td L .
\cr
\eea
In particular 
\beq
 \int_{\curve\setminus \cup_\alpha D_\alpha} |y|^2 \ d^2 x   \leq \pi \td L .
\eeq

Let $U$ an open subset of $\curve_{\mathcal P} \setminus \cup_\alpha D_\alpha$, that excludes some small disks around all ramification points of $\mathcal P$.
There exists some $K>0$ such that
\beq
\min_{x\in U, \ i\neq j} |\mathcal Y_i(x)-\mathcal Y_j(x)| \geq K >0,
\eeq
where $\mathcal Y_i(x)$ are the roots of $\mathcal P(x,y)=0$.

Let also $r>0$ small enough so that there is $V$ an open subset of $U$, such that for all $x_0\in X(V)$, the ball $X^*D(x_0,r)$ is contained in $U$.
Let $r$ small enough, so that there are at least $2\#\Newt+1$ disjoint discs of radius $r$ in $V$, denoted $D_1,\dots,D_{2\#\Newt+1}$, of respective centers $q_1,\dots,q_{2\#\Newt+1}$.

Let
\bea
||Q||
&=& \left(\int_{U} d^2 x \ \frac{|Q(x,\mathcal Y)|^{2/d}}{|\mathcal P'_y(x,\mathcal Y)|^{2/d}} \right)^{d/2}, \cr
\eea
where $\ii^{-1}(x,\mathcal Y)\in U $ designates a point on the curve $\mathcal P(x,\mathcal Y)=0$, i.e. $\mathcal Y = \mathcal Y_i(x)$ for some $i$.

Remark that $Q(x,\mathcal Y) = \mathcal P(x,\mathcal Y)+Q(x,\mathcal Y)=P(x,\mathcal Y) = \mathcal P_d(x)\prod_{i=1}^d (\mathcal Y-Y_i(x))$, and $\mathcal P'_y(x,\mathcal Y)= \mathcal P_d(x)\prod_{i=2}^{d-1} (\mathcal Y-\mathcal Y_i(x))$, where we labeled $\mathcal Y_1=\mathcal Y$.
This gives
\bea
||Q||^{2/d}
& = & \int_{U} d^2 x \ \frac{|Q(x,\mathcal Y)|^{2/d}}{|\mathcal P'_y(x,\mathcal Y)|^{2/d}}  \cr
& \leq & \frac{1}{ K^{2\frac{d-1}{d}}} \int_{U} d^2 x \ |\prod_{i=1}^d (\mathcal Y-Y_i(x))|^{2/d}  \cr
& \leq & \frac{1}{d \ K^{2\frac{d-1}{d}}} \int_{U} d^2 x \ \sum_{i=1}^d |\mathcal Y-Y_i(x)|^2  \qquad \leftarrow \text{ AM-GM inequality}\cr
& \leq & \frac{2}{d \ K^{2\frac{d-1}{d}}} \int_{U} d^2 x \ \sum_{i=1}^d (|Y_i(x)|^2+|\mathcal Y|^2) \cr
& \leq & \frac{2}{d \ K^{2\frac{d-1}{d}}} \left(\pi \td L + d \int_{U} d^2 x \ |\mathcal Y|^2\right) .
\eea

This implies that $||Q||$ is bounded on the level sets of $F$.

However, $||Q||$ is not a norm (it would be the H\"older norm if $d/2\leq 1$ but here we have $d/2\geq 1$), so we can not yet conclude.

Let us show the following lemma:
\bl
For all $x_0\in X(V)$ and $\ii^{-1}(x_0,\mathcal Y(x_0))\in V$, there exists $x\in D(x_0,r)$ such that $\left|\frac{Q(x,\mathcal Y(x))}{\mathcal P'_y(x,\mathcal Y(x))}\right| \leq \left(\pi r^2\right)^{-d/2} ||Q||  $.
There exist at least $\#\Newtint$  points among $q_1,\dots,q_{2\#\Newt+1}$, for which $\left|\frac{Q(q_i,\mathcal Y(q_i))}{\mathcal P'_y(q_i,\mathcal Y(q_i))}\right| \leq \left(\pi r^2\right)^{-d/2} ||Q||  $.
\el
\proof{
For all $\ii^{-1}(x_0,\mathcal Y(x_0))\in V$, we have either:
\begin{itemize}

\item $\left|\frac{Q(x_0,\mathcal Y(x_0))}{\mathcal P'_y(x_0,\mathcal Y(x_0))}\right| \leq \left(\pi r^2\right)^{-d/2} ||Q||  $  
\item or $\left|\frac{Q(x_0,\mathcal Y(x_0))}{\mathcal P'_y(x_0,\mathcal Y(x_0))}\right| > \left(\pi r^2\right)^{-d/2} ||Q||  $  

\end{itemize}
In this second case, $Q(x_0,\mathcal Y(x_0))\neq 0 $.
Assume that there is no $x$ in $D(x_0,r)$ such that $Q(x,\mathcal Y(x))= 0 $, we then have
\bea
\left(\frac{Q(x_0,\mathcal Y(x_0))}{\mathcal P'_y(x_0,\mathcal Y(x_0))}\right)^{2/d}
&=& \Res_{x\to x_0} \frac{dx}{x-x_0} \left(\frac{Q(x,\mathcal Y(x))}{\mathcal P'_y(x,\mathcal Y(x))}\right)^{2/d} \cr
&=& \frac{1}{2\pi} \int_0^{2\pi} d\theta \left(\frac{Q(x_0+\td r e^{i\theta},\mathcal Y(x_0+\td r e^{i\theta}))}{\mathcal P'_y(x_0+\td r e^{i\theta},\mathcal Y(x_0+\td r e^{i\theta}))}\right)^{2/d}  \qquad \forall \td r \leq r \cr
&=& \frac{1}{\pi r^2}  \int_{0}^r \td r d\td r \int_0^{2\pi} d\theta \left(\frac{Q(x_0+\td r e^{i\theta},\mathcal Y(x_0+\td r e^{i\theta}))}{\mathcal P'_y(x_0+\td r e^{i\theta},\mathcal Y(x_0+\td r e^{i\theta}))}\right)^{2/d}.   \cr
\eea
This implies that
\bea
\left|\frac{Q(x_0,\mathcal Y(x_0))}{\mathcal P'_y(x_0,\mathcal Y(x_0))}\right|^{2/d}
&\leq & \frac{1}{\pi r^2} \int_{U\cap X^{-1}(D(x_0,r))} d^2 x \left|\frac{Q(x_0+\td r e^{i\theta},\mathcal Y(x_0+\td r e^{i\theta}))}{\mathcal P'_y(x_0+\td r e^{i\theta},\mathcal Y(x_0+\td r e^{i\theta}))}\right|^{2/d}   \cr
&\leq & \frac{1}{\pi r^2} \ ||Q||^{2/d}.   \cr
\eea
This contradicts our hypothesis. Therefore there exists $x\in D(x_0,r)$ such that $Q(x,\mathcal Y(x))=0$. 

Then, notice that $Q(x,\mathcal Y(x))$ can have at most $\#\Newt$ zeros on $\curve$, therefore, among the discs $D_1,\dots,D_{2\#\Newt+1}$, $Q(x,\mathcal Y(x))$ can have a zero in at most half of them, and therefore has no zero in the others, and thus is bounded by $\left(\pi r^2\right)^{-d/2} ||Q||$ in at least $\#\Newtint$ of them.
This proves the lemma.
}

Then, consider the $\#\Newtint$  points $u_l=(q_l,\mathcal Y(q_l))$ for $l=1,\dots,\#\Newtint$.
By definition we have
\beq
A \vec Q = \vec B
\eeq
with $\vec Q=(Q_{i,j})_{(i,j)\in \Newtint}$ the $\#\Newtint$ dimensional vector of coefficients of $Q$, $\vec B=(Q(u_l)/\mathcal P'_y(u_l)  )_{l=1,\#\Newtint}$ the $\#\Newtint$ dimensional vector of evaluations,
and  $A$ the $\#\Newtint\times \#\Newtint$ square matrix $A_{l;(i,j)} = q_l^i \mathcal Y(q_l)^j $.
The matrix $A$ is invertible, and is independent of $Q$.
This gives
\beq
\vec Q = A^{-1} \vec B.
\eeq
With the sup-norm this gives
\beq
||Q||_{sup} \leq ||A^{-1}|| \ ||B||_{sup},
\eeq
which shows that all coefficients $Q_{i,j}$ are bounded.

Therefore the level sets are compact.
}

\bt[Minimum]
\label{th:minF}
$F$ admits a minimum

\et

\proof{
$F$ is continuous, it is bounded from below, and its level sets are compact.
The intersection of all level sets
\beq
\cap_{L>\inf F} \{ Q \ | \ F(\mathcal P+Q)\leq L\}
\eeq
is a decreasing intersection of non-empty compacts, therefore it is a non-empty compact.
Let $Q$ an element of this compact.
We have 
\beq
F(\mathcal P+Q) = \inf F,
\eeq
so it is a minimum.
}

\subsection{Derivatives}

As a corollary of proposition \ref{prop:cellMgfromdimnopole}, we have

\bl[Cotangent space]
The cotangent space of an affine space is isomorphic to the underlying vector space
\beq
T^*\modsp \sim \CC[\Newtint].
\eeq
It is isomorphic to $H'^1(\curve)$:
\bea
T_P^*\modsp  & \to & H'^1(\curve_P) \cr
\delta P_{k,l} & \mapsto & \omega_{k,l} = - \frac{x^k y^l dx}{P'_y(X,Y)}.
\eea
\el
\proof{
Since $P(X,Y)=0$, we have
\beq
P'_y(X,Y)\delta Y + \sum_{k,l} \delta P_{k,l} X^k Y^l =0,
\eeq
and thus
\beq
\delta Y dx  = - \sum_{k,l} \delta P_{k,l} \ \frac{x^k y^l dx}{P'_y(X,Y)}.
\eeq
If $(k,l)\in \Newtint$, then $\omega\in H'^1(\curve)$ has no poles at punctures, it means that $\delta t_{\alpha,k}=0$ for all 2nd kind and 3rd kind times.
}

\bp[Derivative]
Let us consider a deformation $\delta  \in T_P^*\modsp$.
Let
\beq
\omega = \sum_{k,l} \omega_{k,l} \delta P_{k,l} = \delta YdX,
\qquad \text{ where} \qquad
\omega_{k,l} = - \frac{x^k y^l dx}{P'_y(X,Y)}.
\eeq
We have
\beq
4 \delta F 
= \frac{1}{\pi} \Im \int_{\curve \setminus \cup_\alpha D_\alpha}  \overline{Y dX} \wedge \omega 
 + \sum_\alpha \frac{1}{\pi} \Im \int_{D_\alpha}  (\overline{Y dX}-\overline{dV_\alpha}) \wedge \omega -2\sum_\alpha\Re \Res_\alpha V_\alpha \omega.
\eeq
This is independent of the radius $R_\alpha$.
Therefore it has a limit as $R_\alpha\to 0$.
Since $Y dX-dV_\alpha$ can have at most a simple pole at $\alpha$ and $\omega$ is holomorphic at $\alpha$, the last integral tends to 0 as $R_\alpha\to 0$, and therefore the first integral has a limit.
We write:
\beq
 \frac{1}{4\pi} \Im \int_{\curve }  \overline{Y dX} \wedge \omega 
= \lim_{R_\alpha\to 0} \frac{1}{4\pi} \Im \int_{\curve \setminus \cup_\alpha D_\alpha}  \overline{Y dX} \wedge \omega .
\eeq

\ep

\proof{
Stokes theorem.
}

\bp[Second Derivative]
Let $\delta\in T^*\modsp$.
Define
\beq
\Omega = \sum_{(k,l)} \delta P_{k,l} \sum_{(i,j)} \delta P_{i,j}
 \frac{\omega_{k,l}\omega_{i,j}}{Y dX} \left( \frac{y P''_{y,y}(x,y)}{P'_y(X,Y)} - l-j\right).
\eeq
The Hessian, is the Hermitian quadratic form $H(\omega,\omega)$:
\beq
H(\omega,\omega) 
= \frac{1}{4\pi}\Im \int_{\curve }  \overline{\omega} \wedge \omega  + \frac{1}{4\pi}\Im \int_{\curve }  \overline{Y dX} \wedge \Omega
-\frac12 \Re \sum_\alpha \Res_{\alpha} V_\alpha \Omega
.
\eeq

\ep

\proof{Simple computation.}

\subsection{Minimization with constrained prescribed Integrals }

Let $\mathcal L$ an algebraic 3-dimensional sub-manifold of $\CC\times \CC$, with boundaries at most above the punctures.
Generically, $\mathcal L\cap \td\curve$ will be a 1-dimensional algebraic submanifold of $\td\curve$, that we can write as a finite union of smooth Jordan arcs.
These arcs may end at the punctures or not.
Up to homotopic deformations, they can be moved to integer linear combinations of cycles $\mathcal A_i, \ i=1,\dots,2\genus $, or small circles around the punctures, or arcs that end at the punctures.
Therefore
\bea
\int_{\ii^{-1}(\mathcal L\cap\td\curve)}
Y dX 
&=& \sum_{i=1}^{2\genus} c_i(\mathcal L) \oint_{\acycle_i} Y dX + \sum_{\alpha} c_\alpha(\mathcal L) \tilde t_{\alpha,0} + \sum_{\alpha} \td c_\alpha(\mathcal L) 2\pi\ii \Res_\alpha Y dX \cr
&=& \sum_{i=1}^{2\genus} c_i(\mathcal L) 2\pi\ii \eta_i + \sum_{\alpha} c_\alpha(\mathcal L) \tilde t_{\alpha,0} + \sum_{\alpha} \td c_\alpha(\mathcal L) 2\pi\ii t_{\alpha,0}, \cr
\eea 
with $c_i(\mathcal L) , c_\alpha(\mathcal L), \td c_\alpha(\mathcal L) $ integers.
If we take the real part, due to hypothesis of \ref{hyp:realresidues}, we have
\bea
\Re \int_{\ii^{-1}(\mathcal L\cap\td\curve)}
Y dX 
&=& \sum_{i=1}^{2\genus} c_i(\mathcal L) 2\pi \epsilon_i + \sum_{\alpha} c_\alpha(\mathcal L) \tilde t_{\alpha,0}  .
\eea 

Remark that the integers $c_i(\mathcal L) , c_\alpha(\mathcal L), \td c_\alpha(\mathcal L) $ are locally constant, but they can be discontinuous over $\modsp$.

\bd[Moduli space with prescribed integrals]

Let $\mathcal L_1,\dots,\mathcal L_N$ be given. Let 
 $\ell_1,\dots,\ell_N$ be $N$ real numbers.
 Let
\beq
\modsp(\mathcal P;\mathcal L_i,\ell_i) = \left\{ P=\mathcal P+Q\in \modsp \ | \ \forall i=1,\dots,N \ \quad \Re \int_{\ii^{-1}(\mathcal L_i\cap\td\curve)} Y dX = 2\pi\ell_i \right\}.
\eeq
\ed

\bt
If $\modsp(\mathcal P;\mathcal L_i,\ell_i)$ is a non-empty closed subset of $\modsp$, then
the restriction $F$ is continuous on it, and it has an infimum
\beq
\inf_{\modsp(\mathcal P;\mathcal L_i,\ell_i)} F \geq \inf_{\modsp} F >-\infty.
\eeq
It has a minimum.

\et

\proof{
The maps $P \mapsto \ell_i$ are continuous, so $\modsp(\mathcal P;\mathcal L_i,\ell_i)$ is closed.


The continuity of $F$ on it, and the infimum are trivial.

Since $F$ is continuous on $\modsp(\mathcal P;\mathcal L_i,\ell_i)$, the level sets are closed.
We have already seen that level sets are bounded, so they are compact.
The intersection of all level sets, is the intersection of a decreasing sequence of non-empty compacts, so is a non-empty compact.
A point in the intersection is a minimum.
}
\subsection{Energy from Prepotential}

Here we shall see another definition of the energy $F$.
We shall define a function $\check F$ on $\modsp$, from the prepotential $F_0$, and we shall then prove that $\check F$ and $F$ are equals.
The advantage is that this expression of $F$ will be expressed in local period coordinates, and will allow to see how a minimum is related to the Boutroux condition.

Let us choose some genus $\genus\leq \dim\CC[\Newtint]$.
Let $U\subset \modsp^{(\genus)}$ a  simply connected open domain of $\modsp^{(\genus)}$, in which we choose a continuous symplectic Jordan cycles marking (lemma \ref{Lemmacontinuouscylces}), and we choose a fundamental domain $D$ as in Definition \ref{def:fondDomain}, continuous on $U$.

This allows to have period coordinates well defined over $U$, as well as puncture-times:
\beq
\begin{split}
t_{\alpha,k} = \Res_\alpha \zeta_\alpha^k\,Y dX, 
&\qquad \alpha=\text{punctures}\ \ k=0, \dots, r_\alpha\cr
\eta_i = \frac{1}{2\pi\ii}\oint_{\acycle_i} Y dX, 
&\qquad i=1,\dots\genus \cr
\end{split}
\eeq
as well as their conjugate times
\beq
\begin{split}
\td t_{\alpha,k} = \frac{1}{k}\Res_\alpha \zeta_\alpha^{-k}\,Y dX, 
&\qquad \alpha=\text{punctures}\ \ k=1, \dots, r_\alpha\cr
\td t_{\alpha,o} = g(z)-g_\alpha(z), &\qquad \text{independent of }z \in D  \cr
\td\eta_i = \frac{1}{2\pi\ii}\oint_{\bcycle_i} Y dX, 
&\qquad i=1,\dots\genus 
\end{split}
\eeq

\bd[Free energy]
We define the prepotential $F_0$ as (see \cite{EO07, E2017})
\bea
F_0 &:=& \frac12  \left( \sum_\alpha \sum_{k=1}^{r_\alpha} t_{\alpha,k} \td t_{\alpha,k} + \sum_\alpha t_{\alpha,0} \td t_{\alpha,o} + 2\pi\ii \sum_{i=1}^\genus \eta_i \td\eta_i \right) .
\eea

We also define
\beq
\hat F := \frac12  \left( \sum_\alpha \sum_{k=1}^{r_\alpha} t_{\alpha,k} \td t_{\alpha,k} + \sum_\alpha t_{\alpha,0} \td t_{\alpha,o}  \right) 
 = F_0 - \pi\ii \sum_{i=1}^\genus \eta_i \td\eta_i.
\eeq

In addition, define
\beq
\check F := -\Re \hat F + \pi (\td\zeta^t \epsilon - \zeta^t \td\epsilon)
= -\Re \hat F + \pi \zeta^t E^{-1} \epsilon,
\eeq
where $E$ is the symplectic matrix  of size $2\genus$
\beq\label{eq:Esympl}
E=\begin{pmatrix}
0 & \text{Id} \cr
- \text{Id} & 0
\end{pmatrix}.
\eeq

\ed

\br
Notice that $F_0$ depends on the choice of fundamental domain $D$, and of the symplectic Jordan loops marking. 
It is not a function of $P$ alone. In other words $F_0$ is not defined as a function on $\modsp$.

\er

\bp
$\check F$ and $\hat F$ are independent of a choice of marking of Jordan loops.
Moreover, we have in the cotangent space $T^*U$, the following differentials
\beq
dF_0 = 2\pi\ii \sum_{i=1}^{\genus} \td\eta_i d\eta_i,
\eeq
\beq
d\hat F = \pi\ii \left(\sum_i \td\eta_i d\eta_i - \eta_i d\td\eta_i\right), 
\eeq
and
\beq\label{eq:dF}
d\check F = 2\pi\left( \sum_{i=1}^{\genus} \td\zeta_i d\epsilon_i - \zeta_i d\td\epsilon_i \right).
\eeq

If we choose an arbitrary basis of cycles, not necessarily symplectic, then we have
\beq
d\hat F = \pi\ii \ \eta^t E^{-1} d\eta,
\eeq
\beq
d\check F = 2\pi\ \zeta^t E^{-1} d\epsilon,
\eeq
where $E$ is the $2\genus\times 2\genus$ intersection matrix $E_{i,j} = \acycle_i\cap \acycle_j=-E_{j,i} $.

\ep

\proof{
The fact that $\hat F$ is independent of a choice of marking is obvious, because we subtracted from $F_0$ the part that depends on it.
Then, consider a change of Jordan loop marking, by taking linear combinations of them. 
This implies that $\eta$ changes to $C\eta$ where $C$ is an invertible matrix with integer coefficients, so that $\epsilon$ changes to $C\epsilon$ and $\zeta\to C\zeta$, and $E$ changes to $C E C^t$. 
This shows that the  $\zeta^t E^{-1} \epsilon$ is  invariant under a change of basis, and therefore $\check F$ is independent of a choice of marking.

\medskip

The relation
\beq
dF_0 = {2\pi\ii} \sum_i \td\eta_i d\eta_i
\eeq
comes from the fact that $F_0$ is the Seiberg-Witten prepotential, this was proved for instance in \cite{EO07,bertola2007boutroux}.

The expression for $d\hat F$ follows immediately, and since we are in a symplectic basis with $E$ of the form eq \eqref{eq:Esympl}, this takes the form 
\beq
d\hat F = \pi\ii \ \eta^t E^{-1} d\eta,
\eeq
which is clearly invariant under any change of basis of cycles.

Then, taking the real part we have
\beq
\begin{aligned}
d\Re\hat F 
&=&& -\pi \left( \zeta^t E^{-1} d\epsilon + \epsilon^t E^{-1} d\zeta \right) \cr
&=&& -\pi \left( \zeta^t E^{-1} d\epsilon - d\zeta^t E^{-1} \epsilon\right), \cr
\end{aligned}
\eeq
and thus
\beq
d \check F  =-d\Re \hat F + \pi d(\zeta^t E^{-1} \epsilon) = 2\pi \zeta^t E^{-1} d\epsilon .
\eeq
}

\bp
The map $\hat F$:
\bea
\modsp &\to \mathbb C \cr
P &\mapsto \hat F
\eea
is well defined, and is holomorphic  in each $\modsp^{(\genus)}$ (with respect to the complex structure of  period coordinates $\eta_i$).

The map $\Re \hat F$ :
\bea
\modsp &\to \mathbb R \cr
P &\mapsto \Re\hat F
\eea
is well defined, and is locally harmonic in each $\modsp^{(\genus)}$.

The map $\check F$ :
\bea
\modsp &\to \mathbb R \cr
P &\mapsto \check F
\eea
is well defined (it is not harmonic).

\ep

\proof{
There is no continuous section of Jordan loops marking over the full $\modsp $, and this is why we used some open set $U\subset \modsp^{(\genus)} $ to define $\check F$. However, we have seen that $\check F$ and $\hat F$ are in fact independent of the choice of marking, so they are well defined over the full $\modsp^{\genus} $ and also other their disjoint union $\modsp$.
}

\bt[Hessian of $\check F$]\label{th:Hessian}
Let $\genus\leq \#\Newtint$, and let $U\subset \modsp^{(\genus)} $ an open  domain, in which we choose a continuous marking of Jordan cycles (lemma \ref{Lemmacontinuouscylces}).
In $U$ we use the period coordinates $\epsilon_1,\dots,\epsilon_{2\genus}$.
We consider the real and imaginary parts of the Riemann matrix of periods
\beq
\tau=R+\ii \ I
 , \quad
I>0,\quad  R=R^t, \quad I=I^t.
\eeq
The $2\genus\times 2\genus$ Hessian matrix of $\check F$ is
\bea
\frac{1}{2\pi}\frac{\partial^2 \check F}{\partial\epsilon_i \partial\epsilon_j} 
&=& \begin{pmatrix}
I+R I^{-1} R & -R I^{-1} \cr
-I^{-1} R &  I^{-1}
\end{pmatrix}  \cr
 &=& \begin{pmatrix}
\mathbf{1} & -R \cr
0 & \mathbf 1 \end{pmatrix}\,
\begin{pmatrix}
 I & 0 \cr
0 &  I^{-1} \end{pmatrix}\,
\begin{pmatrix}
\mathbf{1} & 0 \cr
-R & \mathbf 1
 \end{pmatrix}\,
,\eea
which is symmetric (as any Hessian matrix) and positive definite.
$\check F$ is strictly convex in any convex subdomain of $U$ (convex in period coordinates).

\et

\proof{
This is a simple computation.
Moreover, since $I>0$ this matrix is clearly positive definite, and thus invertible. It is also clearly symmetric.
}

\bc
\label{cor:minFBoutroux}
$\check F$ is strictly convex in $U$ (in the real coordinates $\epsilon$).
If $\check F$ has a minimum in a convex $U$, then it is unique and is a Boutroux curve:
\beq
\forall \ \gamma\in H_1(\curve\setminus\text{punctures},\ZZ)
 , \quad \Re\oint_\gamma Y dX=0.
\eeq
\ec

\proof{
Strict convexity comes from the fact that the Hessian is positive definite.
From strict convexity, it is clear that the minimum if it exists is unique.
(A minimum doesn't necessarily exist in $U$, it could be at the boundary of $U$ and we shall discuss that issue later below).

Consider that a minimum is reached in $U$, this implies that $d\check F=0$, i.e. 
\beq
\forall \ i=1,\dots,2\genus \ , \qquad \zeta_i=0 = \frac{1}{2\pi}\Re \oint_{\acycle_i} Y dX,
\eeq
where we used the convention that $\bcycle_i=\acycle_{\genus+i}$.

Also, if $\gamma$ is a small circle around a puncture $\alpha$ we have
\beq
\Re \int_\gamma Y dX = \Re 2\pi\ii \Res_\alpha Y dX = -2\pi \Im t_{\alpha,0} = 0,
\eeq
by our assumption \ref{hyp:realresidues}.

Since any closed Jordan loop $\gamma$ on  $\curve\setminus\text{puncture}$ is homotopic to an integer linear combination of cycles $\acycle_i$s and circles around punctures, this implies the Boutroux condition.
}

\subsection{Uniqueness of the energy}

\bt
The two definitions of $F$ coincide
\beq 
F = \check F.
\eeq
\et

\proof{
We recall that we chose a fundamental domain $D$, with its symplectic marking of cycles bordering $D$, and we have defined $g(z) = \int_{o_i}^z Y dX $ analytic in $D$.

On $D$ we have
\bea
\frac{1}{2\pi\ii}\int_{\curve\setminus \cup_\alpha D_\alpha} \overline{Y dX} \wedge Y dX
&=& \frac{1}{2\pi\ii}\int_{\partial(\curve\setminus \cup_\alpha D_\alpha)} \overline{g} \ Y dX \cr
&=& \frac{-1}{2\pi\ii} \sum_\alpha  \int_{\mathcal C_\alpha} \overline{g} \ Y dX \cr
&& + \frac{1}{2\pi\ii} \sum_\alpha  \int_{o_i}^{p_\alpha}  \overline{2\pi \ii t_{\alpha,0}} \ Y dX \cr
&& + \frac{1}{2\pi\ii} \sum_{i=1}^\genus  \int_{\acycle_i} \ Y dX   \left( \overline{-\int_{\bcycle_i} \ Y dX} \right) \cr
&& + \frac{1}{2\pi\ii} \sum_{i=1}^\genus  \int_{\bcycle_i} \ Y dX   \left( \overline{\int_{\acycle_i} \ Y dX} \right) \cr
&=& \frac{-1}{2\pi\ii} \sum_\alpha  \int_{\mathcal C_\alpha} \overline{(\td t_{\alpha,0}+g_\alpha)} \ Y dX \cr
&& - \sum_\alpha  t_{\alpha,0} (g(p_\alpha)-g(o_i)) \cr
&& + \frac{1}{2\pi\ii} \sum_{i=1}^\genus  2\pi\ii \eta_i \left( \overline{-2\pi\ii \td \eta_i} \right) \cr
&& + \frac{1}{2\pi\ii} \sum_{i=1}^\genus  2\pi\ii \td\eta_i \left( \overline{2\pi\ii \eta_i} \right) \cr
&=& -\sum_\alpha \overline{\td t_{\alpha,0}} \ t_{\alpha,0} -  \frac{1}{2\pi\ii} \sum_\alpha  \int_{\mathcal C_\alpha} \overline{g_\alpha} \ Y dX \cr
&& - \sum_\alpha  t_{\alpha,0} \ g(p_\alpha) \cr
&& + 2\pi\ii \sum_{i=1}^\genus  \eta_i \overline{\td \eta_i} - \td \eta_i \ \overline{\eta_i}  \cr
&=& -\sum_\alpha \overline{\td t_{\alpha,0}} \ t_{\alpha,0} -  \frac{1}{2\pi\ii} \sum_\alpha  \int_{\mathcal C_\alpha} \overline{g_\alpha} \ Y dX \cr
&& - \sum_\alpha  t_{\alpha,0} \ g(p_\alpha) 
 +4\pi \Im \sum_{i=1}^\genus \td \eta_i \ \overline{\eta_i}  \cr
&=& -\sum_\alpha \overline{\td t_{\alpha,0}} \ t_{\alpha,0} -  \frac{1}{2\pi\ii} \sum_\alpha  \int_{\mathcal C_\alpha} \overline{g_\alpha} \ Y dX \cr
&& - \sum_\alpha  t_{\alpha,0} \ g(p_\alpha) 
 +4\pi  \sum_{i=1}^\genus \td \zeta_i \epsilon_i - \zeta_i \td \epsilon_i\cr
\eea

Then we use lemma \ref{lem:intCalpha} in appendix \ref{apx:lemmaintCalpha}:
\beq
\frac{1}{2\pi\ii} \int_{\mathcal C_\alpha} \overline{g_\alpha} \ Y dX
=
- \sum_{k=1}^{r_\alpha} \frac{|t_{\alpha,k}|^2}{k} R_\alpha^{-2k}
+ \sum_{k=1}^{\infty} k |\td t_{\alpha,k}|^2 R_\alpha^{2k} + 2 t_{\alpha,0}^2 \ln R_\alpha  -t_{\alpha,0} g_\alpha(p_\alpha)+\pi \ii t_{\alpha,0}^2 .
\eeq

This implies
\bea
&& \frac{1}{2\pi\ii}\int_{\curve\setminus \cup_\alpha D_\alpha} \overline{Y dX} \wedge Y dX
 -\sum_{\alpha}\sum_{k=1}^{r_\alpha} \frac{1}{k} |t_{\alpha,k}|^2 R_{\alpha}^{-2k} + \sum_{\alpha}\sum_{k=1}^{\infty} k |\td t_{\alpha,k}|^2 R_{\alpha}^{2k} 
 +2\sum_{\alpha}|t_{\alpha,0}|^2 \ln R_\alpha \cr
&=& -\sum_\alpha \overline{\td t_{\alpha,0}} \ t_{\alpha,0} 
- \sum_\alpha  t_{\alpha,0} \ g(p_\alpha) 
 +4\pi  \sum_{i=1}^\genus \td \zeta_i \epsilon_i - \zeta_i \td \epsilon_i  + \sum_{\alpha}t_{\alpha,0} g_\alpha(p_\alpha) \cr
&=& -\sum_\alpha \overline{\td t_{\alpha,0}} \ t_{\alpha,0} 
- \sum_\alpha  t_{\alpha,0} \ \td t_{\alpha,0}
 +4\pi  \sum_{i=1}^\genus \td \zeta_i \epsilon_i - \zeta_i \td \epsilon_i  \cr
&=& -2 \Re \left( \sum_\alpha t_{\alpha,0} \ \td t_{\alpha,0}
 -2\pi  \sum_{i=1}^\genus \td \zeta_i \epsilon_i - \zeta_i \td \epsilon_i  \right),  \cr
\eea
and thus
\bea
4F &=& 
 \frac{1}{2\pi\ii}\int_{\curve\setminus \cup_\alpha D_\alpha} \overline{Y dX} \wedge Y dX
 -\sum_{\alpha}\sum_{k=1}^{r_\alpha} \frac{1}{k} |t_{\alpha,k}|^2 R_{\alpha}^{-2k} + \sum_{\alpha}\sum_{k=1}^{\infty} k |\td t_{\alpha,k}|^2 R_{\alpha}^{2k} 
 +2\sum_{\alpha}|t_{\alpha,0}|^2 \ln R_\alpha \cr
&&  -2 \Re  \sum_\alpha \sum_{k=1}^{r_\alpha} t_{\alpha,k} \td t_{\alpha,k} \cr
&=& -2 \Re \left( \sum_\alpha t_{\alpha,0} \ \td t_{\alpha,0} + \sum_{k=1}^{r_\alpha} t_{\alpha,k} \td t_{\alpha,k}
 -2\pi  \sum_{i=1}^\genus \td \zeta_i \epsilon_i - \zeta_i \td \epsilon_i  \right)  \cr
&=& -4 \Re \hat F  +4\pi   \left( 
\sum_{i=1}^\genus \td \zeta_i \epsilon_i - \zeta_i \td \epsilon_i  \right)  \cr
&=& 4 \check F.
\eea
}

\section{Boutroux Curves}
\label{sec:BC}
This is the main theorem
\bt[Boutroux Curve]
\label{th:existBoutroux}
There exists at least one Boutroux curve in $\modsp$.
Boutroux curves are isolated in $\modsp$.

\et

\proof{
Because of theorem~\ref{th:minF}, $F$ admits at least one minimum on $\modsp$. Let $P=\mathcal P+Q$ a minimum of $F$.
It belongs to some $\modsp_\genus$, with $\genus\leq \dim\CC[\Newtint]$.

It may happen that $\genus=0$, in which case, the Boutroux condition is trivially satisfied (all cycles are contractible or reduce to small circles around punctures and we have $\Re \oint_{\mathcal C_\alpha} YdX = \Re( 2\pi \ii\Res_\alpha YdX) =\Re( 2\pi\ii t_{\alpha,0})=0 $ ).

Otherwise, eq \eqref{eq:dF} i.e. corollary \ref{cor:minFBoutroux} implies that $dF=0$ in $\modsp_\genus$, which implies $\zeta_i=0$ for all $i=1,\dots,2\genus$, and therefore we get Boutroux condition.

Boutroux curves are isolated, because in the period coordinates, $F$ is locally strictly convex.
}
\bt
If $\mathcal P+Q$ is a Boutroux curve we have
\beq
F(\mathcal P+Q) = -\Re F_0(\mathcal P+Q).
\eeq

\et

\proof{
We have for any $P$
\beq
F = -\Re F_0 -2\pi \zeta^t \td\epsilon,
\eeq
and $\zeta$ vanishes for Boutroux curves.
}

\section{Spectral network of first kind}
\label{sec:spnetwork1}

A Boutroux curve has canonically some graphs associated to it, often called \textbf{spectral networks}.
However, there is two versions used in many applications.
For hyperelliptic curves (degree two in $y$) the two versions almost coincide as we shall see in subsection \ref{sec:spnetwk12}.

Let's denote $\curve\setminus \text{punctures}$ by $\curve^*$.

\bt[Harmonic function]
 If $P\in\modsp $ is a Boutroux curve, let $o_i\in\curve_i$ a generic point in each connected component $\curve_i$ of $\curve$.
The following function:
\bea
\phi : & \curve \to & \RR \cr
& p \mapsto & \phi(p) = \Re \int_{o_i}^p Y dX
\eea
is well defined and harmonic on $\curve^*$.

\br For $x\leftrightarrow y $ symmetry, we have the following function:
\bea
\td\phi : & \curve \to & \RR \cr
& p \mapsto & \phi(p) = \Re \int_{o_i}^p XdY
\eea
which is also well defined and harmonic on $\curve^*$.
\er
\et

\proof{
The integration path from $o_i$ to $p$ is not unique, but two different paths differ by a closed Jordan loop $\gamma$, and $\Re \int_\gamma Y dX=0 $, so $\phi(p)$ is independent of the chosen path.
This makes it a well defined function on $\curve^*$.
It is the real part of a locally analytic function, so it is harmonic.

For $\td\phi$, notice that by integration by parts, on any Jordan loop one has $\int_\gamma XdY = -\int_\gamma Y dX$, and also $\Res_\alpha XdY = -\Res_\alpha Y dX=-t_{\alpha,0} $.
}

\bd[Spectral Network]
\label{def:SpNcl}
For each  $a\in\curve $ that is a ramification point or a zero of $Y dX$ (in particular this includes ramification points, and all zeros of $Y$), let 
\beq
\check\Gamma_a :=  \text{connected component of } \{p \ | \ \phi(p)=\phi(a)\} \text{ that contains } a,
\eeq
and
\beq
\check\Gamma := \cup_a \check\Gamma_a.
\eeq
$\check\Gamma_a$ is called the ``vertical trajectory" passing through $a$.

\ed

\bt
Each  $\check\Gamma_a$ is a finite union of smooth Jordan arcs.

Except at zeros or poles of $Y dX$, these arcs have in the $x$-chart, a tangent in the direction $e^{-i\arg Y(x)}$.

These arcs can cross only at points where $Y dX=0$ or at punctures.
Let the arcs that end at punctures be called ``non-compact", and arcs that don't end at punctures be called ``compact".

If $a$ is a zero of $Y dX$, possibly a ramification point with canonical local coordinate $\zeta_a = (x-X_a)^{-1/a_a}$, and where $y\sim \eta_a \zeta_a^{-1/b_a}$, so that $Y dX \sim -a_a \eta_a \zeta_a^{-a_a-b_a-1}d\zeta_a$. With $a_a+b_a+1<0$, the arcs of $\check\Gamma_a$ start from $a$ at angles
\beq
e^{\ii\frac{-\arg\eta_a +\frac{\pi}{2}+ k\pi }{-a_a-b_a}}
\qquad k=1,\dots, 2|a_a+b_a|.
\eeq

\et
\proof{There is a well defined tangent $Y dX$ at each point, in a direction given by $\arg(Y dX) \in \frac{\pi}{2}+\pi \ZZ$, which implies $\arg dx = \frac{\pi}{2} -\arg Y + \pi \ZZ$.
The only points where this is not the case is when $Y dX=0$ or $Y dX$ has a pole, and at these points we use the local coordinate $\xi$.
}
\br
Since the spectral networks are described by algebraic equations the number of these trajectories is always finite.
\er
\bd[Cellular decomposition]
The complement
\beq
\curve\setminus\check\Gamma = \cup_{i=1}^m \check{\mathcal D}_i
\eeq
is a finite union of disjoint connected open sets $\check{\mathcal D}_i\subset \curve$, not containing any zero nor pole of $Y dX$.
$\phi$ is a harmonic function on each of them.
The boundaries of $\check{\mathcal D}_i$ are arcs of $\check\Gamma$, and  must contain at least a zero $a$ of $Y dX$.
Let $a$ a zero of $Y dX$ at the boundary of $\check{\mathcal D}_i$.
Let
\bea
g_a : & \check{\mathcal D}_i \to & \CC \cr
& p \mapsto & g_a(p) = \int_{a}^p Y dX.
\eea
The map $g_a$ is well defined in a neighborhood of $a$.
The real part $\Re g_a(p) = \phi(p)-\phi(a)$ is globally well defined.

\ed

\bt[Elementary pieces]
\label{th:sp1Elpieces}
The image $g_a(\check{\mathcal D}_i)\in \CC $, is a domain of $\CC$, or of $\CC/\ii\td c \ZZ$ for some $\td c\in \RR^*$, whose boundaries are (if several) vertical lines. One of the boundaries is the imaginary axis.
Since there are only two types of domains bounded by vertical lines in $\CC$ and only two types of domains bounded by vertical lines in $\CC/\ii\td c \ZZ$, only four possibilities can occur:
\begin{itemize}

\item $g_a(\check{\mathcal D}_i)$ is a half plane in $\CC$, either $\Re z>0$ or $\Re z<0$.

\item $g_a(\check{\mathcal D}_i)$ is a vertical strip in $\CC$ bounded by two lines $\Re z=0$ and $\Re z=c$ where $c\neq 0$ is some real constant. $c$ must be of the form $c=\phi(b)-\phi(a)$ for some $b$ another zero of $Y dX$.

\item $g_a(\check{\mathcal D}_i)$ is a half-cylinder, i.e. a half-plane quotiented by $z\to z+\ii \td c$ for some $\td c\in \RR^*$.

\item $g_a(\check{\mathcal D}_i)$ is a cylinder (or annulus) a vertical strip bounded by two lines $\Re z=0$ and $\Re z=c$ where $c=\phi(b)-\phi(a)$, and quotiented by $z\to z+\ii \td c$ for some $\td c\in \RR^*$.

\end{itemize}

In all cases, the map $p\mapsto g_a(p)$ (resp. $p\mapsto g_a(p)\mod \ii\td c $ if cylinder or half-cylinder), is a conformal isomorphism between $\check{\mathcal D}_i$ and its image.

In the first two cases (strip or half-plane), $\check{\mathcal D}_i$ is simply connected, and in the last two cases (cylinder or half-cylinder), $\check{\mathcal D}_i$ is not simply connected.

In all cases except cylinder, $\check{\mathcal D}_i$ has a puncture on its boundary.

Together, the $g_a(\check{\mathcal D}_i)$ form charts of an atlas of $\curve$, with transition maps that are translations $g_a(p)=g_b(p) + c+\ii \td c$.
The charts are either half-planes, strips, cylinders or half-cylinders.

For strips or cylinders, $c=\phi(b)-\phi(a) $ is called the \textbf{width}.

For cylinders and half-cylinders, $\tilde{c}$ is called the \textbf{perimeter} of the cylinder.

\et
\proof{
The image of local patches of $\check{\mathcal D}_i$ must be patches of $\CC$ bounded by vertical lines.
Since $dg_a=Y dX$ never vanishes nor has poles in $\check{\mathcal D}_i$, $g_a$ is locally a holomorphic isomorphism.
$g_a$ is not globally defined, it is defined only up to additive constants, which means that transition maps must be translations.
Since the transition maps must match the boundary $\Re z=0$, they  must be vertical translations $g\mapsto g+\ii \td c$ with $\td c\in \RR$.

The only connected domains of $\CC$ that have only vertical lines as boundaries, can only be a half-plane or a strip.
If $\td c\neq 0$, the transition maps being $g\mapsto g+\ii \td c$ imply that the image can be a half-plane or a strip quotiented by a vertical translation, i.e. a cylinder or a half-cylinder. These are the only possibilities.

If $\td c=0$ then $g_a$ is an isomorphism, and if $\td c\neq 0$ then $g_a \mod \ii \td c $ is an isomorphism.
}

\bt
In each $\check{\mathcal D}_i$, the map $X:\check{\mathcal D}_i \to \CC$ is a conformal isomorphism to its image.

\et

\proof{
If $\check{\mathcal D}_i$ is a cylinder or half-cylinder and contains a non-contractible loop $\gamma$, the projection $X(\gamma)$ in $\CC$ is contractible (because $\CC$ is simply connected). Hence, $\int_{X(\gamma)} dX =0$
and, if $\check{\mathcal D}_i$ is a half-plane or strip, it is simply connected. 

By definition $\check{\mathcal D}_i$ contains no ramification point, so $X$ is a conformal isomorphism.

}

\bt[Metric and geodesics]
The restriction of the metric $|Y dX|^2$ of $\CC\times \CC$, to $\curve$ is equal to $\frac{1}{2\ii} \overline{dg_a} \wedge dg_a$. It is thus the canonical Euclidean metric of $g_a(\check{\mathcal D}_i)$.
The geodesics are fixed angles lines $\arg dg_a = \text{constant}$, i.e. Eulcidian straight lines in the charts $g_a(\check{\mathcal D}_i)$.
As a consequence, the vertical trajectories $\phi = \text{constant}$, and therefore the edges of $\check\Gamma$ are geodesic.

\et

\bt[Half-cylinder=Fuchsian]
\label{Th:halfcylFuchsian}
Half-cylinders have some puncture $\alpha$ at their boundary, and their perimeter is $\td c = 2\pi t_{\alpha,0}$.
The puncture $\alpha$ is then necessarily a simple pole of $Y dX$.
We call it a ``Fuchsian" puncture. Therefore, half-cylinders are Fuchsian punctures.

\et

\proof{
Near $\alpha$ we have $YdX\sim t_{\alpha,r_\alpha} \zeta_\alpha^{-r_\alpha-1} d\zeta_\alpha$, and thus:

- If $r_\alpha=0$ we have $\phi\sim t_{\alpha,0} \ln{|\zeta_\alpha|}$ whose vertical trajectories are circles around $\alpha$.

- If $r_\alpha>0$ we have $\phi\sim - \Re \left( \frac{t_{\alpha,r_{\alpha}}}{r_\alpha} \zeta_\alpha^{-r_\alpha} \right)$ whose vertical trajectories can't be circles around $\alpha$.

- If we have a half-cylinder, we see that there is a foliation of circles as vertical trajectories surrounding $\alpha$, and this can be compatible only with $r_\alpha=0$, i.e. a simple pole.

}

\bt[No cylinders]
\label{th:Spntwk1NoCylynders}
There is no cylinders on the graph $\check\Gamma$ of a Boutroux curve.
\et

\proof{
The proof uses combinatorics of graphs to compute the Euler characteristics.
The Euler characteristics of $\curve$ can be computed from the number of vertices, edges and faces of $\check\Gamma$.
Let:
\begin{itemize}
\item $v=$ number of zeros of $ydx$. Each zero of $ydx$ is of some degree $v_i$.
\item $N=$ number of poles of $ydx$. Each pole of $ydx$ is of some degree $d_\alpha$.
\item $c_{1/2} = $number of Fuchsian poles, i.e. with $d_\alpha=1$.
\item $f=h+s+c+c_{1/2} = $the number of faces, where $h=$ number of half-planes, $s=$number of strips, $c=$number of cylinders and $c_{1/2}=$ number of half-cylinders. 
\item $e_c=$ number of compact edges of $\check\Gamma$, i.e. going from a zero of $ydx$ to a zero of $ydx$.
\item $e_{nc}=$ number of non-compact edges of $\check\Gamma$, i.e. going from a zero of $ydx$ to a pole of $ydx$.
\item $e=e_c+e_{nc}=$ the total number of edges.
\end{itemize}

We have the following relations:
\begin{itemize}

\item Since non-compact edges can end only on half-planes and on strips, and each half plane has 2 non-compact edges and each strip has 4, and all are doubly counted:
\beq
2 e_{nc} = 2h + 4s.
\eeq

\item Since from a zero of degree $v_i$ of $ydx$ we have $2(v_i+1)$ half edges, and half edges can be either compact or non-compact we have
\beq
2e_c+e_{nc} = \sum_i 2(v_i+1) = 2v + 2 \deg_{zeros} ydx.
\eeq

\item Since from a pole of degree $d_\alpha$ of $ydx$ we have $2(d_\alpha-1)$ half-planes, we have
\beq
h = \sum_{\alpha} 2(d_\alpha-1) = -2N + 2\deg_{poles} ydx.
\eeq

\item Together these relations imply that the total number of edges is
\bea
 e & = & e_c + e_{nc} \cr
& = & v + \deg_{zeros} ydx + s -N + \deg_{poles} ydx. \cr
\eea

\item The Euler characteristic is thus:
\bea
2-2 g
&=& f-e+ (v+N-c_{1/2}) \cr
&=& f-s-c_{1/2} -\deg_{zeros} ydx - \deg_{poles} ydx +2N \cr
&=& f-s-c_{1/2}-h -\deg_{zeros} ydx + \deg_{poles} ydx  \cr
&=& c -\deg_{zeros} ydx + \deg_{poles} ydx.  \cr
\eea
Every meromorphic 1-form on a Riemann surface of genus $g$ satisfies
\beq
\deg_{poles} ydx - \deg_{zeros}ydx = 2-2g,
\eeq
this implies that
\beq 
c=0.
\eeq
There is no cylinders on a Boutroux curve.

\end{itemize}

}

\bd[Tiles]
We can further subdivide each 
\beq
\check{\mathcal D}_i = \cup_j \check{\mathcal D}_{i,j}
\eeq
by cutting along horizontal trajectories emanating from the vertices of $\check\Gamma$ that are on the boundary of $\check{\mathcal D}_i$.

Each $\check{\mathcal D}_{i,j}$ can have two, three or four sides, that cross at right angles:
\begin{itemize}
\item If it has two sides, we call it a ``corner tile" or ``L tile", it has infinite width and height.
\item If it has three sides, we call it a ``U tile", it has either finite width, infinite height (vertical ``U") or infinite width, finite height (horizontal ``U").
\item If it has four sides, we call it a ``rectangle tile" or ``R tile", it has finite width and finite height. In particular it has a finite area = width $\times$ height.

\end{itemize}
Each tile is simply connected.

\ed

\bt
Let $\alpha$ a puncture.
Consider the union of all tiles that have $\alpha$ at their boundary.
Let $\check{\mathcal D}_\alpha$ the union of these tiles, all horizontal and vertical trajectories (their interior) ending at $\alpha$, and $\alpha$ itself.
$\check{\mathcal D}_\alpha$ is topologically a disc.

The discs $\check{\mathcal D}_\alpha$ are disjoint.

The complement
\beq
\curve \setminus \cup_\alpha \check{\mathcal D}_\alpha
\eeq
is the union of  a graph (all compact horizontal and vertical lines ) and all rectangle tiles.

\begin{figure}
\begin{center}
\includegraphics[scale=.2]{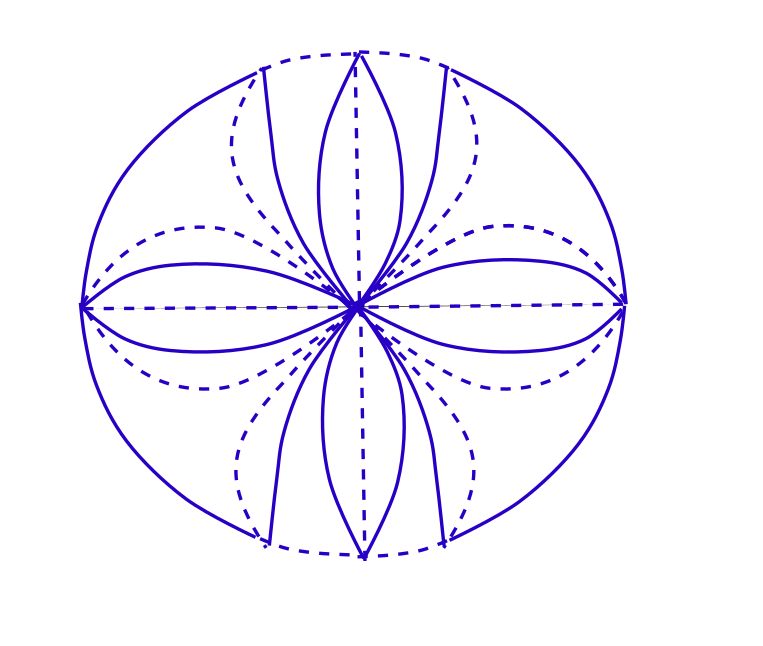}
\end{center}
\caption{Example of a domain $\check{\mathcal D}_\alpha$, it is a gluing of L and U tiles. Vertical edges are continuous and horizontal edges are dashed.}
\label{fig:Dalpha}
\end{figure}

See Fig.\ref{fig:Dalpha}.
\et

\proof{

For each puncture $\alpha$, let $D_\alpha$ a disc of radius $R_\alpha$ small enough around $\alpha$, such that $D_\alpha$ contains no zero of $YdX$ nor ramification or nodal point nor other punctures.
Consider all tiles that intersect $D_\alpha$. They must be L or U tiles.
Moreover, no tile can touch $\alpha$ without intersecting $D_\alpha$, therefore $\check{\mathcal D}_\alpha$ is precisely the union of all tiles that intersect $D_\alpha$, and they must be U or L tiles.

Since each U and L tiles have exactly two edges going to $\alpha$, 
the gluing of U and L tiles around $\alpha$ has necessarily the topology of a disc around $\alpha$.

Every L or U tile is connected, and touches at most one puncture.
This implies that $\check{\mathcal D}_\alpha$ are disjoint.

The complement is the set of edges that don't touch punctures, i.e. all compact vertical and horizontal edges, and also all rectangle tiles.
}


\section{Spectral network of second kind}
\label{sec:spnetwork2}

There is another way of defining the spectral network that is very useful in applications, in particular in WKB analysis.

We mention that for hyperelliptic curves (of the form $P(x,y) = y^2-R(x)$ with $R(x)\in \CC(x)$), the first and second kinds are closely related as we shall see in subsection \ref{sec:spnetwk12}.

\medskip

So here, we define the spectral network as:
\bd[Spectral network]
\label{def:spnetwk1C}

We define $\Gamma\subset \mathbb C$:
\beq
\Gamma:=\{x\in \mathbb C\,|\, \exists\, p\neq p',\, X(p)= X(p')=x\,\text{and}\,\phi(p)=\phi(p')\}
\eeq
(this set is independent of the choice of basepoint ``$o$" in the definition of $\phi$). $\Gamma$ is a graph embedded in $\mathbb C$. We complete $\Gamma$ by adding the vertices, i.e. we take the closure of $\Gamma$.

\ed

\bd[Spectral network on $\curve$]
\label{def:spnetwk1Sigma} The set 
\beq
\td\Gamma:=\{p\in\curve\,|\, \exists\, p'\neq p,\, X(p)= X(p')\,\text{and}\,\phi(p)=\phi(p')\}
\eeq
forms a graph $\td\Gamma$ on  $\curve$. We complete $\td\Gamma$ by adding the vertices and the punctures, i.e. we take the closure of $\td\Gamma$. 
We have $X(\td\Gamma)=\Gamma$, and $\td\Gamma\subset X^{-1}(\Gamma)$.
\ed

\br
For an arbitrary generic $x\in\mathbb C$, $X^{-1}(\{x\})=\{p_1(x),\dots,p_d(x)\}$ contains $d$ points where $d=\deg_y P$.
In addition, $\{\phi(p_1(x)),\phi(p_2(x)),\dots,\phi(p_d(x))\}$ are all distinct.
\er
For a point $x\in\mathbb C\setminus\Gamma $, let us order the $p_i$'s  by the ordering of the $\phi(p_i(x))$.
\beq
\forall\,x \ \text{generic}\, , \qquad
\phi(p_1(x))<\phi(p_2(x))<\dots<\phi(p_d(x))
\eeq

\bd
For a point $p\in \curve\setminus X^{-1}(\Gamma) $, we call the index of $p$ the integer $i(p)$, such that:
\beq
p_{i(p)}(x) = p.
\eeq
For a point $p$ that belongs to $X^{-1}(\Gamma) $ and doesn't belong to $\td\Gamma$, we define the index by continuity from its neighborhood in $\curve$. Thus, the index is defined on $\curve\setminus\td\Gamma$. We have
\beq
1\leq i(p)\leq d=\deg X.
\eeq
\ed

\bd[Domains of given index]
Let:
\beq
\td{\mathcal D}_i := \{p\in \curve\setminus\td\Gamma\,|\,\, i(p)=i\} = \cup_j \td{\mathcal D}_{i,j}
\quad , \,\, \td C_{i,j}  = X(\td {\mathcal D}_{i,j})
\eeq
i.e. $\td{\mathcal D}_i\subset \curve$ is the open set of points of index $i$, and $\td{\mathcal D}_{i,j}$ are its connected components.
Let 
\beq
\td \Gamma_i = \partial{\td {\mathcal D}_i} 
, \qquad
\Gamma_i = X(\td\Gamma_i).
\eeq
\ed
\bp
For any $(i,j)$, the map $X:\td {\mathcal D}_{i,j}  \rightarrow \td C_{i,j}$ is an analytic bijection, whose inverse is analytic.

\ep

\proof{
the map $X:\td {\mathcal D}_{i,j}  \rightarrow \td C_{i,j}$ is surjective by definition. It is a bijection whose inverse is $x\mapsto p_i(x)$.
The only point where $X$ or its inverse would be non-analytic can only be punctures or ramification points, which are vertices of the graph, and are at the boundaries of $\tilde{\mathcal D}_{i,j}$s, they are outside.

}

\bp
For any fixed index $i$, the disjoint union of all $\td C_{i,j}$ with index $i$ is the complex plane itself (except the graph $\Gamma_i$).
\beq
\mathbb C\setminus \Gamma_i=\sqcup_{j} \td C_{i,j}, 
\eeq
and thus
\beq
\overline{\cup_{j} \td C_{i,j}} = \mathbb C.
\eeq
\ep
In other words, we have $d$ copies of the complex plane, cut along the spectral network graph.

\proof{ for each $x\in\mathbb C\setminus \Gamma$, $p_i(x)$ belongs to $\td{\mathcal D}_i$, and thus is necessarily in some $\td {\mathcal D}_{i,j}$, and thus $x\in \td C_{i,j}$.

Moreover, the $\td C_{i,j}$ are disjoints, indeed, imagine that there exists some $x\in  \td C_{i,j}\cap \td C_{i,j'}$, that means that $p_i(x)\in \td {\mathcal D}_{i,j}\cap \td {\mathcal D}_{i,j'} = \emptyset$, so this is impossible.
We thus have
\beq
\sqcup_{j} \td C_{i,j} = \mathbb C\setminus \Gamma_i.
\eeq

}

Those $d$ copies of $\mathbb C$ with cuts, provide an atlas of $\curve$, whose charts are the $\tilde C_{i,j}$s.
The transition maps are obtained by gluing the charts along edges and at vertices of the graph, with transition function $x\mapsto x$.
\bl
For every domain of given index ${\mathcal D}_i$, there is a finite number of connected components ${\mathcal D}_{i,j}$.
\el
\proof{The edges of the graph $\td \Gamma_i$ are algebraic lines, therefore the number of connected components is finite.}
\medskip

{\bf Edges:}

\bd[Edges permutations]

Each edge $e$ of $\td\Gamma$ is at the intersection of two domains, whose index differ by one, 
$\td{\mathcal D}_{i,j}$ and $ \td{\mathcal D}_{i+1,j'}$.
The edge $X(e)$ of $\Gamma$ is also at the intersection of  domains $\td C_{i,j}$ and $\td C_{i,j''}$,  with the same index $i$.
The domain $\td {\mathcal D}_{i,j''}$ has an edge $e'$ such that $X(e')=X(e)$, and on the other side on $\curve$, there is a domain $\td {\mathcal D}_{i+1,j'''}$.

We define the permutations (these are products of two transpositions):
\beq
\sigma_e: 
((i,j) \leftrightarrow (i+1,j') ) \ 
((i,j'') \leftrightarrow (i+1,j'''))
\eeq
\beq
\tau_e: 
((i,j) \leftrightarrow (i+1,j''') ) \ 
((i,j'') \leftrightarrow (i+1,j'))
\eeq

$\sigma_e$ permutes the domains that analytically continue each other across $e$.

$\tau_e$ permutes the two domains on top of each other along $e$.

\begin{figure}
\begin{center}
\includegraphics[scale=0.4]{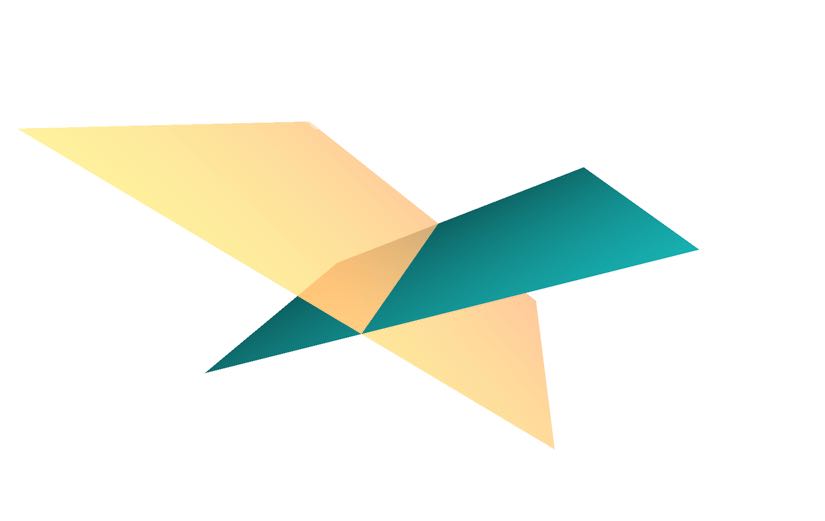}
\caption{two sheets meet along an edge $e$. They have adjacent index $i,i+1$. $\sigma_e$ permutes the domains that analytically continue each other across $e$ (it permutes the half planes of the same color).
$\tau_e$ permutes the two domains on top of each other along $e$ (it permutes the colors).
 }
\label{fig:Edge}
\end{center}
\end{figure}

See Fig.\ref{fig:Edge}.

\ed

\bl
Along each edge $e$ we have
\beq
\sigma_e^2 = \text{Id},\quad
\tau_e^2 = \text{Id},\quad
\tau_e \sigma_e =  \sigma_e \tau_e .
\eeq

\el

\proof{
simple computation.
}

\medskip

{\bf Vertices:}

\smallskip

$\bullet$ Branch points.

At a regular branch point we have $p_i(x)\to p_{i+1}(x)$. Let $z=\sqrt{x-a}$.
We have $y\sim y(a) + y'(a) z + \frac{1}{2}y''(a)z^2 + \dots$, and thus $\phi(p_i)-\phi(p_{i+1})\sim \frac{4}{3}\,\Re(y'(a) (x-a)^{3/2}) + \dots $, and thus $a$ is a trivalent vertex.

$\bullet$ Higher Branch points.

At a branch point of order $r$ we have $p_i(x)=p_{i+1}(x)=\dots=p_{i+r-1}(x)$. Let $z=(x-a)^{1/r}$.
We have $y\sim y(a) + y'(a) z + \frac{1}{2}y''(a)z^2 + \dots$, and thus $\phi(p_i)-\phi(p_{i+1})\sim \frac{r}{r+1}\,\Re((1-\e^{2i\pi/r})\,y'(a) (x-a)^{(r+1)/r}) + \dots $, and thus $a$ is a $r+1$ valent vertex.

$\bullet$ Nodal points.

It may happen that vertices can be nodal points (but not all nodal points are vertices).
These are points where $y(p_i)=y(p_{i+1})$, with $p_i\neq p_{i+1}$, and in fact where $(y(p_i)-y(p_{i+1}))$ vanishes at an order $k$.
At those points we have
\beq
\phi(p_i)-\phi(p_{i+1}) \sim \Re(y'(d)\,(x-d)^{k+1}),
\eeq
and thus they give $2(k+1)$ valent vertex.

\smallskip

$\bullet$ Virtual vertices.

They are points where $\phi(p_i)= \phi(p_{i+1}) = \phi(p_{i-1})$, but $p_i\neq p_{i-1}\neq p_{i+1}$, and the differentials $d(\phi(p_i)-\phi(p_{j}))$ do not vanish. In other words, the boundaries of the $\td{\mathcal D}_{i,j}$ have smooth tangents there.

\begin{figure}
\begin{center}
\includegraphics[scale=0.26]{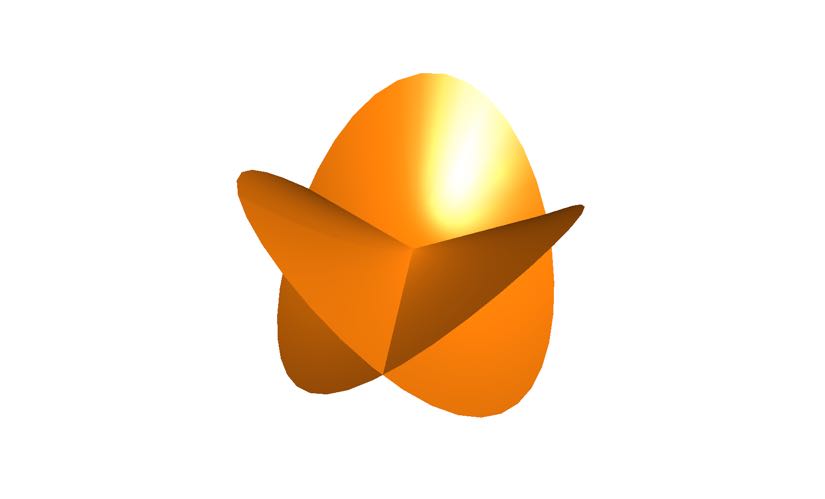}
\includegraphics[scale=0.25]{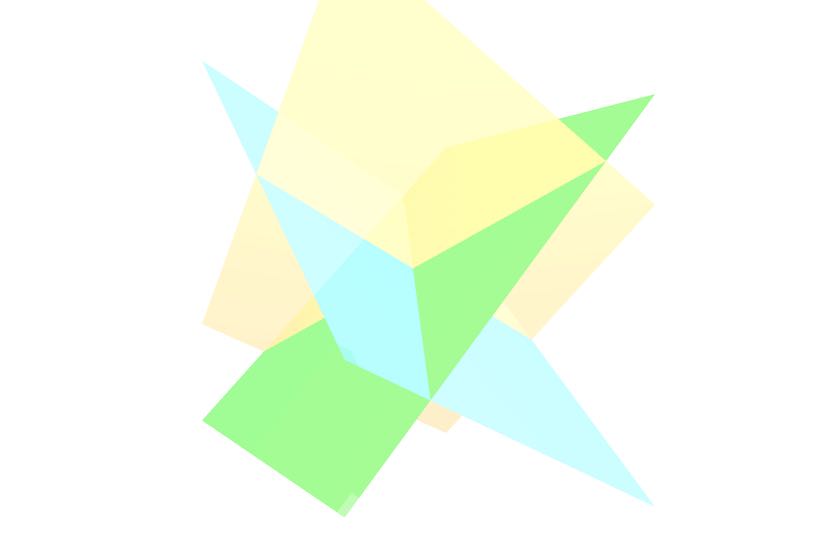}
\caption{On the left a regular ramification point, it is a trivalent vertex, two sheets meet. On the right, a virtual vertex. three sheets meet and all have different normal vectors.}
\label{fig:vertices}
\end{center}
\end{figure}

See Fig.\ref{fig:vertices}.

\bl
For any domain $\td{\mathcal C}_{i,j}$, there is no edge at the boundary of $\td{\mathcal C}_{i,j}$, such that both sides are in $\td{\mathcal C}_{i,j}$. We say that the boundary of $\td{\mathcal C}_{i,j}$ has no self-edge.

Equivalently, for every $e\in \td\Gamma$, $\sigma_e\circ\tau_e$ has no fixed point.

\el

\proof{
Assume that there would exist an edge $e\subset \partial\td{\mathcal C}_{i,j}$, such that both sides are in $\td{\mathcal C}_{i,j}$.
Let $x\in e$, and $x_{+}$ and $x_-$ some points very close to $x$ on each side.
By definition of $\td{\mathcal C}_{i,j}$, we have $p_i(x_+)\in \td{\mathcal D}_{i,j}$ and $p_i(x_-)\in \td{\mathcal D}_{i,j}$, and thus
\beq
\lim_{x_+\to x} p_i(x_+) = \lim_{x_-\to x} p_i(x_-) =p_x.
\eeq
Let $e'=\{p_x \ | \ x\in e\}$, such that $X(e')=e$. $e'$ is a boundary edge of $\td{\mathcal D}_{i,j}$, in fact it is a self-edge of $\td{\mathcal D}_{i,j}$.
This implies that $\td{\mathcal D}'_{i,j}=\td{\mathcal D}_{i,j}\cup e'$ is an open connected domain of $\curve$.
This implies that $YdX$ is holomorphic on a neighborhood of $e'$.
Thus, $\sigma_{e'}((i,j))=(i,j) $, which is impossible because $\sigma_e$ must always move the index by one, i.e $\sigma_{e'}((i,j))=(i\pm 1,j') $.
This is a contradiction, so the assumption that $ \partial\td{\mathcal C}_{i,j}$ has a self edge was impossible.
}

\bd[Admissible and maximal domains]
Let $\mathcal D$ a union of domains $\td{\mathcal D}_{i,j}$s, and edges of $\td\Gamma$.

$\mathcal D$ is called admissible iff:
\begin{itemize}

\item $\mathcal D$ is open,

\item $\phi$ is harmonic on $\mathcal D$,

\item $X$ is injective on $\mathcal D$. 

\end{itemize}
Let 
\beq
\mathcal C = X(\mathcal D)= \mathop{\cup}_{\td{\mathcal D}_{i,j}\subset\mathcal D} \td{\mathcal C}_{i,j} \mathop{\cup}_{e\subset\mathcal D} X(e).
\eeq

We say that $\mathcal D$ is maximal iff there is no $\mathcal D'$ admissible such that $\mathcal D\subset \mathcal D'$ and $\mathcal D\neq \mathcal D'$.

\ed

The following lemmas are immediate:
\bl
Every single domain $\td{\mathcal D}_{i,j}$ is admissible.
For every admissible domain, there exists a maximal admissible domain that contains it.
\el

\bl
If $\mathcal D$ is admissible, then $\mathcal C=X(\mathcal D) $ is an open  domain of $\CC$.
Its boundary is a graph (possibly empty).
Its complement is a graph and possibly a finite union of open sets  of $\CC$.
The map $X:\mathcal D\to \mathcal C$ is a conformal bijection.

\el

\bt
\label{thm:maxdomain}
Let $\mathcal D$ a maximal admissible domain, and $\mathcal C=X(\mathcal D)$.

Then  the complement of $\mathcal C$ can contain no open set, it must be a graph $\hat\Gamma$.

$X:\mathcal D \to \mathcal C$ is a conformal bijection. 
$\Phi=\phi\circ X^{-1}$ is harmonic on $\mathcal C=\CC\setminus\hat\Gamma$. $\Phi$ can be extended by continuity to $\CC$. $\Phi$ is then continuous on $\CC$, harmonic on $\CC\setminus\hat\Gamma$, and the places where it is not harmonic is  exactly on $\hat\Gamma$.

\et

\proof{
Assume that the complement of $\mathcal C$ contains an open domain.
Let us choose $x$ in this open domain.
$x$ is not in $\Gamma$, so it has $d$ distinct preimages $p_1(x),\dots p_d(x)$.
Each of them is in some $\td{\mathcal D}_{i,j}$.

If we assume that there exists some $(i,j)$ such that $\td{\mathcal C}_{i,j}\cap\mathcal C=\emptyset $, then we would have $\td{\mathcal D}_{i,j}\cap\mathcal D=\emptyset $, and we could add $\td{\mathcal D}_{i,j}$ to $\mathcal D$ to obtain an admissible domain. This would contradict the maximality of $\mathcal D$.

Therefore, for every $(i,j)$ we have $\td{\mathcal C}_{i,j}\cap\mathcal C\neq\emptyset $, which implies that there is some $(i,j)$ such that $\td{\mathcal D}_{i,j}\cap\mathcal D\neq\emptyset $.
This means that $\td{\mathcal D}_{i,j}\subset \mathcal D$, which contradicts our hypothesis that $x\notin X(\mathcal D)$.

This implies that the complement of $\mathcal C$ can not contain any open domain, it can contain only edges.

By definition $\phi$ is harmonic on $\mathcal D$, and $\Phi=\phi\circ X^{-1}$ is harmonic on $\mathcal C=\CC\setminus\hat\Gamma$. 

The only places where it could be non-harmonic could be a subgraph of $\hat\Gamma$.

Let $e$ and edge of $\hat\Gamma$. It is a boundary of $\mathcal C$, and thus there is an $e'$ boundary of $\mathcal D$ for which $X(e')=e$.
If we assume that $\Phi$ is harmonic on $e$, this would imply that $\phi$ is harmonic on $e'$. This implies that $\mathcal D\cup e'$ would be admissible. This would contradict the maximality of $\mathcal D$.
Therefore $\Phi$ must be non-harmonic on the edges of $\hat\Gamma$.

}

\bd
Each edge $e$ of $\hat\Gamma$ has two sides $e_+$, $e_-$ on $\partial\mathcal D$, oriented such that $\mathcal D$ is on their left.
They border two domains, $e_+$ borders $\mathcal D_{e_+}$, and $e_-$ borders $\mathcal D_{e_-}$.
We must have
\beq
\mathcal D_{e_-} = \sigma_e\circ\tau_e(\mathcal D_{e_+}).
\eeq
On $e_+$ (resp. $e_-$), let
\beq
d\rho_e := \frac{1}{2\pi}\Im (dg - \tau_e^* dg).
\eeq
Adding the measure of all edges:
\beq
d\rho := \sum_{e = \text{ edges of }\hat\Gamma} \chi_e \, d\rho_e ,
\eeq
where $\chi_e$ is the characteristic function of $e$.

$d\rho$ is a real measure on $\hat\Gamma$.
\ed

\bp
The only places where $d\rho$ can vanish, are ramification or nodal points.
\ep

\proof{
By definition, on any edge $e$ of $\hat\Gamma$, which is at the intersection of sheets $Y_i(x)$ and $Y_j(x)$, we have that $\Re (Y_i(x)-Y_j(x))dx$ vanishes.
$d\rho_e=0$ implies that the imaginary part is also vanishing, i.e. that $(Y_i(x)-Y_j(x))dx=0$. This implies that either $dx=0$ (ramification point) or $Y_i(x)-Y_j(x)=0$ (nodal point).
}

\bp
At generic ramification points, there are either one or three edges of $\hat\Gamma$.
If there are three edges, the sign of $d\rho$ is the same on all three edges.

\ep

\proof{At generic ramification points we have $(Y_i(x)-Y_j(x))dx \sim C(x-a)^{\frac12}dx$ and thus
$g_i(x)-g_j(x)\sim \frac23 C (x-a)^{\frac32}$.
There are three lines where $\Re C (x-a)^{\frac32}=0$ (See Fig.\ref{fig:genericbpspnetwork2}).
If we consider the case where $\hat\Gamma$ has three lines meeting, then we choose the branches of the square root discontinuous accross each of them, and it is easy to see that the sign of $\Im(g_i-g_j)$ is the same on the three edges.
}

\begin{center}
\begin{figure}
\begin{center}
\begin{tikzpicture}[scale=2.2]
\def\R{1cm};
\begin{scope}[shift={(0,0)},rotate=120]
\node[inner sep=0pt] (bp) at (0,0) {};
\draw (bp) -- (0:\R) ;
\draw[line width=3pt] (bp) -- (120:\R) ;
\draw (bp) -- (240:\R);
\node (B) at  ++(60:\R/2)  {$-$};
\node (B) at  ++(180:\R/2)  {$-$};
\node (B) at  ++(-60:\R/2)  {$+$};
\end{scope}
\begin{scope}[shift={(2.5,0)},rotate=0]
\node[inner sep=0pt] (bp) at (0,0) {};
\draw[line width=3pt] (bp) -- (0:\R) ;
\draw[line width=3pt] (bp) -- (120:\R) ;
\draw[line width=3pt] (bp) -- (240:\R);
\node (B) at  ++(60:\R/2)  {$-$};
\node (B) at  ++(180:\R/2)  {$-$};
\node (B) at  ++(-60:\R/2)  {$-$};
\end{scope}

\end{tikzpicture}

\includegraphics[scale=.5]{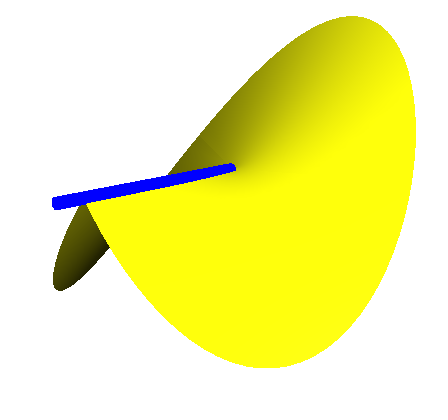}
$\qquad$
\includegraphics[scale=.5]{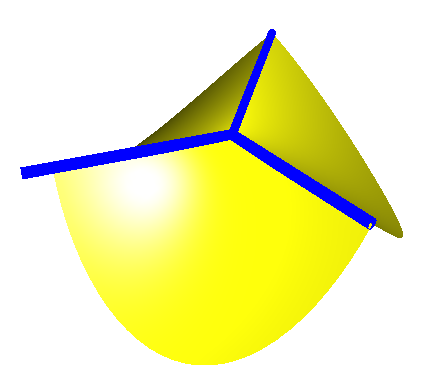}

\caption{Generic branch point. An edge belongs to $\hat\Gamma$ only if the sign of $Y_i-Y_j$ is the same on both sides. If $--$ then $d\rho_e$ is positive and if $++$ then $d\rho_e$ is negative. Left: only one edge belongs to $\hat\Gamma$. Right: three edges belong to $\hat\Gamma$, and $d\rho_e$ is of the same sign on the three edges.}
\label{fig:genericbpspnetwork2}
\end{center}
\end{figure}
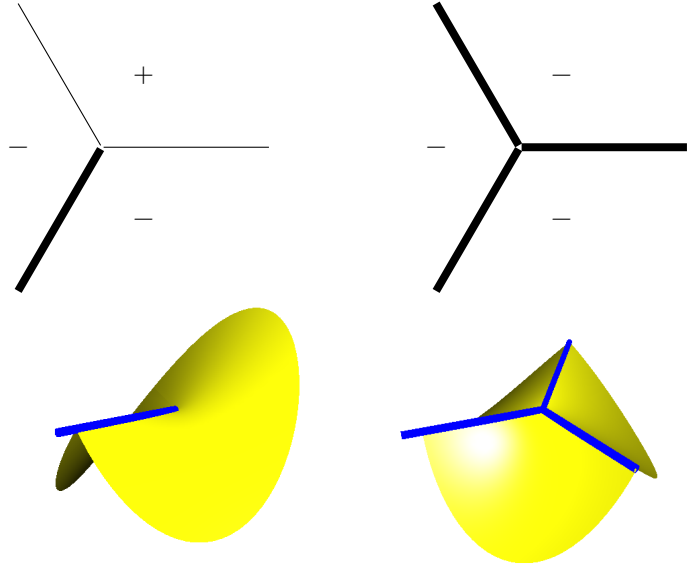
\end{center}

\bp
At generic nodal points, there are either 0, 2 or 4 edges of $\hat\Gamma$.
If there are four edges, the sign of $d\rho$ is the same on all four edges.

When dealing with two edges, if they are adjacent, $d\rho$ has the same sign along both edges, but if they are aligned, the sign of $d\rho$ becomes opposite.

\ep

\begin{center}
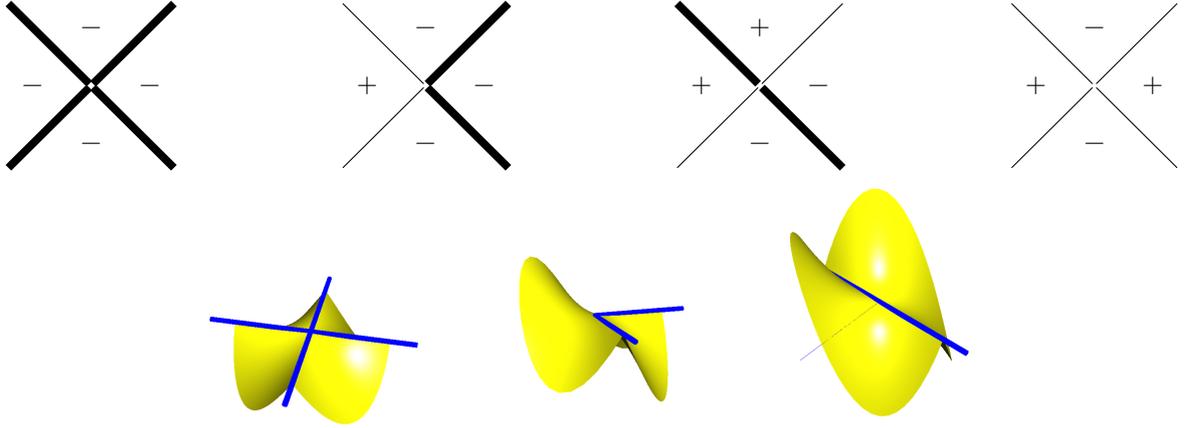
\begin{figure}
\begin{center}
\begin{tikzpicture}[scale=2.2]

\def\R{1cm};

\begin{scope}[shift={(0,0)},rotate=45,scale=0.7]
\node[inner sep=0pt] (bp) at (0,0) {};
\draw[line width=3pt] (bp) -- (0:\R) ;
\draw[line width=3pt] (bp) -- (90:\R) ;
\draw[line width=3pt] (bp) -- (180:\R);
\draw[line width=3pt] (bp) -- (270:\R);
\node (B) at  ++(45:\R/2)  {$-$};
\node (B) at  ++(135:\R/2)  {$-$};
\node (B) at  ++(225:\R/2)  {$-$};
\node (B) at  ++(315:\R/2)  {$-$};
\end{scope}

\begin{scope}[shift={(2,0)},rotate=-45,scale=0.7]
\node[inner sep=0pt] (bp) at (0,0) {};
\draw[line width=3pt] (bp) -- (0:\R) ;
\draw[line width=3pt] (bp) -- (90:\R) ;
\draw (bp) -- (180:\R);
\draw (bp) -- (270:\R);
\node (B) at  ++(45:\R/2)  {$-$};
\node (B) at  ++(135:\R/2)  {$-$};
\node (B) at  ++(225:\R/2)  {$+$};
\node (B) at  ++(315:\R/2)  {$-$};
\end{scope}

\begin{scope}[shift={(4,0)},rotate=-45,scale=0.7]
\node[inner sep=0pt] (bp) at (0,0) {};
\draw[line width=3pt] (bp) -- (0:\R) ;
\draw (bp) -- (90:\R) ;
\draw[line width=3pt] (bp) -- (180:\R);
\draw (bp) -- (270:\R);
\node (B) at  ++(45:\R/2)  {$-$};
\node (B) at  ++(135:\R/2)  {$+$};
\node (B) at  ++(225:\R/2)  {$+$};
\node (B) at  ++(315:\R/2)  {$-$};
\end{scope}

\begin{scope}[shift={(6,0)},rotate=45,scale=0.7]
\node[inner sep=0pt] (bp) at (0,0) {};
\draw (bp) -- (0:\R) ;
\draw (bp) -- (90:\R) ;
\draw (bp) -- (180:\R);
\draw (bp) -- (270:\R);
\node (B) at  ++(45:\R/2)  {$-$};
\node (B) at  ++(135:\R/2)  {$+$};
\node (B) at  ++(225:\R/2)  {$-$};
\node (B) at  ++(315:\R/2)  {$+$};
\end{scope}

\end{tikzpicture}
\includegraphics[scale=.3]{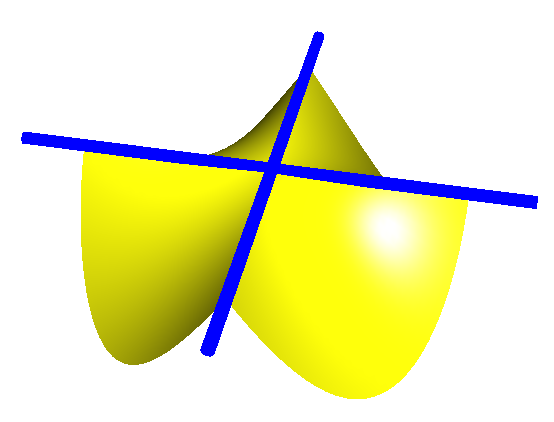}
$\qquad$
\includegraphics[scale=.3]{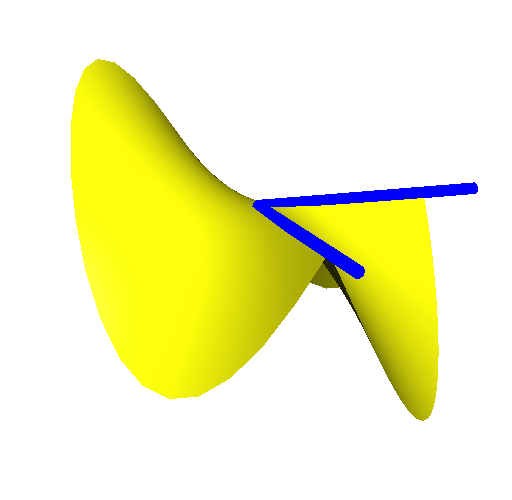}
$\qquad$
\includegraphics[scale=.3]{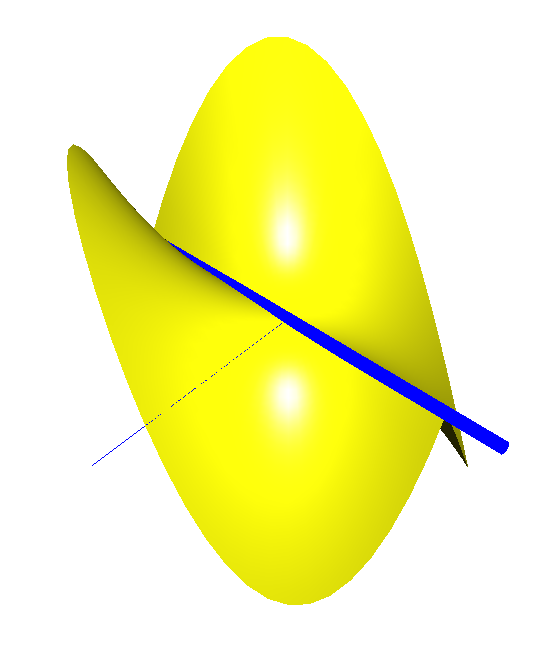}

\caption{Generic nodal point. An edge belongs to $\hat\Gamma$ only if the sign of $Y_i-Y_j$ is the same on both sides. If $--$ then $d\rho_e$ is positive and if $++$ then $d\rho_e$ is negative. Left: four edges, the sign is the same on the four edges. Middle: two adjacent edges on $\hat\Gamma$, same sign. Right: two aligned edges on $\hat\Gamma$, opposite sign. Last case, no edge on $\hat\Gamma$.}
\label{fig:genericnodalpnetwork2}
\end{center}
\end{figure}
\end{center}

\proof{At generic nodal points we have $(Y_i(x)-Y_j(x))dx \sim C(x-a)dx$ and thus
$g_i(x)-g_j(x)\sim \frac12 C (x-a)^{2}$.
There are four lines where $\Re C (x-a)^{2}=0$ (See Fig.\ref{fig:genericnodalpnetwork2}).
The signs are easily computed in each case.
}

\bt
\label{th:measureDeltaPhi}
The measure $d\rho$ on the edges of $\hat\Gamma$ coincides with the measure $\frac{1}{2\pi\ii} \Delta\Phi$ on $\CC$.
Since $\Delta\Phi = 0$ on $\CC\setminus\hat\Gamma$, the measure is localized on $\hat\Gamma$.

\et

\proof{
This follows from Stokes theorem.
Let $f:\CC\to \RR$ a real valued $C^\infty$ bounded function.
We have
\bea
\int_{\CC} f \Delta \Phi
&=& \int_{\CC\setminus \hat\Gamma} f \Delta \Phi \cr
&=& \int_{\CC\setminus \hat\Gamma} f \bar{d} d \Phi \cr
&=& \int_{\hat\Gamma}  f \  d\Phi\cr
&=& \sum_{e= \text{edges of }\hat\Gamma} \int_e  f \ (Y_{left}-Y_{right})dx \cr
&=& \int_{\hat\Gamma}  f \ 2\pi\ii d\rho .
\eea
}

\bl
On $\curve$ we have
\beq
\Delta \td \phi = -\Delta \phi.
\eeq
\el

\proof{
Remember that
\beq
\td\phi = -\phi + \Re (X Y),
\eeq
and $X Y$ is an analytic function on $\curve$, whose Real part is then harmonic, and therefore
\beq
\Delta \td\phi = -\Delta \phi.
\eeq
}

Moreover, we mention the following theorem that can be very useful:

\bt[Change of functions that don't affect the spectral network]
The spectral network $\Gamma$ is unchanged if we change $Y\to Y+V'(x)$ where $V(x)\in \CC(x)$.
In particular the measure is unchanged
\beq
\Delta (\Phi(x)+V(x)) = \Delta \Phi(x).
\eeq
In other words, we change
\beq
P(x,y) \to P(x,y+V'(x)).
\eeq
This changes the Newton's polygon, and the moduli space, but to an isomorphic one.

\et

\proof
{
This is obvious since $V(X(z^i(x))) = V(X(z^j(x)))$, so this change of function doesn't change the spectral network, the indices and the domains.
}

\subsection{Hyperelliptical case, comparison of the two kinds of spectral networks}
\label{sec:spnetwk12}

An hyperelliptical plane curve is a plane curve with a Newton's polygon of the form:
\beq
\mathcal P(x,y) = y^2 D(x) - U(x),
\eeq
where $D$ and $U$ are polynomials of $x$.
The moduli space $\modsp=\mathcal P+\CC[\Newtint]$ is an affine vector space of polynomials of $x$:
\beq
\modsp \subset \mathcal P+\CC[x], \qquad
\dim \modsp \leq \frac12(\deg U+\deg D)-1.
\eeq
But notice that if $D$ has multiple zeros, the dimension may be smaller, here we give only an upper bound.

\bd[Hyperelliptic involution]
There exists an involution $\sigma:\curve\to\curve$, such that $X\circ\sigma=X$ and $Y\circ\sigma=-Y$.

The fixed points must have $y=0$, i.e. $U(x)=0$, and $x=\infty$ if $\deg U-\deg D$ is odd.

The fixed points of $\sigma$ are the ramification points and odd punctures.

Nodal points are pairs $(p,\sigma(p))$, they are invariant as a pair, but $p$ itself is not invariant.

\ed

\subsubsection{Geometry of hyperelliptic curves}

\begin{itemize}

\item Punctures are zeros of $D(x)$, and $X^{-1}(\infty)$ if $\deg U>\deg D-4$.

\item Ramification points are the odd zeros of $U(x)$.

\item Nodal points are the even zeros of $U(x)$.

\item The genus of $\curve$ is
\beq
\genus = -1+\left\lfloor\frac12 \#\text{ramification points} \right\rfloor .
\eeq

\item Let
\beq
U(x) = U_-(x) U_+(x)^2,
\eeq
where $U_-(x)$ has only odd zeros, and $U_+(x) = \sqrt{U(x)/U_-(x)} $ contains all the even zeros, chosen so that $U_-$ and $U_+$ have no common zeros.

\item Let $r$ the number of ramification points, let $n$ the number of nodal points.
\bea
r= \#\text{zeros of } U_- \cr
n= \#\text{zeros of } U_+. \cr
\eea

\end{itemize}

From now on, we choose a Boutroux curve in $\modsp$.

We choose the origin for defining $\phi$, to be a ramification point, i.e. a point invariant under the hyperelliptic involution.

\bl
$\phi$ is odd under the involution:
\beq
\phi\circ\sigma=-\phi.
\eeq
In particular, all ramification points have $\phi=0$.
\el

\proof{
We have $\sigma^* YdX = -YdX$, and therefore $\sigma^*d\phi=-d\phi$.
This implies that $\phi+\sigma^*\phi$ must be a constant, and since it vanishes at one point, it must be zero.
}

\subsubsection{Spectral Networks of 1st kind}

The spectral network is the graph whose edges are horizontal trajectories starting from every branch points or nodal points.

\bl
\label{lemma:Spntw1fund}
The spectral network graph $\td\Gamma$ of Definition \ref{def:SpNcl} is invariant under the involution:
\beq
\sigma(\td\Gamma)=\td\Gamma.
\eeq
Moreover edges emanating from ramification points must have $\phi=0$.

The graph cuts the curve into domains that are either half-planes, strips and half-cylinders (if Fuchsian punctures).
\el
\proof{If a line is a vertical line, i.e. $d\phi=0$, then its image by the involution is $-d\phi=0$ and is also a vertical line.}

\subsubsection{Spectral Networks of 2nd kind}

\bd
\label{lemma:Spntw2fund}
Let 
\beq
\td \Gamma_0 := \phi^{-1}(\{0\}), \quad
\Gamma_0 := X(\td\Gamma_0).
\eeq
Let
\beq
\mathcal D_+ := \{ p \ | \ \phi(p)>0\}, \quad
\mathcal D_- := \{ p \ | \ \phi(p)<0\},
\eeq
and $\mathcal C_\pm = X(\mathcal D_{\pm})$.
We have $\partial \mathcal D_+ = \partial \mathcal D_- = \td\Gamma_0$.
\ed

\bl
Let $\mathcal D_{\pm,j}$ the connected components of $\mathcal D_+$.
Each domain $\mathcal D_{\pm,j}$ is a finite union of strips, half-planes and half-cylinders with vertical boundaries, which are the connected components of $\mathcal D_{\pm,j}\setminus \check\Gamma $ where $\check\Gamma$ is the spectral network of 1st kind.
\el

\proof{
Each connected components of $\mathcal D_{\pm,j}\cap \check\Gamma $ is obtained by cutting the connected components of $\curve \setminus \check\Gamma $ by the vertical line $\phi=0$.
In all cases cutting a vertical half-plane, a vertical strip or a vertical half-cylinder by the line $\phi=0$ gives 1 or 2 half-plane,  strip or half-cylinder.
}

\bp
\label{prop:Sp2hyperD-max}
$\mathcal D_{-}$ is a maximal admissible domain.
\ep

\proof{
It is admissible because $\phi$ is harmonic in each connected component, and $X$ is $1:1$ on $\mathcal D_-$.
It is maximal, because $X(\mathcal D_-)=\CC P^1 \setminus \Gamma_0 $, its complement is a graph. Therefore, no other open domain can be added to $\mathcal D_-$.
Moreover, since the sign of $\phi$ is the same on both sides of each edge, $\phi$ is not harmonic on edges. 
Adding an edge to $\mathcal D_-$ would make it not admissible.
}

The following proposition is an immediate consequence 
\bp
\label{th:Graphmaxdomainhyperelliptic}
Let $\mathcal D$ a maximal admissible domain.
It is a finite union of vertical half-plane, vertical strip or vertical half-cylinder (if Fuchsian puncture).

Its boundary $\hat\Gamma=\partial X(\mathcal D)$ is a subgraph of $X(\phi^{-1}(\{0\}))$, all edges have $\Phi=0$.
The measure $\frac{1}{2\pi\ii}\Delta\Phi$ is a real measure, localized on the boundary $\hat\Gamma$.

\ep

\section{Applications and examples}
\label{sec:apps}
There are many applications of Boutroux curves and their spectral networks.
Most famous examples are vertical trajectory foliations of the moduli space of Riemann surfaces by Strebel graphs, and eigenvalues equilibrium density for random matrices.

\subsection{Example: Weierstrass curve}

Let us exemplify all the method for the Weierstrass curve.
Let
\beq
\mathcal P(x,y) = y^2-x^3+g_2 x + g_3
\eeq
whose moduli space is 
\beq
\modsp = \{g_3\} \sim \CC ,\qquad  \ \dim\modsp=1.
\eeq
In other words we shall keep $g_2$ fixed and take $g_3$ as a coordinate of $\modsp$.

For $4g_2^3-27g_3^2\neq 0$, the curve has genus $\genus=1$, 
we have $g_2 = 15\nu^4 G_4(\tau)$, $g_3 = -35\nu^6 G_6(\tau)$, in other words we can view $g_3$ as a function of $\tau$
\beq
g_3 = -35 g_2^{\frac32} \ \frac{G_6(\tau)}{(15 G_4(\tau))^{\frac32}},
\eeq
and we can moreover view $g_3$ as a function of $q=e^{2\pi\ii\tau}$.
In other words we can parametrize $\modsp$ by the coordinate  $q$.

The degenerate curve $4g_2^3-27g_3^2=0$, will be considered to correspond to $\tau=\ii\infty$, i.e. $q=0$.

For $q\neq 0$ we have
\beq
t_{\infty,5}=-2, \quad
t_{\infty,1}=g_2, \quad
\td t_{\infty,5}=\frac{1}{10}g_2 g_3, \quad
\td t_{\infty,1}=g_3.
\eeq
\beq
\eta = 3\ii \nu^5 G'_4(\tau), \quad
\td\eta = \tau\eta +12\ii \nu^5 G_4(\tau).
\eeq
This gives
\beq
\hat F = \frac{2}{5}g_2 g_3 = -210 \nu^{10} G_4(\tau)G_6(\tau),
\eeq
and thus
\bea
F &=& -\frac{2}{5}\Re g_2 g_3 + \pi(\td\zeta\epsilon-\zeta\td \epsilon) = -\frac{2}{5}\Re g_2 g_3 + \pi \Im \  \td \eta \bar\eta \cr
&=&  210 \ \Re \left(\nu^{10}G_4(\tau)G_6(\tau)\right) + 9 \pi |\nu|^{10}  \ \Im \Big( 
 \tau |G'_4(\tau)|^2 + 4  \overline{ G'_4(\tau)} G_4(\tau)
\Big).
\eea
When $4g_2^3-27g_3^2=0$, we have $F = -\frac{2}{5}\Re g_2 g_3$, which is the limit when $q\to 0$, i.e. $F$ is continuous at $q=0$.

The Boutroux curve, i.e. the minimum of $F$ is reached at $\zeta=\td\zeta=0$, i.e. when $F = -\frac{2}{5}\Re g_2 g_3$.

\begin{figure}
\begin{center}
\includegraphics[scale=0.3]{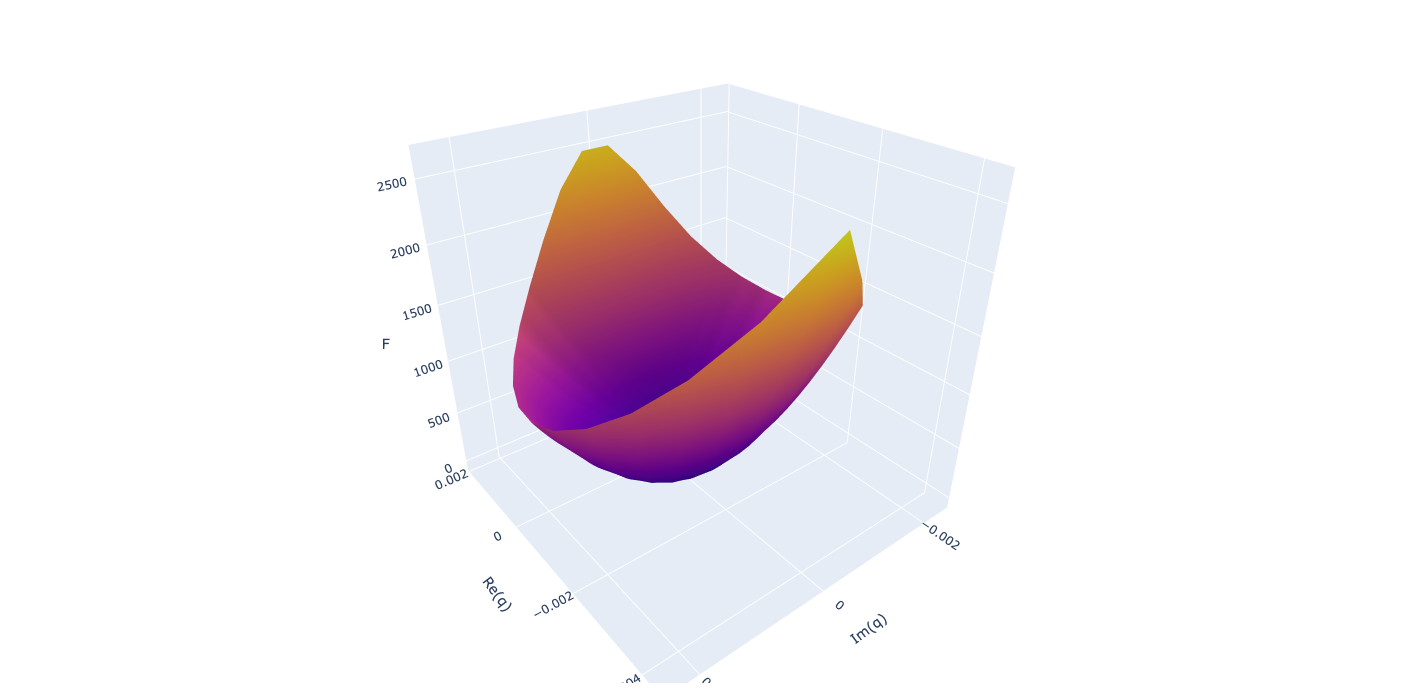}
\end{center}
\caption{$F(q)$ for the Weierstrass curve. We plot $F$ as a function of $q=e^{2\pi\ii\tau} $ rather than a function of $g_3= -35 g_2^{\frac32} \ G_6(\tau) \ (15 G_4(\tau))^{-\frac32}$. We plot for $q$ in the fundamental domain, i.e. $-\frac12< \Re\tau\leq \frac12$ and $|\tau|\geq 1$. The minimum is reached at $q=0$.}
\label{fig:FWeierstrass}
\end{figure}

$F$ as a function of $q=e^{2\pi\ii\tau}$ is plotted in Fig.\ref{fig:FWeierstrass}.
Let us admit that if $g_2 \in \RR$, the minimum is reached at the degenerate curve $q=0$ (this is obvious on Fig.\ref{fig:FWeierstrass}).

\begin{figure}
\begin{center}
\includegraphics[scale=0.7]{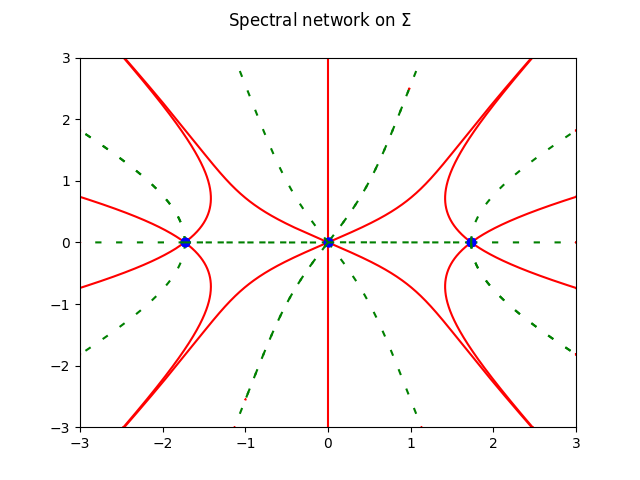}
\end{center}
\caption{Spectral network in the $z$ plane, for the Weierstrass curve. Continuous lines are vertical trajectories, Dashed lines are horizontal trajectories. Vertical trajectories make 10 half-planes and 2 strips.}
\label{fig:spnetworkWierstrass}
\end{figure}

We parametrize the degenerate curve as
\beq
\left\{
\begin{array}{l}
g_2 = -3 u^2, \quad g_3 = 2 u^3 \cr
X(z) = z^2-2u \cr
Y(z) = z^3 -3 u z
\end{array}
\right.
\eeq
We have
\beq
g(z) = \frac25 z^5 - 2 u z^3.
\eeq
The 1st kind of spectral network has 10 half-planes and 2 strips.
See Fig.\ref{fig:spnetworkWierstrass}.

\subsection{Strebel graphs}

Another major example is the following.

Let $z_1,\dots,z_N$ fixed points in $\CC^N$ with $N\geq 3$, and let $L_1,\dots,L_N$ fixed positive real numbers.
Let
\beq
D(x) = \prod_{\alpha=1}^N (x-z_\alpha)
\eeq

\subsubsection{Newton's polygon}

Let
\beq
\mathcal P(x,y) = y^2 D(x)^2  - \sum_{\alpha=1}^N L_\alpha^2 D'(z_\alpha) \frac{D(x)}{x-z_\alpha}.  
\eeq
We have 
\beq
\modsp = \CC[\Newtint] = D(x) \{ Q \ | \ \deg Q\leq N-4\}, \qquad
\dim\modsp = N-3.
\eeq

\subsubsection{Boutroux curve and Strebel graph}

 Theorem \ref{th:existBoutroux} implies that one can find $Q\in \modsp $ such that this is a Boutroux curve.
 
Thus, let the Boutroux curve
\beq
P(x,y) = y^2 D(x)^2  - \sum_{\alpha=1}^N L_\alpha^2 D'(z_\alpha) \frac{D(x)}{x-z_\alpha}  +D(x)Q(x).
\eeq

We choose the origin $o$ for computing $\phi$ to be a point invariant under the involution $Y(o)=-Y(o)$, i.e. $Y(o)=0$.

There are $2N$ punctures, which are simple poles, let us denote them $z_{\alpha,\pm}$, and at which we have
\beq
y \mathop{\sim}_{z_{\alpha,\pm}} \frac{\pm L_\alpha}{x-z_\alpha} + \text{hol}.
\eeq
Near the puncture $z_{\alpha,\pm}$ we have
\beq
\phi \sim \pm L_\alpha \ln{|x-z_\alpha|},
\eeq
the vertical trajectories near the punctures are circles surrounding the punctures.

\subsubsection{First Kind}

Let $\check \Gamma$ the graph of Definition \ref{def:SpNcl} of all vertical trajectories starting from all zeros of $y$ (this includes ramification points and possibly nodal points).
They cut $\curve$ into connected domains.
From  theorem \ref{th:sp1Elpieces}, connected domains can be only half-planes, strips, cylinders or half-cylinders, and from theorem \ref{th:Spntwk1NoCylynders} there is no cylinders.
Moreover, since all punctures are Fuchsian, there are no edges ending at the punctures, and thus there is no half-planes neither strips.
The only faces are half-cylinders ending at the punctures.

Moreover, since there is no strip, this implies that all vertical edges must have the same value of $\phi$.
Since we chose $\phi$ such that $\phi$ vanishes at a branch point, then $\phi$ must be zero on all the graph $\check\Gamma$:
\beq
\check\Gamma = \phi^{-1}(\{0\}).
\eeq

$\check\Gamma$ is a cellular graph on $\curve$, whose faces are discs around the punctures, and its projection $\Gamma=X(\check\Gamma)$ is a cellular graph on $\CC P^1$  whose edges are vertical trajectories, and whose faces are discs around the  points $z_\alpha$, of perimeter $2\pi L_\alpha$.

This is the \textbf{Strebel graph}.

\begin{figure}
\begin{center}
\includegraphics[scale=0.76]{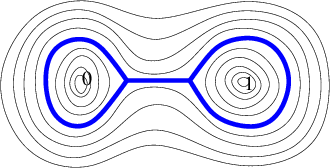}
\caption{Strebel graph for $N=3$. There are 3 faces of prescribed perimeters.}
\label{fig:Strebel03}
\end{center}
\end{figure}

\subsubsection{Second Kind}

Let $\mathcal D_- = \{p\in \Sigma \ | \ \phi(p)<0  \} $ (resp. $\mathcal D_+ = \{p\in \Sigma \ | \ \phi(p)>0  \} $).

$\mathcal D_-$ (resp. $\mathcal D_+$) is a union of connected components, and $X$ is 1:1 on each connected component.
From proposition \ref{prop:Sp2hyperD-max}, we know that
$\mathcal D_-$ is a maximal admissible domain. Each connected component contains exactly one puncture. There are $N$ connected components. The two kinds of spectral networks coincide.

\subsubsection{Strebel differential and Strebel Graph}

We have thus found, that if we have chosen a Boutroux curve
\beq
y^2 = \frac{R(x)}{\prod_{\alpha=1}^N (x-z_\alpha)^2},
\eeq
with $\deg R\leq 2N-4$ and $R(z_\alpha) = L_\alpha^2 D'(z_\alpha)^2$, then the following  quadratic differential
\beq
\Omega =
(YdX)^2 =  \left( \sum_{\alpha=1}^N \frac{L_\alpha^2}{(x-z_\alpha)^2} - \frac{R(x)}{D(x)}\right) dx^2
\eeq
is such that the vertical trajectories of $\sqrt{\Omega}$ form a cellular graph whose faces are discs surrounding the $z_\alpha$s, and with perimeter (in the metric $\sqrt{\Omega}$) $2\pi L_\alpha$.

$\Omega$ is called a \textbf{Strebel differential}, and the cellular graph $\Gamma=\Gamma_0$ of its vertical trajectories is called the \textbf{Strebel graph}.

\medskip

One can verify that the Strebel graph is left invariant by M\"obius transformations $x\to (ax+b)/(cx+d)$ and $y\to (cx+d)^2 y$, with $ad-bc=1$. In other words we have a map:
\bea
({\CC P^1} ^N/ \operatorname{Aut}(\CC P^1)) \times \RR_+^N   & \to  \text{Quadratic differentials}  & \to  \text{Graphs} \cr
(z_i,L_i)_{i=1,\dots,N} \mod \text{M\"obius} & \mapsto \Omega=(YdX)^2_{\text{Boutroux}}\quad \quad &  \mapsto \text{Strebel Graph}\quad
\eea
and we notice that
\beq
({\CC P^1} ^N/ \operatorname{Aut}(\CC P^1)) = \modsp_{0,N}
\eeq
is the moduli space of Riemann surfaces of genus $0$ and $N$ marked points.

Strebel's theorem extends this to $\modsp_{g,n}$ for every $g$, and our method above can be extended to that case.

\subsection{1 Matrix model}

Let $V(x)\in \CC[x] $ a polynomial of degree $d\geq 2$, written as
\beq
V(x) = \sum_{k=1}^d \frac{t_{k}}{k} x^{k}.
\eeq

\subsubsection{Newton's polygon}

Let
\beq
\mathcal P(x,y) = y^2 - \frac14 V'(x)^2 +   t_d x^{d-2} .
\eeq
It is an hyperelliptic curve.

There are exactly two punctures, that we denote $\infty_+$ and $\infty_-$.
We have at $\infty_\pm $, $a_\pm=a_{\infty_\pm}=1$, $b_\pm=b_{\infty_\pm}=\deg V'=d-1$, $r_\pm=r_{\infty_\pm}=\deg V = d $.
We have
\beq
\zeta_{\infty_+}=\zeta_{\infty_-} = \zeta = x^{-1},\quad
Y \sim_{\infty_\pm} \pm (\frac12 V'(X) - \frac{1}{X}) + O(X^{-2}).
\eeq
The times at $\infty_\pm$ are 
\beq
t_{\infty_\pm,k} = \Res_{\infty_\pm } \zeta^{k} Y dX  = \Res_{\infty_\pm } X^{-k} Y dX = \mp \frac12 t_k,
\eeq
and
\beq
t_{\infty_\pm,0} = \pm 1.
\eeq

We have
\beq
\Newtint = \{(i,0) , \ i=0,\dots,d-3\}, \ \ \#\Newtint=d-2,
\eeq
and
\beq
\modsp = \CC[\Newtint] = \{ y^0 Q(x) \ | \ Q(x)\in \CC[x]  , \  \deg Q\leq d-3\}, \qquad
\dim\modsp = d-2.
\eeq

\subsubsection{Boutroux curve}

 Theorem \ref{th:existBoutroux} implies that there exists $Q\in\modsp$, such that the curve is Boutroux:
\beq
P(x,y) = y^2 - \frac14 V'(x)^2 +   t_d x^{d-2} +  Q(x),
\eeq
with $\deg Q\leq d-3$.

\subsubsection{First Kind}

Let $\check \Gamma$ the graph of Definition \ref{def:SpNcl}.

Since punctures are not Fuchsian, there is no half-cylinder, and it follows from theorem \ref{th:sp1Elpieces}, theorem \ref{th:Spntwk1NoCylynders}
 and theorem \ref{Th:halfcylFuchsian} that
\bp
The faces of $\CC\setminus \Gamma$, where $\Gamma=X(\check \Gamma)$, are half-planes and strips.
All ramifications points are on $\check \Gamma \cap \phi^{-1}(\{0\})$.
\ep

Let us further subdivide $\curve$ by cutting along $\phi^{-1}(\{0\})$. 
\bp
\label{prop:sp110MMhfstrips}
The faces of $\CC\setminus X(\phi^{-1}(\{0\}) $ are half-planes and strips.
All branch points are on $\Gamma\cap X(\phi^{-1}(\{0\}))$.
\ep
\proof{
The faces of   $\CC\setminus X(\phi^{-1}(\{0\}) $ are obtained by further cutting the faces of  $\CC\setminus\Gamma  $, along the trajectories $\phi=0$.
Since all faces of $\CC\setminus\Gamma  $ are half-planes and strips, cutting them along $\phi=0$ can only produce also half-planes and strips.
}
\begin{figure}
\begin{center}
\includegraphics[scale=0.7]{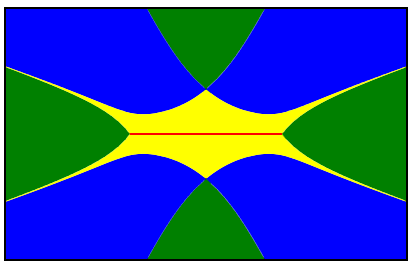}
\end{center}
\caption{Example of spectral network for 1-Matrix model. One branch cut, two strips and eight half-planes. We have thirteen vertical trajectories in total.}
\label{fig:spnetwork1MM}
\end{figure}

\subsubsection{Second Kind}

Let us consider the second kind of the spectral network.
We cut $\curve$ into domains of index $1$ and domains of index $2$, that we rename index $-$ and index $+$.
Thanks to the hyperelliptical involution $y\to -y$, we always have that $\phi(p_+(x))=-\phi(p_-(x))$, and therefore the 2 domains are:
\beq
\td{\mathcal D}_+ = \{ p \ | \ \phi(p)>0\} = \cup_j \td{\mathcal D}_{+,j}, \quad
\td{\mathcal D}_- = \{ p \ | \ \phi(p)<0\} = \cup_j \td{\mathcal D}_{-,j}.
\eeq
Their common boundary is the graph $\phi^{-1}(\{0\})$.
\bl
Every domain $\td{\mathcal D}_{\pm,j}$ is a finite union of half-planes and strips (half-planes and strips of proposition \ref{prop:sp110MMhfstrips}).

As a consequence, every $\td{\mathcal D}_{\pm,j}$ 
 must have $\infty_\pm$ at its boundary.
Every $X(\td{\mathcal D}_{\pm,j})$ is non-compact.
\el

\bp
Let $\mathcal D_{\infty_+}$ the interior of the union of all closed domains $\overline{\td{\mathcal D}_{\pm,j}}$ that have $\infty_+$ at their boundary.

\begin{itemize}
\item $\mathcal D_{\infty_+}$ is a maximal admissible domain.

\item $\mathcal D_{\infty_+}$ is a finite union of half-plane and strips bounded by vertical trajectories.
\end{itemize}
\ep

\proof{
$X$ has a simple pole at $\infty_+$, it is $1:1$ in a neighborhood of $\infty_+$, which implies that $X$ is injective in $\mathcal D_{\infty_+}$.
Moreover $\phi$ is harmonic in a neighborhood of $\infty_+$, so it is harmonic in $\mathcal D_{\infty_+}$.
This implies that $\mathcal D_{\infty_+}$ is admissible.

If it were not maximal, it would be possible to add another domain to it.
Since all domains either touch $\infty_+$ or $\infty_-$, and we have already taken all the domains that touch $\infty_+$, the only possibility would be to add a domain that contains $\infty_-$. But this is impossible because $X$ would then not be injective in a neighborhood of $\infty$.
}

Let 
\beq
\hat\Gamma = X(\partial \mathcal D_{\infty_+}) .
\eeq

\bp
$\hat\Gamma$ is a graph, whose edges are vertical trajectories $\phi=0$
\beq
\hat\Gamma \subset X(\phi^{-1}(\{0\})).
\eeq
All branch points are on $\hat\Gamma$.
\ep

\proof{
Since we included in $\mathcal D_{\infty_+}$ all the edges that end at $\infty_+$ and none of the edges that end at $\infty_-$, then all the compact edges of $\partial\mathcal D_{\infty_+}$ must be on $\phi^{-1}(\{0\})$.

Then, remark that a branch point, is a vertex of $X(\phi^{-1}(\{0\}))$ of odd valency, therefore it is impossible that $\Phi$ is analytic around a branch point.
This implies that all branch points must be on $\hat\Gamma$.

}

\bd
Let us define, for $n\in \ZZ/2d\ZZ$:
\beq\label{def:1MMThetan}
\theta_n := -\frac{1}{d}\arg{t_d} + \frac{\pi}{2d} +  \frac{n\pi}{d}
.
\eeq
\ed

\bl
\label{lemma:thetarays}
All vertical trajectories $\Phi=0$ arrive to $\infty$ at one of these angles, and there is exactly one non-compact half-edge of the graph $\Phi=0$ ending at angle $\theta_n$.

For $r$ large enough $\Phi(r e^{i\theta})$ is an increasing function of $\theta$ when $\theta$ is close to $\theta_n$ with $n$ odd and decreasing if $n$ even.
\el

\proof{
Consider a disc neighborhood of $\infty$, on which $\Phi$ is harmonic.
Consider a small circle inside the disc, parametrized by an angle $\theta\in \RR / 2\pi\ZZ$.
We have $\Phi(x)\underset{\infty}{\sim}\frac12\Re V(x)\underset{\infty}{\sim}\Re \frac{t_d}{2d} x^d $.
This implies that the lines $\Phi=0$ approach $\infty$ in directions $\theta_n$ for all $n\in \ZZ/2 d \ZZ$.

Moreover $\Phi(r e^{i\theta})\underset{\infty}{\sim} (-1)^{n+1} \frac{|t_d| r^d}{2d} \sin(d(\theta-\theta_n))$  is an increasing function of $\theta$ when $\theta$ is close to $\theta_n$ with $n$ odd and decreasing if $n$ even.
}

\bp
\label{prop:Graph1MM}
We have the following properties:

\begin{itemize}

\item $\CC\setminus \hat \Phi^{-1}(\{0\})$ is a finite union of connected domains bounded by vertical trajectories $\Phi=0$.
We write
\beq
\CC\setminus \hat \Phi^{-1}(\{0\}) = \bigcup_{j=1}^{m} C_j.
\eeq
\item
Each connected domain $C_j$ is a finite union of strips and half-planes of the 1st kind spectral network, and the vertical trajectories with $\phi\neq 0$ are strictly inside the connected components.

\item Every half plane reaches $\infty$ in a sector of argument $\theta\in ]\theta_n,\theta_{n+1}[$.

\begin{itemize}
\item If $n$ is odd, this is a half-plane $\phi\to +\infty$.
\item If $n$ is even, this is a half-plane $\phi\to -\infty$.
\end{itemize}

\item Every strip reaches $\infty$ at two angles $\theta = \theta_{n_1}$ and $\theta_{n_2}$, such that $n_1-n_2$ is odd.

\item Let us orient the boundary $\partial C_j$ such that $C_j$ sits on the left of its boundary.
$\partial C_j$ can have several connected components, let us say $m_j$, each of them is a vertical trajectory $\Phi=0$:
\beq
\partial C_j =  \bigcup_{i=1}^{m_j} \partial C_{j,i}, \quad \partial C_{j,i}\subset \Phi^{-1}(\{0\}).
\eeq
$C_j$ can reach $\infty$ in $m_j$ distinct angular sectors of some angles of argument $\theta\in ]\theta_{n},\theta_{n+1}[$, with parity $\epsilon$ depends on the angular sectors in which $C_{j,i}$ goes to $\infty$.

\item If $\epsilon=1$, the component $C_{j,i}$ reaches $\infty$ in domain $\Phi> 0$ , and if $\epsilon=-1$, the component $C_{j,i}$ reaches $\infty$ in domain $\Phi<0$.

\end{itemize}

\ep

\proof{
From proposition \ref{prop:sp110MMhfstrips}, $C_j$ is a finite union of strips and  half-planes. Lemma \ref{lemma:thetarays} says that half planes reach $\infty$ in sector of argument $\theta \in (\theta_n,\theta_{n+1})$, each strip ($\Phi=0$ and $\Phi=c$) reaches $\infty$ in two sectors with different arguments, either in domains $\Phi>0$ or $\Phi<0$, which implies two angles such that $n_1-n_2$ is odd. In addition, each angular sector has different parity which depend on the choices of connected components.
}

\bp
Consider the connected components of the graph $\Upsilon = \Phi^{-1}(\{0\})$. It is a tree.

\begin{itemize}

\item Each connected component $\Upsilon_i$ of $\Phi^{-1}(\{0\})$ is a tree.

\item Each connected component $\Upsilon_i$ is the union of boundaries of connected domains of $C_{i,j}$ of $\CC\setminus \Phi^{-1}(\{0\})$ adjacent to $\Upsilon_i$.
\beq
\Upsilon_{i,j}  = \Upsilon_i \cap \partial C_{i,j}, \ j =1,\dots,k_i.
\eeq
We order them cyclically around the tree $\Upsilon_i$ (in the trigonometric order) so that  $j+k_i\equiv j$ and $C_{i,j+1}$ is adjacent and follows $C_{i,j}$.

\item Each $\Upsilon_{i,j}$ corresponds to a pair $(\theta_{n_{i,j}},\theta_{n_{i,j+1}})$ with $n_{i,j}$ and $n_{i,j+1}$ of different parities.
This implies  that $k_i$ must be even
\beq
k_i\in 2\ZZ_+,
\eeq
and the signs alternate
\beq
(-1)^{n_{i,j+1}} = - \ (-1)^{n_{i,j}}.
\eeq
\proof{
From proposition \ref{prop:Graph1MM}, the graph is made up of strips and half planes, since there is no cycles, the graph is a tree. Otherwise, the graph will have faces with finite boundaries, this is not possible since the only poles of the potential is at $\infty$.
}
\end{itemize}

\ep

\subsubsection{Measure}

\bd[Measure]
Let the real measure on Borel subsets of $\CC$:
\beq
\mu(E) = \frac{1}{2\pi } \int_E \Delta\Phi.
\eeq
\ed
From theorem \ref{th:measureDeltaPhi} we have
\bl
The measure is supported on $\hat\Gamma$. 
Along edges of $\hat\Gamma$, the measure has density
\beq
d\mu = \frac{1}{\pi \ii } Y dX = \frac{1}{\pi } \Im \ Y dX,
\eeq
and is real.

\el

\bt[Stieltjes transform]
The Stieltjes transform of $\mu$
\beq
W(x) = \int_{\hat\Gamma} \frac{d\mu(x')}{x-x'}
\eeq
is analytic in $\CC\setminus \hat\Gamma$, and is worth
\beq
W(x) = \frac12 V'(x) - Y(x).
\eeq
\et

\proof{
Let $\td W(x) = \frac1{2} \left( V'(x) - 2 Y(x)\right) $, we have
\beq
\td W(x_{\text{left}}) - \td W(x_{\text{right}}) = - (Y(x_{\text{left}}) - Y(x_{\text{right}})) = -2\pi\ii d\mu(x) 
\eeq
Moreover,
\beq
\td W(x) = x^{-1} + O(x^{-2}).
\eeq
These two properties characterize the Stieltjes transform, and imply that $W=\td W $.
}

\bt[Energy]
We have
\beq
F = \int_{\hat\Gamma} \Re V(x) d\mu(x)  -\int_{\hat\Gamma}\int_{\hat\Gamma} \ln{|x-x'|} d\mu(x')d\mu(x) .
\eeq

\et

\proof{
Remark that in $\CC\setminus\hat\Gamma$ we have
\bea
\int_{\hat\Gamma} \ln{(x-x')} d\mu(x')
&=& \ln x + \int_\infty^x (W(x')-1/x') dx' \cr
&=& \frac12 V(x) - g_{\infty_+}(x) \cr
&=& \frac12 V(x) - g(x) + \td t_{\infty_+,0}.
\eea
Indeed both the left and right hand side behave as $\ln x + O(1/x)$ at large $x$, and both have the same derivative $  \int_{\hat\Gamma} \frac{1}{x-x'} d\mu(x') =W(x)$.
Taking the real part this gives
\beq
\int_{\hat\Gamma} \ln{|x-x'|} d\mu(x')
= \frac12 \Re V(x) - \Phi(x) + \Re \ \td t_{\infty_+,0},
\eeq
and since $\Phi=0$ on $\hat\Gamma$:
\beq
\int_{\hat\Gamma}\int_{\hat\Gamma} \ln{|x-x'|} d\mu(x')d\mu(x)
= \frac12 \int_{\hat\Gamma} \Re V(x)d\mu(x)  + \Re \ \td t_{\infty_+,0}
\eeq
and
\beq\label{eq:1mmtdtinftyintln}
 \Re \ \td t_{\infty_+,0} = 
\int_{\hat\Gamma}\int_{\hat\Gamma} \ln{|x-x'|} d\mu(x')d\mu(x)
- \frac12 \int_{\hat\Gamma} \Re V(x)d\mu(x)  .
\eeq

Beside we have
\beq
t_{\infty_+,k} = - t_{\infty_-,k} = -\frac12 t_k ,
\eeq
and
\beq
\td t_{\infty_+,k} = -\td t_{\infty_-,k}
= \frac{1}{k}\Res_{\infty_+} x^k ydx 
= - \frac{1}{k}\Res_{\infty_+} x^k W(x)dx
= \frac{1}{k} \int_{\hat\Gamma} x^k \ d\mu(x).
\eeq
This implies that
\beq
\sum_{k=1}^{\deg V} t_{\infty_+,k} \td t_{\infty_+,k} = \sum_{k=1}^{\deg V} t_{\infty_-,k} \td t_{\infty_-,k}
= -\frac12 \int_{\hat\Gamma} V(x) \ d\mu(x).
\eeq

\bea
2 F_0
&=& \sum_{k=1}^{\deg V} t_{\infty_+,k} \td t_{\infty_+,k}+ \sum_{k=1}^{\deg V} t_{\infty_-,k} \td t_{\infty_-,k} + t_{\infty_+,0} \td t_{\infty_+,0}  + t_{\infty_-,0} \td t_{\infty_-,0}  \cr
&=&  -\int_{\hat\Gamma} V(x) d\mu(x)  + 2 t_{\infty_+,0} \td t_{\infty_+,0}, 
\eea
and thus, and using eq \eqref{eq:1mmtdtinftyintln}:
\bea
F = -\Re F_0
&=&  \frac12 \int_{\hat\Gamma} \Re V(x) d\mu(x)  - t_{\infty_+,0} \Re \ \td t_{\infty_+,0}  \cr
&=&  \frac12\int_{\hat\Gamma} \Re V(x) d\mu(x)  -  \Re \ \td t_{\infty_+,0}  \cr
&=&  \int_{\hat\Gamma} \Re V(x) d\mu(x)  - \int_{\hat\Gamma}\int_{\hat\Gamma} \ln{|x-x'|} d\mu(x')d\mu(x) .
\eea
}
\bt[Energy]
It is possible to choose the Boutroux curve such that 
$\mu$ is a probability measure (positive and total mass 1) on $\hat\Gamma$, and is the extremal measure of the following functional
\beq
F = \inf_{\nu \in \text{probability measures on }\hat\Gamma} \left(  \int_{\hat\Gamma} \Re V(x) d\nu(x)  -\int_{\hat\Gamma}\int_{\hat\Gamma} \ln{|x-x'|} d\nu(x')d\nu(x) \right).
\eeq

\et

\proof{

This is a classical theorem in potential theory.
In the context of random matrices this is done for example in \cite{BAG97}.
We refer to random matrix literature and potential theory literature.

Let us just sketch the main ideas:
Let the functional defined on the space $\operatorname{Pr}(\hat\Gamma)$ of probability measures on $\hat\Gamma$, equipped with the weak topology:
\beq
\mathcal F(\nu) = \int_{\hat\Gamma} \Re V(x) d\nu(x)  -\int_{\hat\Gamma}\int_{\hat\Gamma} \ln{|x-x'|} d\nu(x')d\nu(x).
\eeq

$\bullet$ $\mathcal F$ is bounded from below: let $\nu_0$ any given probability measure on $\hat\Gamma$, and $U_0(x) = \int_{\hat\Gamma} \ln|x-x'| d\nu_0(x')$. We have
\bea
\mathcal F(\nu) 
&=&   -\int_{\hat\Gamma}\int_{\hat\Gamma} \ln{|x-x'|} d(\nu(x')-\nu_0(x')) \ d(\nu(x)-\nu_0(x)) \cr
&& + \int_{\hat\Gamma}\int_{\hat\Gamma} \ln{|x-x'|} d\nu_0(x') \ d\nu_0(x)  \cr
&& +   \int_{\hat\Gamma} (\Re V(x)-2U_0(x)) d\nu(x) \cr
& \geq & \int_{\hat\Gamma}\int_{\hat\Gamma} \ln{|x-x'|} d\nu_0(x') \ d\nu_0(x)  +  \inf_{\hat\Gamma} (\Re V(x)-2U_0(x)), 
\eea
which shows that $\mathcal F$ is bounded from below.

$\bullet$  $\mathcal F$ is lower semi-continuous:
let $\mathcal F_M(\nu) =\int_{\hat\Gamma} \Re V(x) d\nu(x)  -\int_{\hat\Gamma}\int_{\hat\Gamma} \ln_M{|x-x'|} d\nu(x')d\nu(x)  $ where $\ln_M(x) = \max(M,\ln{|x|})$. 
One can verify that $\mathcal F_M$ is Lipschitzien, and thus continuous  (with the weak topology).
$\mathcal F=\limsup_{M\to -\infty} \mathcal  F_M$ is a limsup of continuous functions, therefore is lower semi-continuous.

$\bullet$ Since $\hat\Gamma$ is compact and is a finite union of Jordan arcs, the space $\operatorname{Pr}(\hat\Gamma)$ is a Susslin space (image by a continuous function = the Jordan arc, of a Polish space = here an interval of $\RR$), which implies that it is complete, and every probability measure on $\hat\Gamma$ is tight.
The level sets of $\mathcal F$ are closed (because $\mathcal F$ is lower semi-continuous) and compact by Prokhorov's theorem.

$\bullet$ The intersection of a decreasing sequence of non-empty compacts is a non empty compact, and any element in this intersection is a minimum of $\mathcal F$. Therefore $\mathcal F$ admits at least one minimum.

$\bullet$ $\mathcal F$ is strictly convex, so the minimum is unique.

$\bullet$ The minimum must satisfy Euler-Lagrange equations, and this can be written in the following way:

a probability measure $\hat\mu$ on $\hat\Gamma$ is said to satisfy Euler-Lagrange equations if and only if
\beq
\exists \ell\in \RR \ , \quad
\Phi(x) =\int_{\operatorname{supp}(\hat\mu)} \left(\frac12\Re V(x') - \ln{|x-x'|} \right)d\hat\mu(x')
\quad  \left\{
\begin{array}{l}
=\ell \ \ \text{if } x\in \operatorname{supp}(\hat\mu) \cr
\geq \ell \ \ \text{if } x\in \hat\Gamma\setminus \operatorname{supp}(\hat\mu) \cr
\end{array}
\right. .
\eeq
Let $\hat W(x) = \int_{x'\in \hat\Gamma} \frac{1}{x-x'}d\hat\mu(x')$ the Stieltjes transform of $\hat\mu$.
It is analytic outside $\operatorname{supp}(\hat\mu)\subset\hat\Gamma$.
It may be discontinuous across $\operatorname{supp}(\hat\mu)\subset \hat\Gamma$, with discontinuity
\beq
x\in \operatorname{supp}(\hat\mu) \qquad \implies \quad
\hat W(x_{left})dx-\hat W(x_{right})dx
= 2\pi\ii d\hat\mu(x).
\eeq
On the other hand, the Euler-Lagrange equations in the support imply that
\beq
x\in \operatorname{supp}(\hat\mu) \qquad \implies \quad
\hat W(x_{left})+\hat W(x_{right}) -V'(x) = 0.
\eeq
Let us define $R(x) := V'(x)\hat W(x) - \hat W(x)^2$.
$R(x)$ is analytic outside of $\operatorname{supp}(\hat\mu)$, and on $\operatorname{supp}(\hat\mu)$ it satisfies
\bea
R(x_{left})- R(x_{right})
&=& V'(x)(\hat W(x_{left})-\hat W(x_{right})) \cr
&& - (\hat W(x_{left})+\hat W(x_{right}))(\hat W(x_{left})-\hat W(x_{right})) \cr
&=& 0.
\eea
Thanks to Cauchy-Riemann equations, this shows that $R(x)$ is in fact analytic in the whole complex plane.
Moreover it behaves at $\infty$ as $O(|V(x)/x|)$, so $R(x)$ is an entire function bounded by a polynomial, it must be a polynomial.
We have
\beq
\hat W(x)^2 = V'(x)\hat W(x)-R(x).
\eeq
If we write $\hat y = \frac12 V'(x)-\hat W(x)$, we see that $\hat y$ is an algebraic function of $x$ satisfying an equation
\beq
\hat y^2 = (\frac12 V'(x)-\hat W(x))^2 = \frac14 V'(x)^2 - R(x). 
\eeq
Remark that
$\hat P(x,y) = y^2 - \frac14 V'(x)^2 +R(x) $ is a plane curve that belongs to our moduli space $\modsp$.

The Euler-Lagrange equations imply that $\hat\Phi(x)=\Re \int \hat y dx $ is constant on the support, which implies that $\hat P$ is a Boutroux curve.
Moreover, it is a Boutroux curve with a positive probability measure $\hat \mu$.

This implies that it is possible to choose a Boutroux curve in $\modsp$ such that the Boutroux curve's measure $\mu = \frac{1}{2\pi}\Delta\Phi$ is a positive probability measure.
}

\subsubsection{$g$-function in the Riemann-Hilbert method}

The ingredients of the Steepest-descent method of \cite{deift1992steepest} are a graph, a set of ``jump matrices" associated to each edge of the graph $\Upsilon$, and a function $g$ defined in the complex plane, and that has certain properties near the edges of the graph.

We claim that the following data is the data needed for the Steepest-descent method of \cite{deift1992steepest}:
\begin{itemize}

\item $\Upsilon$ contains all edges and vertices of $\hat\Gamma$.

\item for each connected component of $\hat\Gamma$ (made of vertical edges that border of some half-plane $\phi<0$), let $a$ its highest vertex (highest value of $\Im g$). From $a$ follow a horizontal trajectory where $\phi$ is non-decreasing. Each time you meet a vertex, choose the uppermost horizontal trajectory.
Add this horizontal trajectory to $\Upsilon$.

\item add to $\Upsilon$ some ``lenses" around edges of $\hat\Gamma$. Since edges of $\hat\Gamma$ border domains where $\Phi<0$, we choose the lenses small enough to be entirely in domains $\Phi<0$.

\item add to $\Upsilon$ some small circle around vertices of $\hat\Gamma$,

\item the $g$-function is the function $g$. It's real part is constant and vanishing on the edges of $\hat\Gamma$. It is growing on horizontal trajectories.
$\Re g$ is $<0$ on the lenses. $g$ behaves like $C_a (x-x_a)^{r_a/a_a}$ near a vertex $a$.

\end{itemize}

We then refer to \cite{deift1992steepest,deift1999,bertola2007boutroux}.

\subsubsection{Interpretation as matrix models}

Let $N$ a positive integer.
Consider the following measure on $H_N$ the space of Hermitian matrices of size $N$:
\beq
\frac{1}{Z_N} e^{-N \tr V(M) } dM,
\eeq
where $dM$ is the canonical Lebesgue measure on $H_N$.
(all this can be extended to $H_N(\gamma)$ the set of normal matrices with eigenvalues on $\gamma$, i.e. $H_N(\RR)=H_N$ if $\gamma=\RR$).
The normalization constant is
\beq
Z_N = \int_{H_N(\gamma)} e^{-N \tr V(M) } dM.
\eeq
In the large $N$ limit, the empirical density of eigenvalue of $M$ will tend to a limit, called the ``equilibrium density" $d\mu(x)$.
Equivalently in the large $N$ limit, the Stieltjes transform of the empirical density of eigenvalue of $M$ will tend to a limit $W(x)$.

The conjecture is that 
\beq
W(x) = \frac12 V'(x)-Y(x),\quad
d\mu(x) = \frac{1}{2\pi \ii } \left(Y(x_{left})-Y(x_{right}) \right),
\eeq
where $P(x,y)$ is a Boutroux curve, $\hat\Gamma$ is the cellular graph of a maximal domain containing all tiles adjacent to the puncture $\infty_+$, and $Y(x)$ is the solution of $P(x,y)=0$ in $\CC\setminus\hat\Gamma$, and $d\mu$ is the measure supported on $\hat\Gamma$ given by the discontinuity of $Y$.

In principle this conjecture can be proved by the Riemann--Hilbert method and the Steepest-descent method of \cite{deift1992steepest}.

\subsection{2 Matrix model}

Let $V(x)\in \CC[x], \td V(y)\in \CC[y] $ be two polynomials, of degree at least two. Denote their leading coefficients:
\beq
V(x) = \frac{t_{\td d}}{\td d} x^{\td d} + O(x^{\td d-1}),\quad
\td V(y) = \frac{\td t_{d}}{d} y^{d} + O(y^{d-1}).
\eeq

\subsubsection{Newton's polygon}

Let
\beq
\mathcal P(x,y) = (y-V'(x))(x-\td V'(y)) -  t_{\td d} \td t_d x^{\td d-2} y^{d-2}.
\eeq
There are exactly two punctures, that we denote $\infty_+$ and $\infty_-$.
We have
\begin{itemize}

\item at $\infty_+$, $a_+=a_{\infty_+}=1$, $b_+=b_{\infty_+}=\deg V'=\td d-1$, $r_+=r_{\infty_+}=\deg V = \td d $.
\beq
x = \zeta_{+}^{-1}, \qquad
y \sim \eta_+ \zeta_{+}^{1-\td d}.
\eeq
The times at $\infty_+$ 
\beq
t_{\infty_+,k} = \Res_{\infty_+} \zeta_+^{-k} Y dX,
\eeq
are such that
\beq
V'(x) = -\sum_{k=1}^{\td d} \frac{t_{\infty_+,k}}{k} x^k,
\eeq
in particular
\beq
t_{\infty_+,k} = -t_{\td d}, \qquad
t_{\infty_+,0} = 1.
\eeq

\item at $\infty_-$, $a_-=a_{\infty_-}=\deg \td V' = d-1$,  $b_-=b_{\infty_-}=1$, $r_-=r_{\infty_-}=\deg \td V = d $.
\beq
x = \zeta_{-}^{1-d}, \qquad
y \sim \eta_- \zeta_{-}^{-1}.
\eeq
\end{itemize}
\subsubsection{Boutroux curve and spectral network}

We have
\beq
\modsp = \{\mathcal P(x,y) + Q(x,y) \ | \ 
Q(x,y) = \sum_{i\leq d-2, \ j\leq \td d-2, \ i+j<d+\td d-4} Q_{i,j} x^i y^j \}.
\eeq
\beq
\dim \modsp = (d-1)(\td d-1)-1.
\eeq

Consider now a Boutroux curve in $\modsp$:
\beq
P(x,y) = \mathcal P(x,y) + Q(x,y) \qquad = \text{Boutroux}.
\eeq

The functions $\phi(p)=\Re \int_o^p Y dX$ and $\td\phi(p)=\Re \int_o^p XdY$ are harmonic on $\curve\setminus\{\infty_+,\infty_-\}$.
We use them to define the index $i(p)$ (resp. $\td i(p)$), and the spectral network graphs $\Gamma$ (resp. $\td \Gamma$) of Section \ref{sec:spnetwork2}.

We use the second version of spectral networks.

\bd

For $\phi$ (resp. $\td \phi$), let $\mathcal D_{+}$ (resp. $\mathcal D_{-}$) a maximal admissible domain that contains the union of all domains $\td{\mathcal D}_{i,j}$ (resp. $\td{\mathcal D}_{i,j}$) that have $\infty_+$ (resp. $\infty_-$) at their boundary. 
Let $\mathcal C_{+}=X(\mathcal D_{+})$ (resp. $\mathcal C_{-}=Y(\mathcal D_{-})$), and let $\hat\Gamma_+$ (resp. $\hat\Gamma_-$) its boundary.
\ed
As an immediate consequence of theorem \ref{thm:maxdomain}, we have:
\bt
$\phi$ (resp. $\td \phi$) is harmonic on $\mathcal D_+$ (resp. $\mathcal D_-$).
$\Phi=\phi\circ X^{-1} $ (resp. $\td\Phi=\td \phi\circ Y^{-1} $) is harmonic on $\mathcal C_+$ (resp. $\mathcal C_-$).

The complement $\hat\Gamma_+= \CC\setminus \mathcal C_+$ (resp. $\hat\Gamma_-=\CC\setminus \mathcal C_-$) is a cellular graph.

The locus where $\Phi$ (resp. $\td\Phi$) is not harmonic is exactly on $\hat\Gamma_+$ (resp. $\hat\Gamma_-$).

\et

\subsubsection{Measures}

\bd[Measures]

Let us define the following measures on $\CC$, supported on $\hat\Gamma_+=\CC\setminus \mathcal C_+$ (resp. $\hat\Gamma_-=\CC\setminus \mathcal C_-$)
\beq\label{eq:dmu2MMfromdiscY}
\left\{\begin{array}{ll}
d\mu = \frac{1}{2\pi\ii}(Y(p^i(x))-Y(p^{j}(x)))dx & \qquad \text{along an edge separating index }(i,j) \cr
d\mu = 0 & \qquad  \text{inside open domains}  \cr
\end{array}\right.
\eeq

\beq
\left\{\begin{array}{ll}
d\td\mu = \frac{1}{2\pi\ii}(X(\td p^{\td i}(y))-X(\td p^{\td j}(y)))dy & \qquad \text{along an edge separating index }(\td i,\td j) \cr
d\td\mu = 0 & \qquad  \text{inside open domains}  \cr
\end{array}\right.
\eeq
They are such that
\beq
\mu(E) = \frac{1}{2\pi}\int_E \Delta \Phi
\qquad \left(\text{resp. } \ 
\td\mu(E) = \frac{1}{2\pi}\int_E \Delta \td\Phi \ \right).
\eeq

\ed

\bt[Stieltjes transform]
The Stieltjes transform
\beq\label{eq:defStieltjes2MM}
W(x) = \int_{\operatorname{supp} d\mu } \frac{d\mu(x')}{x-x'},
\eeq
\beq
\td W(y) = \int_{\operatorname{supp} d\td\mu } \frac{d\td\mu(y')}{y-y'}.
\eeq
defined on the complement $\CC\setminus \mathcal C_+$ (resp. $\CC\setminus \mathcal C_-$), is worth
\beq\label{eq:StieltjesW2mmfromY}
W(x) = V'(x)-Y(x),
\eeq
\beq
\td W(y) = \td V'(y)-X(y).
\eeq

\et

\proof{
The discontinuity of eq \eqref{eq:StieltjesW2mmfromY} across the support of $\mu$ is equal to eq \eqref{eq:dmu2MMfromdiscY} times $2\pi\ii$, and it behaves as $1/x+O(1/x^2)$ at large $x$, this characterizes the Stieltjes transform eq \eqref{eq:defStieltjes2MM}.
Idem for $\td W$.
}

\subsubsection{Interpretation as matrix models}

Let $N$ a positive integer.
Consider the following measure on $H_N\times H_N $:
\beq
\frac{1}{Z_N} e^{-N \tr \left(V(M_1)+\td V(M_2)-M_1 M_2\right) } dM_1 dM_2.
\eeq
where $dM$ is the canonical Lebesgue measure on $H_N$.
(all this can be extended to $H_N(\gamma)\times H_N(\td\gamma) $ the set of normal matrices with eigenvalues on $\gamma$ and $\td\gamma$).
In the large $N$ limit, the empirical density of eigenvalue of $M_1$ (resp. $M_2$) will tend to a limit, called the ``equilibrium density" $d\mu(x)$ (resp. $d\td\mu(y)$).
Equivalently in the large $N$ limit, the Stieltjes transform of the empirical density of eigenvalue of $M_1$ (resp. $M_2$) will tend to a limit $W(x)$ (resp. $\td W(y)$).

The conjecture is that 
\beq
W(x) = V'(x)-Y(x), \quad
(\text{resp. } \td W(y) = \td V'(y)-X(y) ),
\eeq
\beq
d\mu(x) = \frac{1}{2\pi \ii } \left(Y(x_{left})-Y(x_{right}) \right), \quad
(\text{resp. } d\td\mu(y) = \frac{1}{2\pi \ii } \left(X(y_{left})-X(y_{right}) \right) ),
\eeq
where $P(x,y)$ is a Boutroux curve, $\hat\Gamma$ (resp. $\td{\hat\Gamma}$) is the cellular graph of a maximal domain containing all tiles adjacent to the puncture $\infty_+$ (resp. tiles of $\td\phi$ adjacent to $\infty_-$), and $Y(x)$ (resp. $X(y)$) is the solution of $P(x,y)=0$ in $\CC\setminus\hat\Gamma$ (resp. $\CC\setminus\td{\hat\Gamma}$), and $d\mu$ (resp. $d\td\mu$) is the measure supported on $\hat\Gamma$ (resp. $\td{\hat\Gamma}$) given by the discontinuity of $Y$ (resp. $X$).

We believe that this should be provable by Deift-Zhou's steepest descent method \cite{deift1992steepest,deift1999}.
This proof was achieved so far in very few examples of low degree.

\subsubsection{Matytsin property}

In the random 2-matrix model, it is conjectured (and proved in some cases \cite{Guionnet2002LargeDA}) that the partition function $Z_N$, or more precisely $\frac{1}{N^2}\ln Z_N$ has a limit at large $N$.
\beq
\exists \lim_{N\to\infty} \frac{-1}{N^2} \ln Z_N = \mathcal F.
\eeq
Since the equilibrium measures $\mu$ and $\td\mu$ are functions of the potentials $V$ and $\td V$, we can locally describe $\mathcal F$ as a functional of two measures:
\beq
\mathcal F = \mathcal F(\mu,\td\mu).
\eeq
We also define
\bea
\mathcal I(\mu,\td\mu)
&=& -\mathcal F(\mu,\td\mu)   + \int_{\operatorname{supp}(\mu)} \Re V(x)d\mu(x)+\int_{\operatorname{supp}(\td\mu)} \Re \td V(y)d\td\mu(y) \cr
&& - \int_{\operatorname{supp}(\mu)\times \operatorname{supp}(\mu)} \ln{|x-x'|} d\mu(x)d\mu(x')\cr
&& - \int_{\operatorname{supp}(\td\mu)\times \operatorname{supp}(\td\mu)} \ln{|y-y'|} d\td\mu(y)d\td\mu(y').\cr
\eea
The interpretation of $\mathcal I(\mu,\td\mu)$ is that
\bea
e^{N^2 \mathcal I} 
& \mathop{\sim}_{N\to \infty} & \mathbb E\left(
e^{N^2 \Tr M_1 M_2}
\ | \ \text{ knowing that } \  \operatorname{sp}(M_1)\to\mu, \operatorname{sp}(M_2)\to\td\mu \right) \cr
& \mathop{\sim}_{N\to \infty} & \int_{U\in U(N)} e^{N^2 \Tr \Lambda U \td\Lambda U^\dagger} dU \ | \ \text{ with } \  \operatorname{sp}(\Lambda)\to\mu, \operatorname{sp}(\td \Lambda)\to\td\mu.  \cr
\eea
i.e. the expectation value of $e^{N^2 \Tr M_1 M_2}$ knowing that the spectrum $\Lambda$ of $M_1$ (resp. $\td\Lambda$ of $M_2$),  empirical spectral measure, tends to the measure $\mu$ (resp. $\td\mu$) at large $N$.

For fixed diagonal matrices $\Lambda$ and $\td\Lambda$ of size $N$, the following integral
\beq
I_N(\Lambda,\td\Lambda) = \int_{U\in U(N)} e^{N^2 \Tr \Lambda U \td\Lambda U^\dagger} dU,
\eeq
with $dU$ the Haar measure on $U(N)$, is known as the Itzykson-Zuber (case $\Lambda,\td\Lambda$ real) or Harish-Chandra integral (case $\Lambda,\td\Lambda$ purely imaginary) and is worth
\beq
I_N(\Lambda,\td\Lambda) = \frac{\det e^{N^2\Lambda_i\td\Lambda_j}}{\prod_{i<j} (\Lambda_i-\Lambda_j)(\td\Lambda_i-\td\Lambda_j)}.
\eeq

In \cite{Matytsin1994}, Matytsin derived heuristically from this exact formula, that in the large $N$ limit, the functional $\mathcal I(\mu,\td\mu)$ should satisfy some functional equation:

Let
\beq
\hat Y(x) = \frac{d}{dx} \frac{\partial}{\partial d\mu(x)} \left( \mathcal  I(\mu,\td\mu) + \frac12\int_{\operatorname{supp}(\mu)\times \operatorname{supp}(\mu)} \ln{|x-x'|} d\mu(x)d\mu(x') \right),
\eeq
\beq
\hat X(y) = \frac{d}{dy} \frac{\partial}{\partial d\td\mu(y)} \left( \mathcal  I(\mu,\td\mu) + \frac12\int_{\operatorname{supp}(\td\mu)\times \operatorname{supp}(\td\mu)} \ln{|y-y'|} d\td\mu(y)d\td\mu(y') \right).
\eeq
Matytsin claimed that they must be functional inverse of one-another:
\beq
\hat X \circ \hat Y = \text{Id}.
\eeq

Let us verify that this is satisfied by the measures we have obtained from the Boutroux curve and its spectral network:

\bt
The measures $\mu$, $\td\mu$  satisfy the Matytsin property 
\et

\proof{
Let here $\hat\mu$ and $\hat{\td\mu}$ be the measures that minimize  the energy $\mathcal F$.
This implies that 
\beq
\frac{\partial}{\partial d\hat\mu(x)} \mathcal F=0 = \frac{\partial}{\partial d\hat{\td\mu}(y)} \mathcal F.
\eeq
Therefore this implies
\beq
\hat Y(x) = \frac{d}{dx} \frac{\partial}{\partial d\hat\mu(x)} \mathcal I = V'(x) - \int_{\operatorname{supp}(\hat\mu)} \frac{d\hat\mu(x')}{x-x'} = V'(x) - \hat W(x),
\eeq
\beq
\hat X(y) = \frac{d}{dy} \frac{\partial}{\partial d\hat{\td\mu}(y)} \mathcal I = \td V'(y) - \int_{\operatorname{supp}(\hat{\td\mu})} \frac{d\hat{\td\mu}(y')}{y-y'} = \td V'(y) - \hat{\td W}(y),
\eeq
where $\hat W(x)$ (resp. $\hat{\td W}(y)$) designates the Stieltjes transform of the measure $\hat\mu$ (resp. $\hat{\td\mu}$).

The Matytsin property is thus formulated as
\beq
\hat X\circ \hat Y=\text{Id}.
\eeq

Here, the measures $\mu$ and $\td\mu$ constructed from the Boutroux curve satisfy:
\beq
Y(x) = V'(x) - W(x), \quad
X(y) =  \td V'(y)-\td W(y),
\eeq
where $W(x)$ (resp. ${\td W}(y)$) designates the Stieltjes transform of the measure $\mu$ (resp. ${\td\mu}$).
And the function $Y(x)$ is the solution of $P(x,Y(x))=0$, while the function $X(y)$ is the solution of the $P(X(y),y)=0$ for the same Boutroux curve $P(x,y)\in \modsp $.
Therefore they satisfy on $\curve$:
\beq
X\circ Y=\text{Id}_\curve.
\eeq
}

\br
Remark that the Matytsin property is a consequence that the Boutroux property is invariant under the exchange $x\leftrightarrow y$.
In other words, if $\Re \oint_\gamma ydx=0$ for all closed $\gamma$, then $\Re \oint_\gamma xdy=0$ as well.

\er

\subsection{Matrix model with external field}

Let $V(x)\in \CC[x]$ be a polynomial of degree $d\geq 2$
\beq
V(x) = \sum_{k=1}^d \frac{t_k}{k} x^k.
\eeq
Let $r$ a positive integer $r\geq 1$, and let $r$ distinct complex numbers $A_1,\dots,A_r \in \CC^r $, and let $\nu_1,\dots,\nu_r \in \RR^r$ be $r$ real numbers.
We let
\beq
t=\sum_{i=1}^r \nu_i.
\eeq
Let
\beq
S(y)=\prod_{i=1}^r (y-A_i).
\eeq

\subsubsection{Newton's polygon}

Let
\beq
\mathcal P(x,y) = \left( y-V'(x) + \sum_{i=1}^r \frac{\nu_i}{y-A_i}\right) S(y).
\eeq
It has $r+1$ punctures, that we denote $\alpha_i$, $i=0,\dots,r$:

\begin{itemize}

\item at $\alpha_0$, we have $X(\alpha_0)=\infty$, $Y(\alpha_0)=\infty$, $a_0=1$, $b_0=\deg V'=d-1$, $r_0=\deg V = d $.
\beq
x = \zeta_{0}^{-1}, \qquad
y \sim \eta_+ \zeta_{0}^{1-d}
\eeq
The times at $\alpha_0$ are
\beq
t_{\alpha_0,k} = \Res_{\alpha_0} \zeta_0^{-k} Y dX = -t_{k}
\qquad , k=1,\dots,d
\eeq
and
\beq
t_{\alpha_0,0} = t  = \sum_{i=1}^r \nu_i.
\eeq

\item at $\alpha_i$ for $i=1,\dots,r$, we have $Y(\alpha_i)=A_i$ and $X(\alpha_i)=\infty$, and $a_i=a_{\alpha_i}=1$,  $b_i=b_{\alpha_i}=-1$, $r_i=r_{\alpha_i}= 0$.
We have
\beq
t_{\alpha_i,k} = -\nu_i \delta_{k,0}.
\eeq

The moduli space is:
\beq
\modsp = \mathcal P + \left\{ S(y)\sum_{k=0}^{d-3}\sum_{i=1}^r c_{k,i} x^k/(y-A_i)  \right\}, \quad
\dim \modsp = (d-2)r.
\eeq

\end{itemize}

\subsubsection{Boutroux Curve}

Consider now the Boutroux curve

\beq
P(x,y) = \mathcal P(x,y) + Q(x,y) \qquad = \text{Boutroux}.
\eeq

The function $\phi(p)=\Re \int_o^p Y dX$ is harmonic on $\curve\setminus\{\alpha_0,\dots,\alpha_r\}$.
We use it to define the index $i(p)\in [0,\dots,r]$, and the spectral network graphs $\Gamma$ of Section \ref{sec:spnetwork2}.

\subsubsection{Interpretation as matrix model}

This is related to a random matrix model as follows: let $N\in \ZZ $ an integer, let $n_i=N \nu_i$. Let
\beq
A=\operatorname{diag}(\overbrace{A_1\dots,A_1}^{n_1},\overbrace{A_2\dots,A_2}^{n_2},\dots,\overbrace{A_r\dots,A_r}^{n_r} )
\eeq
a diagonal matrix with $r$ degenerate eigenvalues, of total size
$ N=\sum_{i=1}^r n_i=Nt $.
This can be interpreted as a diagonal matrix only if all $n_i$ are positive integers. 
In fact for negative integers it could be interpreted as a fermionic diagonal matrix.
However, the formalism presented here works for $\nu_i$ real.

The Boutroux curve below will be associated to the following matrix measure:
\beq
\frac{1}{Z_N(A)} e^{-\frac{N}{t} \tr \left( V(M)-M A\right) } dM,
\eeq
where $dM$ is the canonical measure on $H_N=H_N(\RR)$ (and generalizable to $H_N(\gamma)$ the set of normal matrices with eigenvalues on $\gamma$), with partition function 
\beq
Z_N(A) = \int_{H_N(\gamma)} e^{-\frac{N}{t} \tr \left( V(M)-M A\right) } dM.
\eeq
In the large $N$ limit, the empirical density of eigenvalue of $M$ is conjectured (only proved in very few cases with small $r$ and small $d=\deg V$) to tend to a limit, called the ``equilibrium density" $d\mu(x)$.
Equivalently in the large $N$ limit, the Stieltjes transform of the empirical density of eigenvalue of $M$ will tend to a limit $W(x)$.

The conjecture is that 
\beq
t W(x) = V'(x)-Y(x),
\eeq
\beq
d\mu(x) = \frac{1}{2\pi \ii t} \left(Y(x_{left})-Y(x_{right}) \right), 
\qquad \operatorname{supp}(d\mu)=\hat\Gamma,
\eeq
where $P(x,y)$ is a Boutroux curve, $\hat\Gamma$ is the cellular graph of a maximal domain containing all tiles adjacent to the puncture $\alpha_0$, and $Y(x)$ is the solution of $P(x,y)=0$ in $\CC\setminus\hat\Gamma$, and $d\mu$ is the measure supported on $\hat\Gamma$ given by the discontinuity of $Y$.

We believe that this should be provable by Deift-Zhou's steepest descent method \cite{deift1992steepest,deift1999}, using the $g$-function of the Boutroux curve as the $g$-function of \cite{deift1992steepest,deift1999}.

\section{Conclusion}
\label{sec:conc}
We have proved the existence of Boutroux curves in the moduli space of algebraic plane curves with prescribed punctured asymptotic behaviors.

This has many practical applications, like finding foliations of surfaces (Strebel's theorem), or finding spectral networks, and finding the $g$-functions useful for the Riemann-Hilbert method in asymptotic theory in random matrices, or in potential theory.

\smallskip
\textbf{Possible generalizations}
\begin{itemize}

\item All the method presented here can probably be extended to algebraic curves over a base curve $x\in \curve_0$ not necessarily $\CC$ or $\CC P^1$, but any compact Riemann surface $\curve_0$.

\item Also instead of $Y\in \CC $, the 1-form $YdX$ could take its values in the cotangent space $T^*\curve_0$, and more generally in the adjoint bundle of a Higgs bundle for a Lie group $G$.
In other words, this formalism should be extended to Hitchin spectral curves.

\item Another generalization could replace $\CC\times \CC$ by $\CC^*\times\CC^*$. It would means replace $X\to \ln{X}$ and $Y\to \ln{Y}$, i.e. a polynomial equation $P(e^x,e^y)=0$ rather than $P(x,y)=0$. Many of the tools used here would be adaptable.
Instead of subtracting poles at punctures, we would subtract logarithmic singularities, but almost all the rest would be similar.

\end{itemize}

\section*{Acknowledgments}

This work is supported by the ERC synergy grant \textbf{ERC-2018-SyG  810573}, ``ReNewQuantum''.
We thank A. Boutet de Monvel, G. Borot, M. Kontsevich and F. Zerbini for discussions on this topic.

\par\noindent\rule{\textwidth}{0.4pt}

\appendix

\section{Punctures and Newton's polygon}
\label{Apppunctures}

\subsection{Punctures and slopes}

It is well known that there is a 1-1 correspondence between punctures and minimal integer segments of the convex envelope of $\Newt$.
I.e. a segment $\alpha=[(i_1,j_1),(i_2,j_2)]$ of the envelope, such that $(i_1,j_1)\in \Newt$, $(i_2,j_2)\in \Newt$ and the half plane strictly to the left of the segment $\alpha$ doesn't contain any point of $\Newt$, and ``minimal" means that the segment $\alpha$ contains no other integer point in its interior.

- If the segment is horizontal ($j_2=j_1$), $\alpha$ corresponds to a pole of $Y$.

- If the segment is vertical ($i_2=i_1$), $\alpha$ corresponds to a pole of $X$.

- Otherwise, $\alpha$ is a pole of both $X$ and $Y$ such that in a local variable $\zeta\to 0$
\beq
X(z)\sim \zeta^{j_2-j_1}, \quad
Y(z)\sim \eta \zeta^{i_1-i_2}.
\eeq
If $\alpha$ is not contained in a larger segment of the envelope, in a neighborhood of the puncture we have asymptotically
\beq
P_{i_1,j_1} x^{i_1}y^{j_1} + P_{i_2,j_2} x^{i_2}y^{j_2} = o(x^{i_1}y^{j_1}),
\eeq
i.e.
\beq
y \sim x^{\frac{i_1-i_2}{j_2-j_1}} \ \ \left(-\frac{P_{i_1,j_1}}{P_{i_2,j_2}}\right)^{1/(j_2-j_1)}.
\eeq
More generally, if $\alpha\subset \alpha'$ a maximal segment of the envelope, we have
\beq
\sum_{(i,j)\in \alpha'} P_{i,j} x^{i}y^{j}  = o(x^{i_1}y^{j_1})
\eeq
i.e.
\beq
y \sim x^{\frac{i_1-i_2}{j_2-j_1}} \ \ C,
\eeq
where $C$ is some solution of
\beq
\sum_{(i,j)\in\alpha'} P_{i,j}C^{j-j_1} = 0.
\eeq

If the segment is horizontal, the values of $X$ at the punctures are the zeros of
\beq
\sum_{i | \ (i,j)\in \alpha'} P_{i,j} x^i =0.
\eeq

The canonical local coordinate in the neighborhood of a puncture $\alpha$ is:
\begin{itemize}
\item If $\alpha$ is a pole of $X$ of some degree $-d_\alpha=j_1-j_2>0$:
\beq
\zeta_\alpha = X^{1/d_\alpha}.
\eeq
\item If $\alpha$ is not a pole of $X$, and is such that :
\beq
\zeta_\alpha = (X-X(\alpha))^{1/d_\alpha} \qquad d_\alpha=\operatorname{order}_{\alpha} X-X(\alpha).
\eeq
\end{itemize}

It is well known that 
\bp
There is a 1-1 correspondence between the set of coefficients $P_{i,j}$ such that $(i,j)\in \dNewt$, and the independent times $t_{\alpha,k}$:
\beq
\CC[\dNewt] \sim \CC^{-1\sum_\alpha (m_\alpha+1)}.
\eeq
\ep
In other words, fixing the non-interior coefficients $P_{i,j}$, is equivalent to having fixed the times.

\subsection{Times and exterior coefficients}
\label{apx:reconstructionP}

\bp[Recovering the polynomial $P$ from the  times]
\label{prop:reconstructionP}

First, for each puncture $\alpha$, define a polynomial $P_{\alpha} \in \CC[x,y]$ as follows:

$\bullet$ If $a_\alpha>0$, let $j\in [1,a_\alpha]$ and let
\beq
P_\infty(x,y) = \prod_{\alpha \in X^{-1}(\infty)}  P_{\alpha}(x,y),
\eeq
where
\beq
P_\alpha(x,y):=  \prod_{j=1}^{a_\alpha}  P_{\alpha,j}(x,y),\quad P_{\alpha,j}(x,y):=  \left(y+\sum_{k=0}^{r_\alpha} \frac{t_{\alpha,k}}{a_\alpha} x^{\frac{k}{a_\alpha}-1} \ e^{2\pi\ii \frac{jk}{a_\alpha}}\right).
\eeq

$\bullet$ If $a_\alpha<0$, let $j\in [1,|a_\alpha|]$ and let
\beq
P_{\alpha,j}(x,y):=  \left(y+\sum_{k=0}^{r_\alpha} \frac{t_{\alpha,k}}{a_\alpha} (x-X(\alpha))^{\frac{-k}{a_\alpha}-1} \ e^{2\pi\ii \frac{jk}{a_\alpha}}\right), \quad
P_\alpha(x,y):=  \prod_{j=1}^{|a_\alpha|}  P_{\alpha,j}(x,y)
\eeq
Then, let 
\beq
P_{X_\alpha}(x) = -y^{d} +  \prod_{\alpha\in X^{-1}(X_\alpha)} P_\alpha(x,y) .
\eeq

We have
\beq
P(x,y) 
 = D(x) \Big( P_\infty(x,y) + \sum_{\alpha, \ a_\alpha<0} P_\alpha(x,y)  \Big)  + \CC[\Newtint].
\eeq
where 
\beq
D(x) = \prod_{\alpha, \ X_\alpha\neq \infty}(x-X_\alpha)^{b_\alpha}.
\eeq

\ep

This implies
\bc
The non-interior coefficients $P_{i,j}$ of $P$, are polynomials of the times.

Vice-versa, the times are algebraic functions of the non-interior coefficients of $P$.
\ec

\proof{
Let
\beq
\td P(x,y) = \prod_{k=1}^d (y-Y_k(x)) = y^d + \sum_{l=1}^d (-1)^l y^{d-l} e_l(Y_1(x),\dots,Y_d(x)),
\eeq
where $Y_k(x)$ are the zeros of $P(x,y)=0$, and $e_l$ are the elementary symmetric polynomials:
\beq
\td P_l(x) = e_l(Y_1(x),\dots,Y_d(x)) = \sum_{1\leq i_1<\dots<i_l\leq d} Y_{i_1}(x)\dots Y_{i_l}(x).
\eeq

$\td P_l(x)\in \CC(x)$ is a rational function of $x$, and it can have poles only where at least one of the $Y_{i_j}(x)$ has a pole, i.e. only if $x=X(\alpha)$ for some puncture $\alpha$.

We can decompose $\td P_l(x)$ into simple elements:
\beq
\td P_l(x) = \td P_{\infty,l}(x) + \sum_{q\in X(\text{finite punctures})} \td P_{q,l}(x)
\eeq
where $\td P_{\infty,l}(x)\in \CC[x]$ is a polynomial, and if $q\neq\infty$ $\td P_{q,l}(x)\in \CC[1/(x-q)]$ is a rational function with poles only at $x=q$ or equivalently a polynomial of $1/(x-q)$ with no constant term.

$\bullet$ Consider $q=\infty$, and let $\alpha\in X^{-1}(\infty)$.
For each such $\alpha$ we have
\beq
Y\sim \left( -\sum_{k=0}^{r_\alpha} \frac{t_{\alpha,k}}{a_\alpha} x^{\frac{k}{a_\alpha}-1}   + O(x^{-1-1/a_\alpha}) \right)
\eeq
This implies
\bea
\td P_{\infty,l}(x) 
&=& e_l\left(\left( -\sum_{k=0}^{r_\alpha} \frac{t_{\alpha,k}}{a_\alpha} x^{\frac{k}{a_\alpha}-1} e^{2\pi\ii \frac{jk}{a_\alpha}}  + O(x^{-2}) \right)_{\alpha\in X^{-1}(\infty), \ j=1,\dots,a_\alpha} \right) \cr
&=& e_l\left(\left( -\sum_{k=0}^{r_\alpha} \frac{t_{\alpha,k}}{a_\alpha} x^{\frac{k}{a_\alpha}-1} e^{2\pi\ii \frac{jk}{a_\alpha}}  \right)_{\alpha\in X^{-1}(\infty), \ j=1,\dots,a_\alpha} \right)  (1+O(x^{-2})) \cr
\eea
The  $O(x^{-2})$ are necessarily powers of $x$ that are at least 2 less than the highest power, i.e. they correspond to point in the Newton's polygon that are strictly to the left of the convex envelope, they are interior points.
This means that, up to interior points we can replace $\td P_\infty$ by $P_\infty$.

$\bullet$ We redo the same for finite poles.

$\bullet$ eventually we multiply by the common denominator $D(x)$ so that $\td P(x) D(x)$ is a polynomial of $x$.
This gives
\beq
D(x)\td P(x) = D(x) \left( P_\infty(x,y)+\sum_{\alpha,a_\alpha<0} P_\alpha(x,y)\right) \ \ \mod \CC[\Newtint].
\eeq
}
\section{Integrals over small circles}

\label{apx:lemmaintCalpha}

\bl
\label{lem:intCalpha} We have
\beq
\frac{1}{2\pi\ii} \int_{\mathcal C_\alpha} \overline{g_\alpha} \ Y dX
=
- \sum_{k=1}^{r_\alpha} \frac{|t_{\alpha,k}|^2}{k} R_\alpha^{-2k}
+ \sum_{k=1}^{\infty} k |\td t_{\alpha,k}|^2 R_\alpha^{2k} + 2 t_{\alpha,0}^2 \ln R_\alpha  -t_{\alpha,0} g_\alpha(p_\alpha)  +\pi \ii t_{\alpha,0}^2
\eeq
where $\mathcal C_\alpha$ is the circle $\zeta_\alpha = R_\alpha e^{\ii \theta}$ with $\theta\in ]-\pi,\pi]$, and  $p_\alpha$ is the point of coordinate $\zeta_\alpha=R_\alpha e^{+\ii \pi}$.

\el

\proof{
We use $\zeta_\alpha = R_\alpha e^{\ii \theta}$ with $\theta\in ]-\pi,\pi]$, and
\beq
YdX = \sum_{k=0}^{r_\alpha} t_{\alpha,k} \zeta_\alpha^{-k-1} d\zeta
_\alpha + \sum_{k=1}^\infty k \td t_{\alpha,k} \zeta_\alpha^{k-1} d\zeta _\alpha
\eeq
and therefore
\beq
g_\alpha = - \sum_{k=1}^{r_\alpha} \frac{t_{\alpha,k}}{k} \zeta_\alpha^{-k} + t_{\alpha,0} \ln \zeta_\alpha
+ \sum_{k=1}^\infty \td t_{\alpha,k} \zeta_\alpha^{k} .
\eeq
This gives
\bea
&& \frac{1}{2\pi\ii} \int_{\mathcal C_\alpha} \overline{g_\alpha} \ Y dX  \cr
&=& \frac{1}{2\pi} \int_{-\pi}^\pi 
\Big( -\sum_{k=1}^{r_\alpha} \frac{\overline{t_{\alpha,k}}}{k} R_\alpha^{-k} e^{\ii k \theta} + t_{\alpha,0} (\ln R_\alpha -i\theta) + \sum_{k=1}^\infty \overline{\td t_{\alpha,k}} R_\alpha^k e^{-\ii k \theta}  \Big) \cr
&&  \Big( \sum_{k=1}^{r_\alpha} t_{\alpha,k} R_\alpha^{-k} e^{-\ii k \theta} + t_{\alpha,0}  + \sum_{k=1}^\infty k \td t_{\alpha,k} R_\alpha^k e^{\ii k \theta}  \Big)  d\theta \cr
&=& 
- \sum_{k=1}^{r_\alpha} \frac{|t_{\alpha,k}|^2}{k} R_\alpha^{-2k}
+ \sum_{k=1}^{\infty} k |\td t_{\alpha,k}|^2 R_\alpha^{2k} + t_{\alpha,0}^2 \ln R_\alpha \cr
&& - \frac{\ii t_{\alpha,0} }{2\pi} \int_{-\pi}^\pi 
  \Big( \sum_{k=1}^{r_\alpha} t_{\alpha,k} R_\alpha^{-k} e^{-\ii k \theta} + t_{\alpha,0}  + \sum_{k=1}^\infty k \td t_{\alpha,k} R_\alpha^k e^{\ii k \theta}  \Big)  \theta d\theta \cr
&=& 
- \sum_{k=1}^{r_\alpha} \frac{|t_{\alpha,k}|^2}{k} R_\alpha^{-2k}
+ \sum_{k=1}^{\infty} k |\td t_{\alpha,k}|^2 R_\alpha^{2k} + t_{\alpha,0}^2 \ln R_\alpha \cr
&& - \frac{ t_{\alpha,0} }{2\pi} \int_{-\pi}^\pi 
\theta  \ d \Big( \sum_{k=1}^{r_\alpha} \frac{-1}{k} t_{\alpha,k} R_\alpha^{-k} e^{-\ii k \theta} + t_{\alpha,0} (\ln R_\alpha +\ii\theta)+ \sum_{k=1}^\infty  \td t_{\alpha,k} R_\alpha^k e^{\ii k \theta}  \Big)   \cr
&=& 
- \sum_{k=1}^{r_\alpha} \frac{|t_{\alpha,k}|^2}{k} R_\alpha^{-2k}
+ \sum_{k=1}^{\infty} k |\td t_{\alpha,k}|^2 R_\alpha^{2k} + t_{\alpha,0}^2 \ln R_\alpha \cr
&& - \frac{ t_{\alpha,0} }{2\pi} \int_{-\pi}^\pi 
\theta  \ d g_\alpha(R_\alpha e^{\ii\theta})   \cr
&=& 
- \sum_{k=1}^{r_\alpha} \frac{|t_{\alpha,k}|^2}{k} R_\alpha^{-2k}
+ \sum_{k=1}^{\infty} k |\td t_{\alpha,k}|^2 R_\alpha^{2k} + t_{\alpha,0}^2 \ln R_\alpha \cr
&& - \frac{ t_{\alpha,0} }{2\pi} (\pi g_\alpha(p_\alpha) + \pi (g_\alpha(p_\alpha)-2\pi\ii t_{\alpha,0}) )
+ \frac{ t_{\alpha,0} }{2\pi} \int_{-\pi}^\pi  g_\alpha(R_\alpha e^{\ii\theta})  d\theta \cr
&=& 
- \sum_{k=1}^{r_\alpha} \frac{|t_{\alpha,k}|^2}{k} R_\alpha^{-2k}
+ \sum_{k=1}^{\infty} k |\td t_{\alpha,k}|^2 R_\alpha^{2k} + t_{\alpha,0}^2 \ln R_\alpha \cr
&& -t_{\alpha,0} g_\alpha(p_\alpha) +\pi \ii t_{\alpha,0}^2
+ \frac{ t_{\alpha,0}^2 }{2\pi} \int_{-\pi}^\pi  (\ln R_\alpha+\ii \theta)  d\theta \cr
&=& 
- \sum_{k=1}^{r_\alpha} \frac{|t_{\alpha,k}|^2}{k} R_\alpha^{-2k}
+ \sum_{k=1}^{\infty} k |\td t_{\alpha,k}|^2 R_\alpha^{2k} + 2 t_{\alpha,0}^2 \ln R_\alpha  -t_{\alpha,0} g_\alpha(p_\alpha) +\pi \ii t_{\alpha,0}^2
\eea
}

\printbibliography

@article{gaiotto2011wallcrossing,
  title={Wall-crossing, Hitchin systems, and the WKB approximation},
  author={Gaiotto, Davide and Moore, Gregory W and Neitzke, Andrew},
  eprint = "0907.3987",
    archivePrefix = "arXiv",
    primaryClass = "hep-th",
    doi = "10.1016/j.aim.2012.09.027",
    journal = "Adv. Math.",
    volume = "234",
    pages = "239--403",
    year = "2013"
}

@article{deift1992steepest,
  title={A steepest descent method for oscillatory Riemann-Hilbert problems},
  author={Deift, Percy and Zhou, Xin},
  journal={Bulletin of the American Mathematical Society},
  volume={26},
  number={1},
  pages={119--123},
  year={1992},
  eprint={math/9201261},
      archivePrefix={arXiv},
      primaryClass={math.AP}
}

@article{deift1999,
title={Uniform asymptotics for polynomials orthogonal with respect to varying exponential weights and applications to universality questions in random matrix theory},
  author={Deift, Percy and Kriecherbauer, Thomas and McLaughlin, K T-R and Venakides, Stephanos and Zhou, Xin},
  journal={Communications on Pure and Applied Mathematics},
  volume={52},
  number={11},
  pages={1335--1425},
  year={1999},
  publisher={Wiley Online Library},
doi = {10.1002/(SICI)1097-0312(199911)52:11<1335::AID-CPA1>3.0.CO;2-1},
}

@misc{eynardlecturesRS,
      title={Lectures notes on compact Riemann surfaces}, 
      author={Bertrand Eynard},
      year={2018},
      eprint={1805.06405},
      archivePrefix={arXiv},
      primaryClass={math-ph}
}

@book{farkas2012riemann,
  title={Riemann Surfaces},
  author={Farkas, H.M. and Kra, I.},
  isbn={9781468499308},
  lccn={79024385},
  series={Graduate Texts in Mathematics},
  url={https://books.google.fr/books?id=6pXuBwAAQBAJ},
  year={2012},
  publisher={Springer New York}
}

@book{fay1973theta,
  title={Theta Functions on Riemann Surfaces},
  author={Fay, J.D.},
  isbn={9780387065175},
  lccn={lc73015292},
  series={Lecture Notes in Mathematics},
  url={https://books.google.fr/books?id=-i3vAAAAMAAJ},
  year={1973},
  publisher={Springer}
}

@book{TataLectures,
  title={Tata Lectures on Theta },
  author={Mumford, D.},
  
  series={Modern Birkhäuser Classics },
 volume={I (no. 28), II (no. 43), III (no. 97)},
  publisher={Birkhäuser Boston, MA},
year={2007}
}

@article{EO07,
      title={Invariants of algebraic curves and topological expansion}, 
      author={Bertrand Eynard and Nicolas Orantin},
      eprint={math-ph/0702045},
      archivePrefix={arXiv},
      primaryClass={math-ph},
    reportNumber = "SPHT-07-021",
    doi = "10.4310/CNTP.2007.v1.n2.a4",
    journal = "Commun. Num. Theor. Phys.",
    volume = "1",
    pages = "347--452",
    year = "2007"
}

@misc{E2017,
      title={The Geometry of integrable systems. Tau functions and homology of Spectral curves. Perturbative definition}, 
      author={Bertrand Eynard},
     eprint = "1706.04938",
    archivePrefix = "arXiv",
    primaryClass = "math-ph",
    reportNumber = "IPhT-T17/086, CRM-3360, IPHT-T17-086",
    month = "6",
    year = "2017"
}

@article{BAG97,
    title={Large deviations for Wigner's law and Voiculescu's non-commutative entropy},
  author={Arous, G Ben and Guionnet, Alice},
  journal={Probability theory and related fields},
  volume={108},
  pages={517--542},
  year={1997},
  publisher={Springer},
doi={10.1007/s004400050119},
issn={1432-2064}
}

@article{Kon92,
	author = {Kontsevich, Maxim},
	title = {{Intersection theory on the moduli space of curves and the matrix Airy function}},
	journal = {Comm. Math. Phys.},
	fjournal = {Communications in Mathematical Physics},
	volume = {147},
	year = {1992},
	number = {1},
	pages = {1--23},
	doi = {10.1007/BF02099526},
}

@book{Thurston,
author = {W.~Thurston},
booktitle = {The geometry and topology of $3$-manifolds},
year = {2002},
note = {electronic version \href{http://www.msri.org/publications/books/gt3m/}{\textsf{http://www.msri.org/publications/books/gt3m/}}}
}

@book{Strebel,
author="Strebel, Kurt",
title="Quadratic Differentials",
bookTitle="Quadratic Differentials",
year="1984",
publisher="Springer Berlin Heidelberg",
address="Berlin, Heidelberg",
pages="16--26",
isbn="978-3-662-02414-0",
doi="10.1007/978-3-662-02414-0_2",

}

@article{Zagier1986,
title={The Euler characteristic of the moduli space of curves.},
  author={Harer, John and Zagier, Don},
  journal={Inventiones mathematicae},
  volume={85},
  pages={457--485},
  year={1986},
  publisher={Springer-Verlag, Berlin},
doi={10.1007/BF01390325},
issn={1432-1297}
}

@article{penner2003decorated,
      title={Decorated Teichm\"uller Theory of Bordered Surfaces}, 
      author={R. C. Penner},
      year={2003},
      eprint={math/0210326},
      archivePrefix={arXiv},
      primaryClass={math.GT}
}

@article{penner2003cell,
      title={Cell decomposition and compactification of Riemann's moduli space in decorated Teichm\"uller theory}, 
      author={R. C. Penner},
      year={2003},
      eprint={math/0306190},
      archivePrefix={arXiv},
      primaryClass={math.GT}
}

@article{Matytsin1994,
	doi = {10.1016/0550-3213(94)90471-5},
	
	year = {1994},
	month = Jan,
	publisher = {Elsevier {BV}},
	volume = {411},
	number = {2-3},
	pages = {805--820},
	author = {A. Matytsin},
	title = {On the large-N limit of the Itzykson-Zuber integral},
	journal = {Nuclear Physics B}
}

@article{Guionnet2002LargeDA,
  title={Large Deviations Asymptotics for Spherical Integrals},
  author={Alice Guionnet and Ofer Zeitouni},
  journal={Journal of Functional Analysis},
  year={2002},
  volume={188},
  pages={461-515},
year = {2002},
issn = {0022-1236},
doi = {10.1006/jfan.2001.3833},
}

@article{bertola2007boutroux,
      title={Boutroux curves with external field: equilibrium measures without a minimization problem}, 
      author={Marco Bertola},
      year={2007},
      eprint={0705.3062},
      archivePrefix={arXiv},
      primaryClass={nlin.SI}
}
\end{document}